\documentclass[letterpaper,11pt]{article}
\pdfoutput=1

\usepackage{jheppub} 

\usepackage[printonlyused]{acronym}

\usepackage{color}

\usepackage{verbatim}
\usepackage{amsmath}
\usepackage{subcaption}
\usepackage{physics}
\usepackage{amssymb}
\usepackage{multirow}
\usepackage{slashed}

\newcommand{\be}{\begin{equation}}
\newcommand{\ee}{\end{equation}}

\renewcommand{\arraystretch}{1.3}

\usepackage{color}
\definecolor{mit-red}{rgb}{0.64,.12,0.2}
\definecolor{darkred}{rgb}{1.0,0.1,0.1}
\definecolor{darkgreen}{rgb}{0.1,0.7,0.1}
\definecolor{darkblue}{rgb}{0.1,0.1,1.0}

\DeclareRobustCommand{\Fig}[1]{Fig.~\ref{#1}}

\DeclareRobustCommand{\Eq}[1]{Eq.~(\ref{#1})}

\allowdisplaybreaks

\begin{document}

\count\footins = 1000
\interfootnotelinepenalty=10000
\setlength{\footnotesep}{0.6\baselineskip}

\renewcommand{\arraystretch}{1.3}

\title{The New Physics Case for Beam-Dump Experiments with Accelerated Muon Beams}

\author{Cari Cesarotti,}
\author{Rikab Gambhir}

\affiliation{Center for Theoretical Physics, Massachusetts Institute of Technology,\\ Cambridge, MA 02139, USA}

\emailAdd{ccesar@mit.edu}
\emailAdd{rikab@mit.edu}

\preprint{MIT--CTP 5606}

\abstract{
As the field examines a future muon collider as a possible successor to the LHC, we must consider how to fully utilize not only the high-energy particle collisions, but also any lower-energy staging facilities necessary in the R\&D process. 
An economical and efficient possibility is to use the accelerated muon beam from either the full experiment or from cooling and acceleration tests in beam-dump experiments.
Beam-dump experiments are complementary to the main collider as they achieve sensitivity to very small couplings with minimal instrumentation.
We demonstrate the utility of muon beam-dump experiments for new physics searches at energies from 10 GeV to 5 TeV. 
We find that, even at low energies like those accessible at staging or demonstrator facilities, it is possible to probe new regions of parameter space for a variety of generic BSM models, including muonphilic, leptophilic, $L_\mu - L_\tau$, and dark photon scenarios. 
Such experiments could therefore provide opportunities for discovery of new physics well before the completion of the full multi-TeV collider. 
}
\maketitle

\section{Introduction}
\label{sec:intro}
%
While there has been enormous progress in understanding the Standard Model (SM) over the past few decades, we have yet to directly identify any physics beyond the SM (BSM), despite evidence seen at colliders, cosmology, and beyond. 
It is therefore a priority of the field to work towards discovering the fundamental nature of BSM phenomena. 
Currently, there is no single leading model for the particle content of BSM sectors. 
Thus it is lucrative to consider robust, quasi-model-independent strategies to maximize our discovery potential.
This approach requires not only theoretical advancements, but also the construction of novel experiments. 

In order to unambiguously quantify the particle content of new physics, we need the clean laboratory environment of colliders. 
While the LHC has provided many high-quality data sets---with more to come in the future high-luminosity runs---now is the time to develop the physics case and technical plan for its successor. 
From the strong community support presented at Snowmass 2021 \cite{Narain:2022qud, Maltoni:2022bqs, Aime:2022flm, MuonCollider:2022xlm}, a promising option is a future multi-TeV muon collider (MuC) \cite{Antonelli:2015nla, Long:2020wfp, Delahaye:2019omf, Delahaye:2013jla, Accettura:2023ked, Buttazzo:2018qqp,MuonCollider:2022nsa}.

Such a machine is advantageous due to several properties of the muon: its fundamental structure and its mass.
Firstly, the muon is a fundamental particle, unlike the composite hadrons accelerated at the LHC and Tevatron.
This would enable us to make unprecedented precision measurements in the electroweak and Higgs sector specifically due to the relatively low background, absence of QCD charge, and energy resolution of the beam particles \cite{Han:2021udl, Chakrabarty:2014pja, Kalinowski:2020rmb, Rodejohann:2010jh, Li:2023ksw, Han:2020uid, Han:2020pif, Forslund:2022xjq, Han:2020uak, Bottaro:2021snn, Bottaro:2022one, Asadi:2021gah}. 
Moreover, unlike with composite particles, collisions of fundamental particles have nearly all the center-of-mass energy available, and much higher $\sqrt{s}$ can be achieved without the PDF penalty \cite{AlAli:2021let}.

Thus far we have only collided electrons at energies of several hundred GeV, as linear colliders lose luminosity at high energies and circular colliders generate insurmountable power losses through synchrotron radiation. 
However, we can potentially overcome these issues by using muons in a circular collider. 
The mass of a muon is roughly two-hundred times heavier than the electron, which suppresses the synchrotron radiation emissions (proportional to $m^{-4}$) by a factor of a billion. 
This allows us to accelerate muons to energies orders of magnitude higher than what is achievable with electrons while maintaining high luminosity. 
A future muon collider would therefore enable the expansion of both the precision and energy frontier.
For a more detailed review about the physics potential of a muon collider, see Ref.~\cite{Accettura:2023ked}. 

It must be stated that with the novelty of a future muon collider come nontrivial challenges. 
Since we have never attempted to construct a full-scale muon collider, the necessary technology is not fully matured, and serious R\&D efforts are needed. 
One of the primary accelerator challenges includes focusing the muons into a single collimated bunch before acceleration, known as 6d cooling \cite{MICE:2019jkl, Neuffer:1994cza}. 
Muons are produced as tertiary particles: a proton beam collides with a target to produce copious pions which then decay into muons with a broad distribution in phase space. 
The proton energy for optimal muon extraction is around 5 to 10 GeV \cite{Daniele2023}, thus the muons have boost factors of $\gamma \sim 10$ and decay on the timescale of microseconds. 
Similar problems do not arise with protons and electrons as they are stable particles and easily produced.
However, considerable advances in accelerator and detector technology have been made in the past ten years (when MuC funding was last discussed) that strongly suggests now is the time to invest in R\&D \cite{MuonCollider:2022glg, MuonCollider:2022ded, Mice:2023bdb}. 
The R\&D program necessary to demonstrate the feasibility of the full scale MuC must therefore include the cooling and partial acceleration of a muon beam. 
While any complete accelerator complex will likely require at least 25 years to come online, the R\&D could begin in the next 5-10 years and could utilize existing infrastructure at laboratories like CERN, Fermilab, or ESS \cite{Baussan:2022fer}.
These dedicated moderate energy and luminosity facilities provide an excellent opportunity for physics studies along the way. 

In this paper we illustrate the utility of beam-dump experiments with muon beams at energies of $E_{0} \sim 10$ GeV---5 TeV.
It is imperative to consider auxiliary experiments that can run in parallel to the main collider and R\&D studies to ensure that we are maximizing the physics output of a future muon collider program.
We consider this energy range and several benchmarks to illustrate the new physics reach for a variety of scales that are relevant for a cooling and acceleration program, a moderate-energy staging facility, and the full collider energy scales.
As there does not yet exist a definitive roadmap for the staging and full construction of the MuC, all of these numbers of subject to change and should be taken as order-of-magnitude projections.

Beam dumps are cost-efficient experiments as well as complementary physics probes compared to high-energy colliders.
These experiments are relatively low cost as they only require a single beam and minimal instrumentation. 
They are particularly economical when the beam is already being recycled from other physics purposes, such as being dumped from the main collider or after 6d cooling and acceleration tests. 
The idea of including beam-dump experiments at high-energy future colliders has previously been explored in the literature \cite{Bjorken:2009mm, Kanemura:2015cxa, Chen:2017awl, Sakaki:2020mqb, Asai:2021xtg, Sieber:2021fue}, in particular at muon colliders for a narrow set of energies and physics models \cite{Cesarotti:2022ttv}. 
In this work we will expand upon the results in Ref.~\cite{Cesarotti:2022ttv} by considering the physics reach at lower, more readily accessible energies that can be probed with test and staging facilities, as well as a wider variety of generic models. 
Note that this work is a proof-of-concept study, meant to demonstrate the possibility of immediate experiments on the way to the 10 TeV collider, rather than a fully-optimized technical document for future experiments. 

A compelling reason to build these experiments is not only the relatively low-cost, but also the novel physics reach. 
The community has expressed significant excitement towards exploring moderately low-mass, weakly-coupled new particles. 
These particles could mediate interactions between dark matter (or more generic hidden sectors) and the SM, and are largely unconstrained by other experiments (see Ref.~\cite{Gori:2022vri} as a recent summary). 
With the proposed high-energy second-generation particle beam, we could probe beyond similar experiments for a variety of models, including muonphilic forces, leptophilic scalars, gauged $L_\mu-L_\tau$, and generic dark photons. 
The discovery of a moderate-mass new particle consistent with any of these models could have potential relevance to many open questions in the SM.

The outline of this paper is as follows:
In Sec.~\ref{sec:fund}, we revisit the basics of the analytic beam-dump sensitivity calculation, including the experimental layout, approximation scheme, and new physics models of interest.
In Sec.~\ref{sec:csc} we present the results of our calculations with a variety of models and experimental configurations.
The sensitivity projections are shown in Sec.~\ref{sec:results}.
Finally, we conclude in Sec.~\ref{sec:conc}.
Additionally we discuss more details of target material properties in Appendix \ref{app:materials}, and many supplemental figures for alternative experimental configurations can be found in Appendix \ref{app:plots}. 

\section{Beam-Dump Fundamentals}
\label{sec:fund}
In this section we review how beam-dump experiments are conducted and how to estimate the number of signal events. 
This section is presented for completeness and to define a consistent vocabulary for the remainder of the paper, but readers familiar with the literature and calculation can skip to Sec.~\ref{sec:csc}.
\subsection{Experimental Design}
\begin{figure}
\begin{center}
\includegraphics[width=0.8\textwidth]{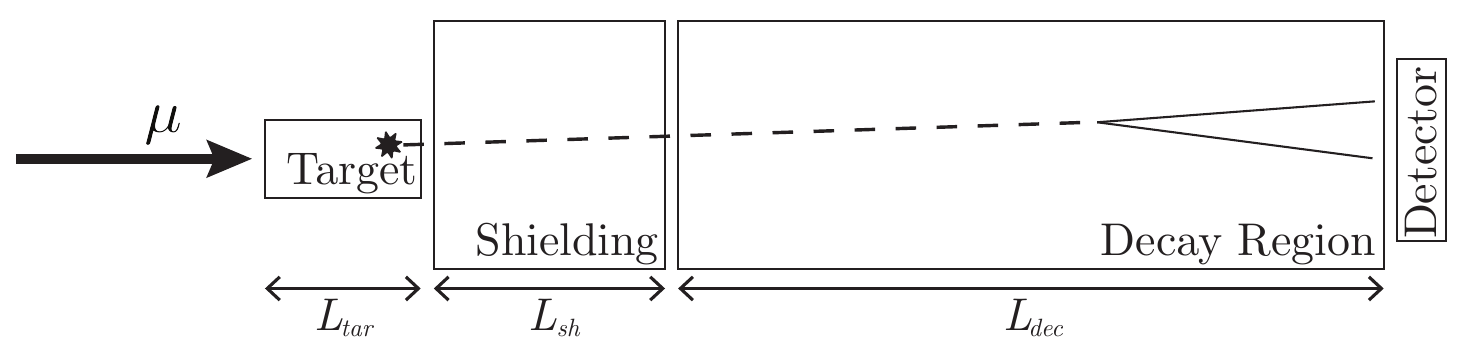}
\end{center}
\caption{A cartoon of the layout of a beam-dump experiment. The incoming beam strikes a dense target. At some point inside the target a new particle may be emitted which propagates to the fiducial region where it decays into visible final states. The new particle must be sufficiently long lived to escape not only the target, but also a shielding region that removes SM events from detection such that the experiment can run with near zero background. }
\label{fig:FTschem}
\end{figure}

We begin by discussing the generic set-up of a beam-dump experiment. 
A schematic diagram of the layout is given in Fig.~\ref{fig:FTschem}.
In the experimental setup of interest, a muon beam strikes a dense material target of length $L_\text{tar}$ such as water or lead. 
As the beam traverses the material, the muon may radiate a particle via the diagram shown in \Fig{fig:2to3}.
Beyond the target is the shielding region of length $L_\text{sh}$ that removes any residual SM particles out of the acceptance of the detector. 
The shielding region includes a powerful magnetic field to deflect the remaining high-energy beam as well as a passive dense material target to stop light particles. 
Note that the residual beam will be truly dumped at this stage, and passive shielding will also be needed to ensure backgrounds from this dump do not interfere with the experiment.
Beyond the shielding is the fiducial decay region of length $L_\text{dec}$ which must be sufficiently long for a majority of the new physics particles to decay back to SM states. 
At the end of the decay region there is a simple detector, such as a tracker or calorimeter, to identify and reconstruct a new physics signal. 

The reach of a beam dump is complementary to that of a high-energy collider.
Instead of colliding two ultra-relativistic particles, a relativistic particle with energy $E_0$ scatters off a nucleon at rest with mass $M$, and the available energy is much less than the nominal beam energy ($\sqrt{\hat{s}}\sim \sqrt{2 E_0 M}$). 
However, because of the density of particles in the target and the extended length of the experimental hall, this set-up can directly probe much smaller couplings than what is accessible at the main collider. 
The reach of these experiments is roughly determined by the size of the experiment, energy of the beam, and the available luminosity: the size of the experiment indirectly sets the sensitivity to couplings as the lifetime of the particle must be roughly the size of the decay region, and the number of signal events produced scales linearly with the luminosity. 
In this work we present order-of-magnitude suggestions for a future experiment, but these parameters must be optimized for the specific experimental considerations. 

We consider a class of new physics scenarios with a single additional boson that couples directly to muons. 
We explore the details of several models: vector, axial vector, scalar, and pseudoscalar new particles. 
To estimate the sensitivity to these models, we consider the lifetime of the particles in the lab frame using the approximate width $\Gamma = \tau^{-1} \sim g^2 m_\text{NP}$:
\begin{equation}
\begin{aligned}
l_\text{NP} &= \gamma \tau_0 \approx \frac{E_0}{m_\text{NP}} \times \frac{1}{g^2 m_\text{NP}}\\
& \approx \left( \frac{E_0}{\text{TeV}} \right) \times \left( \frac{g}{10^{-6}}\right)^{-2} \times \left( \frac{m_\text{NP}}{10 \text{ MeV}}\right)^{-2} \times 100 \text{m}
\end{aligned}
\end{equation}
where $E_0$ is the energy of the incoming beam, $g$ is the coupling, and $m_\text{NP}$ is the mass of the new particle. 
Note that for a lower-energy beam, we can still potentially probe small couplings with a much more compact experiment. 

For this work we consider a visible dilepton signal ($e^+e^-$ or $\mu^+\mu^-$) which can be achieved with some tracking and calorimetry instrumentation. 
While other experiments and proposals have highlighted other methods of detection, such as missing transverse energy, diphotons, and more (e.g. \cite{Bjorken:2009mm, Chen:2017awl, Berlin:2018bsc, Gori:2022vri}), we restrict our interest to the visible signal for simplicity and ease of detection and analysis.

\subsection{Cross Section Estimation with Weiz\"acker-Williams}
In this section, we will follow the prescriptions of Refs.~\cite{Bjorken:2009mm, Liu:2016mqv, Liu:2017htz} to calculate the new physics production cross sections as well as rederive their results. 
Our results match these previous canonical references in the $m_\mu \rightarrow 0$ limit, but we keep additional terms proportional to $m_\mu$ (compared to the electron mass as in the previous references) to be consistent in our approximation scheme.

\begin{figure}
\begin{center}
\includegraphics[width=0.7\textwidth]{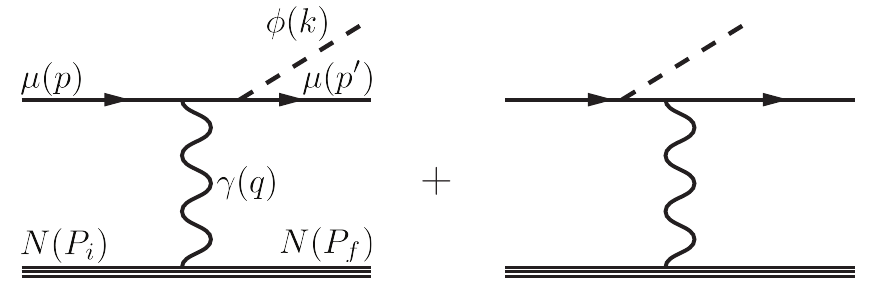}
\end{center}
\caption{Dominant production process of a new particle emitted from the beam on target. A virtual photon is exchanged between the incoming beam particle and a nucleus in the target to induce emission.}
\label{fig:2to3}
\end{figure}

\begin{figure}
\begin{center}
\includegraphics[width=0.7\textwidth]{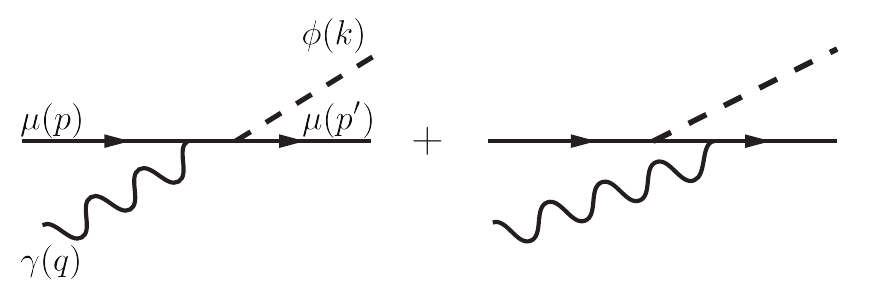}
\end{center}
\caption{The effective 2-to-2 scattering process of \Fig{fig:2to3}. In the limit of an approximately on-shell photon, we can treat these processes as equivalent.}
\label{fig:2to2}
\end{figure}

The dominant production mechanism for all of the new physics models is shown in Fig.~\ref{fig:2to3}. 
In this process, a virtual photon is exchanged from the nucleon to the beam muon, and the muon radiates a new particle $\phi$:
\begin{equation}
\mu(p) + N (P_i) \rightarrow \mu (p') + N(P_f) + \phi (k).
\label{eq:2to3}
\end{equation}
Since calculating the phase space of a 2-to-3 process is difficult and expensive, we instead implement the Weizs\"acker-Williams approximation \cite{Fermi:1924tc, PhysRevD.8.3109, PhysRevD.34.1326, RevModPhys.46.815} where valid. 
In this approximation, a sufficiently relativistic charged particle emits a cloud of on-shell transverse radiation, such that we can effectively reduce the two-to-three process to a simpler 2-to-2 scattering process (shown in Fig.~\ref{fig:2to2})
\begin{equation}
\mu(p) + \gamma (q) \rightarrow \mu (p') + \phi (k)
\label{eq:2to2}
\end{equation}
with the modification of a nuclear form factor on the cross section evaluated at the minimum virtuality of the photon ($t=t_\text{min}$). 
The two differential cross sections are related by \cite{PhysRevD.34.1326, Bjorken:2009mm}
\begin{equation}
\left( \frac{d\sigma_{2\rightarrow3}(p+P_i \rightarrow p' + k + P_f)}{d (p\cdot k) d (k\cdot P_i)} \right)_{\text{WW}} = \left( \frac{\alpha \chi}{\pi} \right) \left( \frac{E_0 x \beta_{\phi}}{1-x}\right) \times 
\left( \frac{d \sigma_{2\rightarrow2} (p + q \rightarrow p' + k) }{d (p\cdot k)} \right)_{t=t_\text{min}} \quad
\end{equation}
where $\alpha \chi / \pi $ is the effective photon flux, $\beta_\phi$ is the standard velocity of $\phi$, and $x \equiv E_\phi / E_0$ is the fraction of energy carried away by the new particle $\phi$. 
The four-momenta are defined in Fig.~\ref{fig:2to3}.
We can write this in a more illuminating form involving the amplitude of the 2-to-2 process as shown in Ref.~\cite{Liu:2016mqv,Liu:2017htz}
\begin{equation}
\left(\frac{d \sigma_{2\rightarrow 3}}{dx d\cos\theta} \right)_\text{WW}=  \frac{g^2}{2\pi} \alpha |\vec{k}| E (1-x) \frac{\mathcal{A}^{22}_{t=t_\text{min}}}{\tilde{u}^2} \chi
\label{eq:WWxsec}
\end{equation}
where $g$ is the coupling, $\mathcal{A}^{22}$ is the 2-to-2 scattering amplitude, and $\tilde u$ is defined with the Mandelstam variable $u$ in the 2-to-2 process
\begin{equation}
\tilde{u} \equiv u - m_\mu^2 = (p - k)^2 - m_\mu^2.
\label{eq:defUtilde}
\end{equation}

The value of $\chi$ used to parameterize the photon flux is computed by integrating the form factors from $t=t_\text{min}$ to $t_\text{max}$ \cite{PhysRevD.8.3109, RevModPhys.46.815, Bjorken:2009mm}
\begin{equation}
\chi \equiv \int_{t_\text{min}}^{t_\text{max}} dt \frac{t-t_\text{min}}{t^2} G(t)
\end{equation}
where the general electric form factor $G(t)$ can be broken into the elastic and inelastic scattering pieces $G(t) = G_\text{in}(t) + G_\text{el}(t)$. 
The form factor is often dominated by the elastic component which scales as the atomic number squared $Z^2$, whereas the inelastic scales linearly. 
The formulas used for evaluating $\chi$ and further details are given in Appendix A of Ref.~\cite{Bjorken:2009mm}.

As $t$ is a function of both the angle of emission $\theta$ and energy fraction $x$, $\chi$ is also dependent on these quantities. 
However, in the regime where both the beam particle and radiated new particle are relativistic, the characteristic angle of emission $\theta_0$ as well as $t_\text{min}$, $t_\text{max}$ become independent of $x$. 
In this limit, $t_\text{min}$ and $t_\text{max}$ are also approximately independent of the angle $\theta$ \cite{Bjorken:2009mm, Liu:2016mqv} which is parametrically small:
\begin{equation}
t_\text{min} \approx \left(\frac{m_\phi^2}{2E} \right)^2 \qquad \qquad t_\text{max} \approx m_\phi^2 + m_\mu^2
\label{eq:IWWt}
\end{equation}

Once the angular and $x$ dependence is removed from the integral bounds of $\chi$, it can be pulled out of the integrand. 
The relationship in Eq.~\ref{eq:WWxsec} is further simplified to 
\begin{equation}
\left(\frac{d \sigma_{2\rightarrow 3}}{dx} \right)_\text{IWW}=  \frac{g^2}{2\pi} \alpha \chi\frac{|\vec{k}|}{E} \frac{(1-x)}{x}\int_{-\infty}^{\tilde{u}_\text{max}} d\tilde{u} \frac{\mathcal{A}^{22}_{t=t_\text{min}}}{\tilde{u}^2} 
\label{eq:ampCross}
\end{equation}
where we have changed variables from $\theta$ to $\tilde u$ and $\tilde{u}_\text{max} = -m_\phi^2 \frac{1-x}{x} - m_\mu^2 x$ \cite{Liu:2016mqv}. 
The additional approximation, called the Improved Weiz\"acker-Williams (IWW) approximation, is in excellent agreement with the WW approximation for sufficiently large beam energy. 
In this work, we will primarily be using IWW for its ease of calculation, but will specify when this approximation breaks down and WW must be used instead.

As previously stated, the WW and IWW approximations are most accurate when both the radiated particle and beam particle are relativistic. 
In the regime of validity, we then expect the new particle to be emitted in the forward direction, thus a reasonably small detector can still have nearly hermetic coverage of signal events.
 The characteristic angle of emission is 
 \begin{equation}
 \theta_0 \lesssim \frac{m_\phi\sqrt{\text{max}\left( \frac{m_\mu}{m_\phi}, \frac{m_\phi}{E_0}\right)}}{E_0}
 \end{equation}
 which we will use to set the scale of the experimental geometry. 
 
We emphasize that these calculations are meant to illustrate proof-of-concept phenomenological studies rather than exact results. 
While we work only in the regime of validity of these approximations, once more is known about the physical experimental considerations, a dedicated simulation for both 
event generation and detector effects should be carried out.

\subsection{Signal Events}
From the production cross section we can now calculate the number of signal events that we expect to observe. 
The differential number of signal events in $x$ and $z$, where $z$ is the distance from production at which the radiated particle decays, is given by \cite{Bjorken:2009mm}
\begin{equation}
\frac{dN}{dx dz} = N_{\mu} \frac{N_0 X_0}{A} \times \mathcal{BR}\times \int^{E_0}_{E_\phi} \frac{d E'}{E'} \int_0^T dt \ I(E'; E_0, t) \times E_0 \frac{d\sigma}{d x'} \bigg\rvert_{x'\equiv E'/E_0} \frac{dP(z-\frac{X_0}{\rho}t)}{dz}
\label{eq:FoM}
\end{equation}
where $N_{\mu}$ is the total number of muons on target, $N_0$ is Avogadro's number, $X_0$ is the radiation length of the target, $A$ is the atomic mass of the target in $g /$mol, $\mathcal{BR}$ is the branching ratio into signal final states, $T$ is the length of the target normalized by the radiation length, and $P(l)$ is the probability that the $\phi$ particle decays at position $l$:
\begin{equation}
\frac{dP(l)}{dl} = \frac{1}{l_0} e^{-l/l_0} 
\label{eq:Pdec}
\end{equation}
where $l_0$ is the lifetime of the particle in the lab frame.
Note that a lot of the complication in Eq.~\ref{eq:FoM} is due to capturing the energy loss of the beam particle as it traverses the target material. 
The radiative losses are paramaterized by the function $I(E'; E_0, t)$. 
In this work, we will be choosing target lengths sufficiently short such that the beam particle retains over $90 \%$ of its energy on average. 
This approximation is often called the thin-target approximation and allows us to set $I(E'; E_0, t) = \delta (E'-E_0)$. 
Further studies of the validity of this approximation can be found in Appendix \ref{app:materials}.

Once we have applied the thin-target approximation, the $t$ dependence is only in the argument of the decay probability function. 
We can see from Eq.~\ref{eq:Pdec} that we can easily factor out this dependence and perform the integral independently of $x$.
This allows us to simplify Eq.~\ref{eq:FoM} to
\begin{equation}
\frac{dN}{dx dz} = N_{\mu} \frac{N_0 \rho l_0}{A} \frac{d\sigma}{d x} \left( e^{\frac{L_\text{tar}}{ l_0}} - 1 \right) \frac{d P(z)}{dz}.
\label{eq:FoMthin}
\end{equation}

Next, we perform the numerical integral over $z$. 
In order for the radiated particle to be detected, it must decay beyond the target and shielding region shown in Fig.~\ref{fig:FTschem}. 
Our expression becomes
\begin{equation}
\frac{dN}{dx} = N_{\mu} \frac{N_0 \rho l_0}{A} \frac{d\sigma}{d x} \left( e^{\frac{L_\text{tar}}{ l_0}} - 1 \right) e^{-(L_\text{tar}+L_\text{sh})} \left( 1-e^{-L_\text{dec}/l_0}\right).
\label{eq:FoMthin}
\end{equation}
Finally, to calculate the final event yield $N$, we must numerically integrate over $x$.
From the exponential dependence on the length scales of the experiment, it is clear that the optimizing the geometry of the experiment is crucial to maximizing the sensitivity to new physics. 
Note that future experimental design should involve further studies, optimization, and simulations to include material effects.
%

\subsection{New Physics Models and Signatures}
\label{ssec:NP}

To illustrate the discovery potential of this robust experimental set up, we consider a variety of new physics models for the various species of new particles. 
The various models motivate the coupling structure of the new particle $\phi$ to SM particles. 
We note that while the UV completions of the models discussed require additional model building, we do not endeavor to fully flush out these models.
The intent of the choice of models we have made is to demonstrate the range of phenomenology we could probe rather than explore a particular motivated scenario. 

With a muon beam-dump experiment, we gain the most sensitivity for models that either have very weak couplings, relatively massive mediators (GeV scale), or couple more strongly to the higher generations. 
The models we consider could potential provide solutions or insight to several persistent questions of the SM, such as the hierarchy problem \cite{Morrissey:2009ur}, the strong CP problem \cite{Dine:1981rt}, the nature of dark matter \cite{Berlin:2018bsc}, or the anomalous magnetic moment of the muon \cite{Muong-2:2023cdq, Lee:2014tba, PhysRevD.93.035006, LEVEILLE197863, Gninenko:2014pea}.
Additionally, these models can arise naturally from string theory \cite{Arkani-Hamed:2006emk}. 

The generic interaction Lagrangian for all species of new particles $\phi$ is
\begin{equation}
\mathcal{L}_\text{int} \supset \frac{1}{2} \partial_\mu \phi ^2 - \frac{1}{2} m_\phi^2 \phi^2 + i g_\psi \mathcal{O}_{\bar \psi \psi \phi }
\end{equation}
where $\mathcal{O}_{\bar \psi \psi \phi }$ is an operator with a fermion-anti-fermion pair of the same flavor and one radiated particle. 
We assume for now a minimal model where the only new particle introduced is $\phi$.
The coupling on this operator is important as it controls not only the rate of particle production, but also the branching ratios and overall lifetime of $\phi$. 

The visible signature of interest for these new particles is a dilepton final state. 
This entails decaying to either $e^+e^-$ or $\mu^+\mu^-$ within the fiducial decay volume as shown in Fig.~\ref{fig:FTschem}. 
We consider this signature as a charged lepton final state provides not only a clean signature, but it is straightforward to instrument the necessary detector as well as perform a standard bump-hunt analysis. 

The models we consider are: 
\begin{itemize}
\item \textbf{Muonphilic.} The new particle only couples to muons. We consider this `best-case scenario' for all species of new particles \cite{Chen:2015vqy, Krnjaic:2019rsv}. 
\item \textbf{Leptophilic.} The new particle is a scalar or pseudoscalar that couples to leptons proportional to their Yukawa coupling \cite{Chivukula:1987py, DAmbrosio:2002vsn, Branco:2011iw, Marshall:2010qi}
\item \textbf{Dark Photon.} The new particle is a dark photon (or heavy $Z'$), meaning a vector that couples to the fermionic charged current of the SM \cite{Holdom:1985ag, Pospelov:2007mp, Arkani-Hamed:2008hhe, Pospelov:2008zw}.
\item \textbf{$L_\mu-L_\tau$.} The new particle is a vector that couples with the gauged $L_\mu - L_\tau$ symmetry \cite{Foot:1990mn, PhysRevD.43.R22, PhysRevD.44.2118, Ma:2001md}. 
\end{itemize}


\subsection{Summary of Assumptions And Parameters}
\label{sec:Summary}
We now briefly summarize the assumptions and parameters that we use to compute the reach over these various models. 
These are motivated in the context of a potential multi-TeV MuC. 
We consider not only the results from a beam-dump experiment done at the various collider energies, but also results from lower-energy beams with the assumption this technology will be developed during the necessary R\&D studies and tests.
Since the specifications and feasibility of a MuC have yet to be determined, these should be taken as benchmarks that illustrate the ballpark discovery potential of such configurations rather than a strict bound on what can be done. 
To span the space of possible beam options, we consider a variety of energies ($E_0 =$ 10 GeV, $m_h/2$, 1.5 TeV, 5 TeV) as well as integrated total muons on target ($N_\mu = 10^{18}, 10^{20}, 10^{22}$).

The low-energy beam options are not intended to be understood as planned run parameters of staging and demonstrator facilities, but rather order-of-magnitude benchmarks. 
We choose 10 GeV as an indicator as what might be possible for cooled muons which will likely have energy $\mathcal{O}(\text{GeV})$.
Additionally, accelerating muons up to several GeV would be a reasonable `small scale'\footnote{`Small scale' muon acceleration would still be a serious, large scale endeavor in terms of construction time and cost.} intermediate step. 
Such an acceleration complex would lend itself well to possible neutrino synergies, e.g. nuSTORM \cite{nuSTORM:2022div}, which propose a $\sim10$ GeV muon storage ring as a means of producing collimated and nearly mono-chromatic neutrino beams for oscillation experiments.
We also consider $E_0 = m_h/2$ as another benchmark even though the full collider is unlikely to run at the Higgs pole.
The small muon Yukawa coupling as well as low-energy beam result in fewer Higgs produced than with vector boson fusion (VBF) at high energy \cite{Accettura:2023ked}.
However, we again take this energy to be indicative of the experimental reach for beam dumps at an order of magnitude beyond the cooled muon energy. 
The multi-TeV energy options are only relevant if the full collider is constructed. 

The number of muons on target ranges from easily achievable over a year of running ($N_\mu = 10^{18}$) to very\footnote{Possibly overly optimistic, even.} optimistic scenarios ($N_\mu = 10^{22}$). 
As luminosity scales with the energy of the beam, a large number of muons on target is more likely achievable for the multi-TeV beam energies.
We ensure that for all the beam energies we consider, the WW approximation (and often the IWW approximation) is still valid and the calculation can be done. 

We must make a few additional assumptions about the experimental set-up. 
The choices that have immediate computable consequence are the material of the target and geometry of the experiment.
For materials, we choose two relatively cheap options: lead and water.
The plots we show in the body of the text will all be with lead targets, but for comparison, water targets are discussed in Appendix \ref{app:materials} as well.
The size of the experiment that we consider is determined by several factors. 
Firstly, we consider targets sufficiently narrow to use the narrow target approximation\footnote{This point is also discussed further in Appendix \ref{app:materials}.}.
This also allows us to be sensitive to slightly larger couplings (which corresponds to shorter-lived particles) than if we were to consider a thick target. 
Secondly, the shielding and veto area is chosen to be as small as possible while still being realistic. 
The magnetic field needed to sweep the un-interacting beam out of the acceptance of the detector is chosen to be strong (roughly 1T) but not unreasonably so. 
Finally, the length of the experimental hall is determined by the size of the detector. 
Having an arbitrarily long detector has exponentially diminishing reach on long lifetimes, as the probability of decaying is suppressed at long distances. 
However, the longer the experiment is, the bigger the detector must be to maintain high signal acceptance. 
Thus we choose an experimental hall length which keeps an $\mathcal{O}(1)$ fraction of signal events. 
We fix the maximal angle of acceptance at the detector to be $\theta = 10^{-2}$, which determines the length of the experiment as well.
\begin{table}[h!]
\begin{center}
\begin{tabular}{ |c|c|c| } 
 \hline
 Target Materials & Beam Energy $E_0$ [GeV] & Muons On Target ($\mu$) \\ 
 \hline
 Lead & 10 & $10^{18}$ \\ 
 Water & $63 \ (m_h/2)$ & $10^{20}$\\ 
  & $1.5 \times 10^3$ & $10^{22}$\\
& $5\times 10^3$ & \\  
 \hline
\end{tabular}
\caption{Summary of experimental parameters.}
\label{tab:Summ}
\end{center}
\end{table}

\section{Preliminary Computations}
\label{sec:csc}
In this section we compute the production mechanisms and decay widths for the various particle species. 
All of the particle species are predominantly produced with the same diagram (shown in Fig.~\ref{fig:2to3}), where the  distinction between models is the exact structure of the interaction vertex, and relative couplings to other SM particles.  
We therefore calculate the production cross section for each species generically, and the model-dependent effects for a given species factorize into the lifetime of the new particle and its branching ratio into the desired dilepton final states. 

To begin, we summarize the relevant interaction Lagrangians for all the species:
\begin{equation}
\begin{aligned}
& \mathcal{L}^S_\text{int}  \supset - ig_S \phi \bar{\psi}\psi \\
&\mathcal{L}^P_\text{int} \supset -  ig_P a \bar{\psi} \gamma^5 \psi \\
& \mathcal{L}^V_\text{int} \supset -  ig_V V_\mu \bar{\psi} \gamma^\mu \psi \\
& \mathcal{L}^A_\text{int} \supset - ig_A A_\mu \bar{\psi} \gamma^\mu \gamma^5 \psi.
\end{aligned}
\label{eq:lagrangians}
\end{equation}
From this we can compute the 2-to-2 scattering amplitude to be used for the evaluation of the cross section using the approximation in Eq.~\ref{eq:WWxsec}:
\begin{align}
    \mathcal{A}^{2\to2}_{S, t=t_{\rm min}}   & \approx \frac{x^2}{1-x} + 2(m_\phi^2  - 4m_\mu^2)\frac{\tilde{u}x + m_\mu^2(1-x) + m_\mu^2 x^2}{\tilde{u}^2}                                     \\
    \mathcal{A}^{2\to2}_{P, t=t_{\rm min}}   & \approx \frac{x^2}{1-x} + 2m_a^2\frac{\tilde{u}x + m_\mu^2(1-x) + m_\mu^2 x^2}{\tilde{u}^2}                                                   \\
    \mathcal{A}^{2\to2}_{V, t=t_{\rm min}}   & \approx 2\frac{2-2x+x^2}{1-x} + 4(m_V^2  + 2m_\mu^2)\frac{\tilde{u}x + m_\mu^2(1-x) + m_\mu^2 x^2}{\tilde{u}^2}                             \\
    \mathcal{A}^{2\to2}_{A, t=t_{\rm min}}   & \approx \frac{4m_\mu^2x^2}{(m_A^2)(1-x)} + 2\frac{2-2x+x^2}{1-x} + 4(m_A^2  - 4m_\mu^2)\frac{\tilde{u}x + m_\mu^2(1-x) + m_\mu^2 x^2}{\tilde{u}^2}
\end{align}
 \label{eq:Amps22}
where 
\begin{equation}
 \tilde{u}(x, \theta) \approx  -xE_0^2 \theta^2 - m_X^2 \frac{1-x}{x} -m_\mu^2 x
\label{eq:defU}
\end{equation}
is the Mandelstam variable (Eq.~\ref{eq:defUtilde}) in the 2-to-2 scattering process evaluated at $t \sim t_{\min}$.
The approximations made in Eq.~\ref{eq:Amps22} are consistent with the forward, boosted bremsstrahlung of the new particle. 
In all cases, we verify that the photon virtuality, $t_{\min}$, is smaller than all other relevant mass scales, which is a fundamental assumption of the WW and IWW approximations. 
We also numerically check that $t_{\min}$ and $t_{\max}$ are independent of the angle of emission $\theta$, which is necessary to use the IWW approximation over the WW approximation, which we find to be the case for $E_0 \gtrsim 100$ GeV.
Thus, we use the full WW approximation (Eq.~\ref{eq:WWxsec}) at $E_0 = 10$ GeV and $m_h/2$, and the IWW approximation (Eq.~\ref{eq:ampCross}) at $E_0 =1.5$ and $3$ TeV.

\begin{figure}[t!]
     \centering
     \begin{subfigure}[b]{0.49\textwidth}
         \centering
         \includegraphics[width=.95\textwidth]{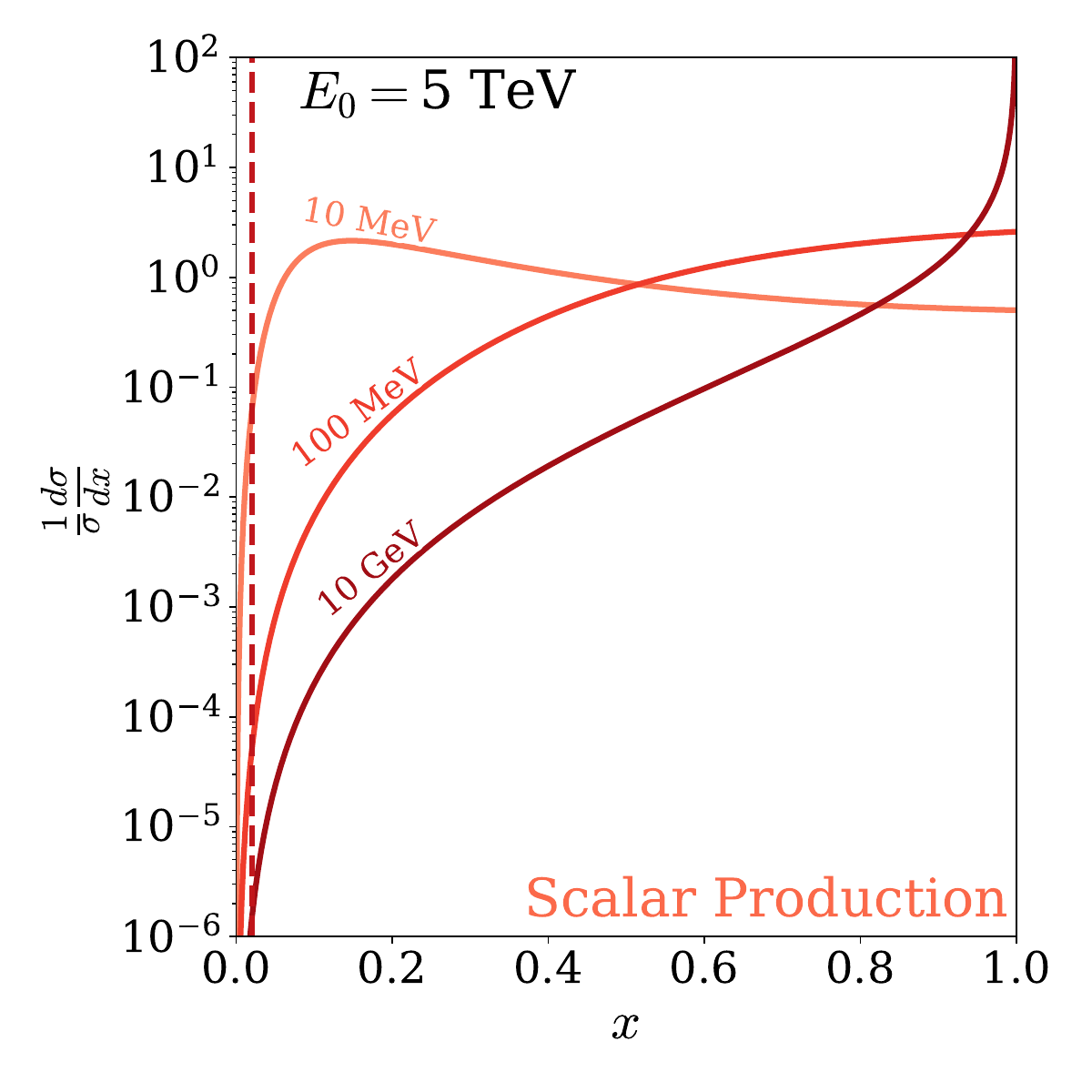}
         \caption{Scalar Production}
         \label{fig:xsecScalar}
     \end{subfigure}
     \hfill
     \begin{subfigure}[b]{0.49\textwidth}
         \centering
         \includegraphics[width=.95\textwidth]{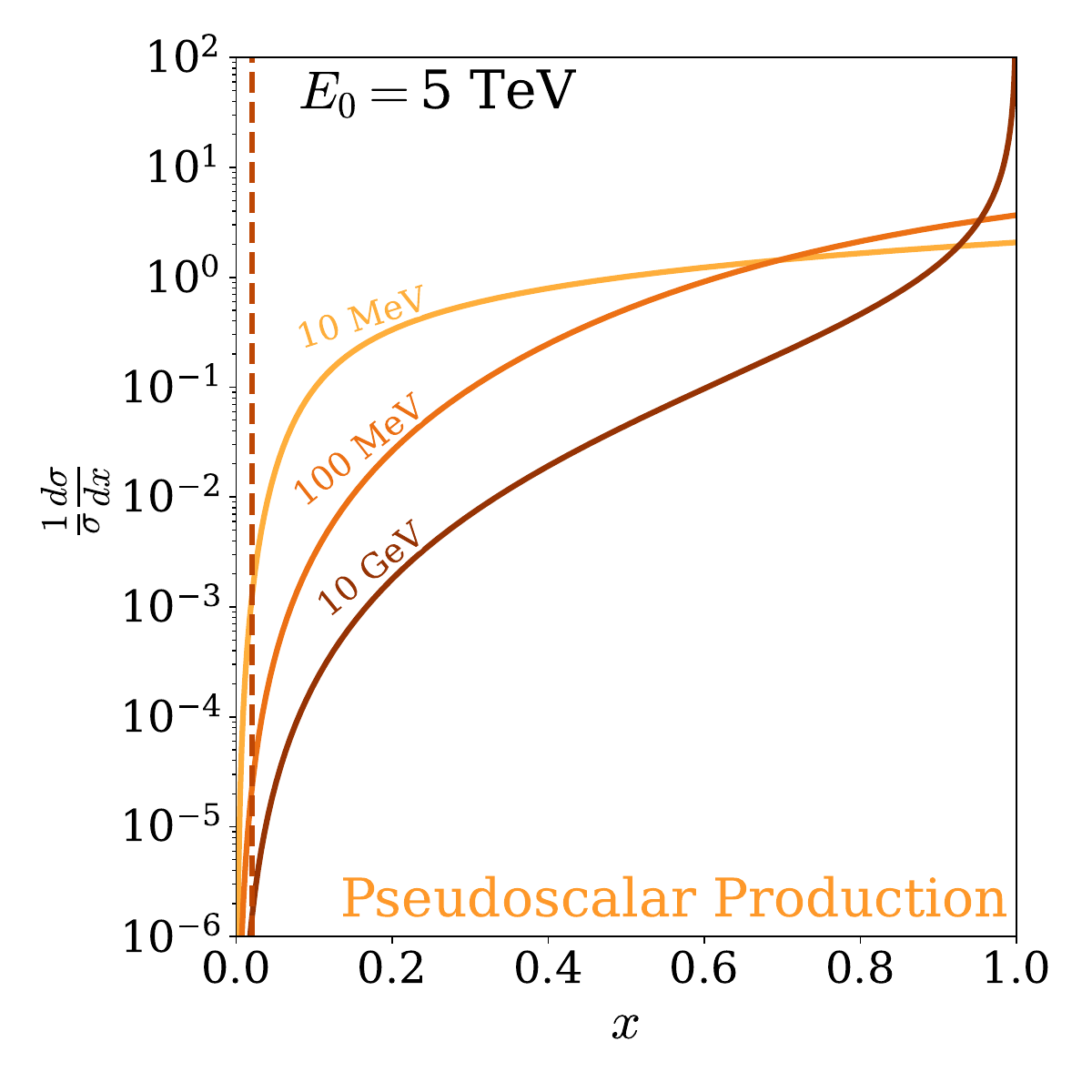}
         \caption{Pseudoscalar Production}
         \label{fig:xSecPS}
     \end{subfigure}
     \hfill
     \begin{subfigure}[b]{0.49\textwidth}
         \centering
         \includegraphics[width=.95\textwidth]{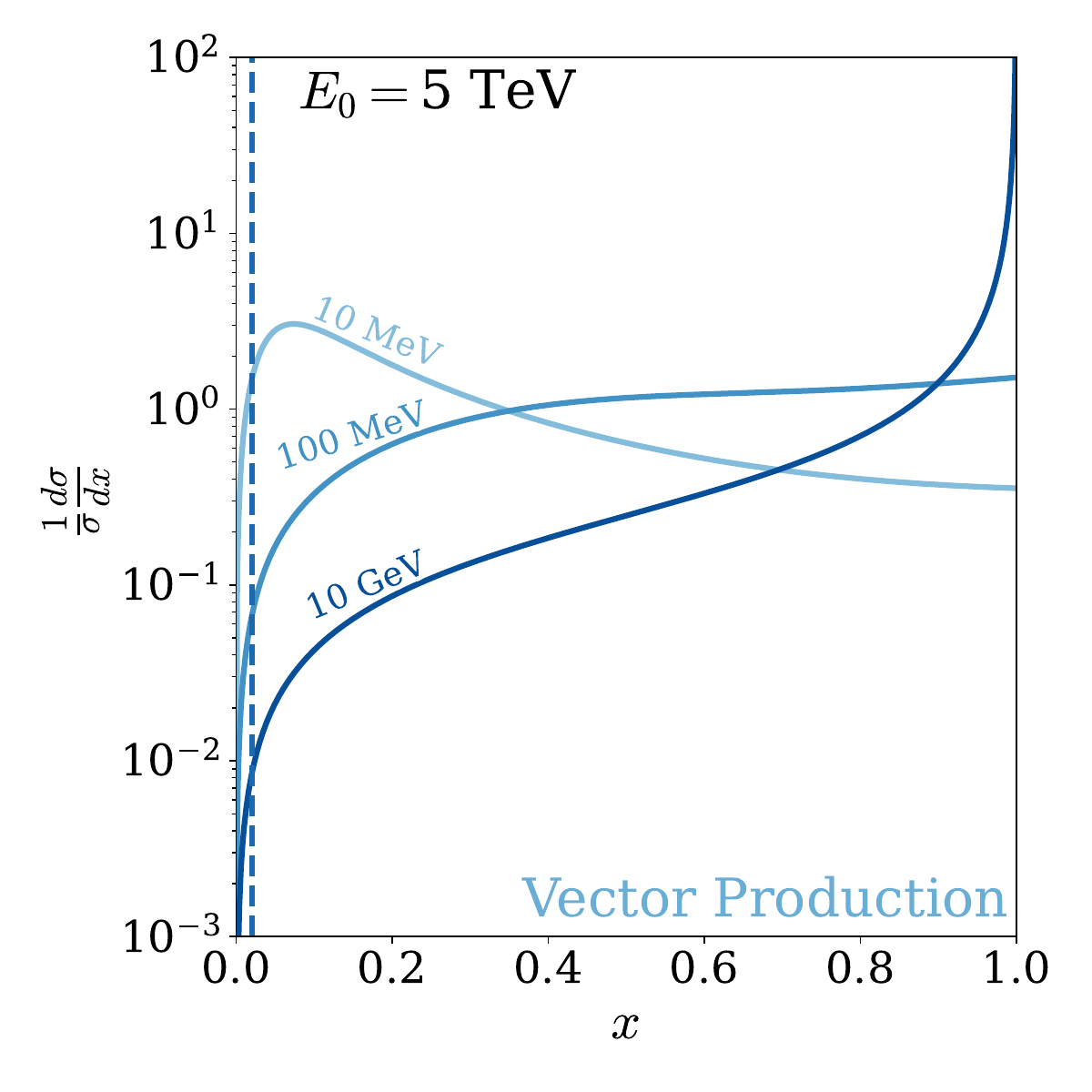}
         \caption{Vector Production}
         \label{fig:xSecV}
     \end{subfigure}
     \begin{subfigure}[b]{0.49\textwidth}
         \centering
         \includegraphics[width=.95\textwidth]{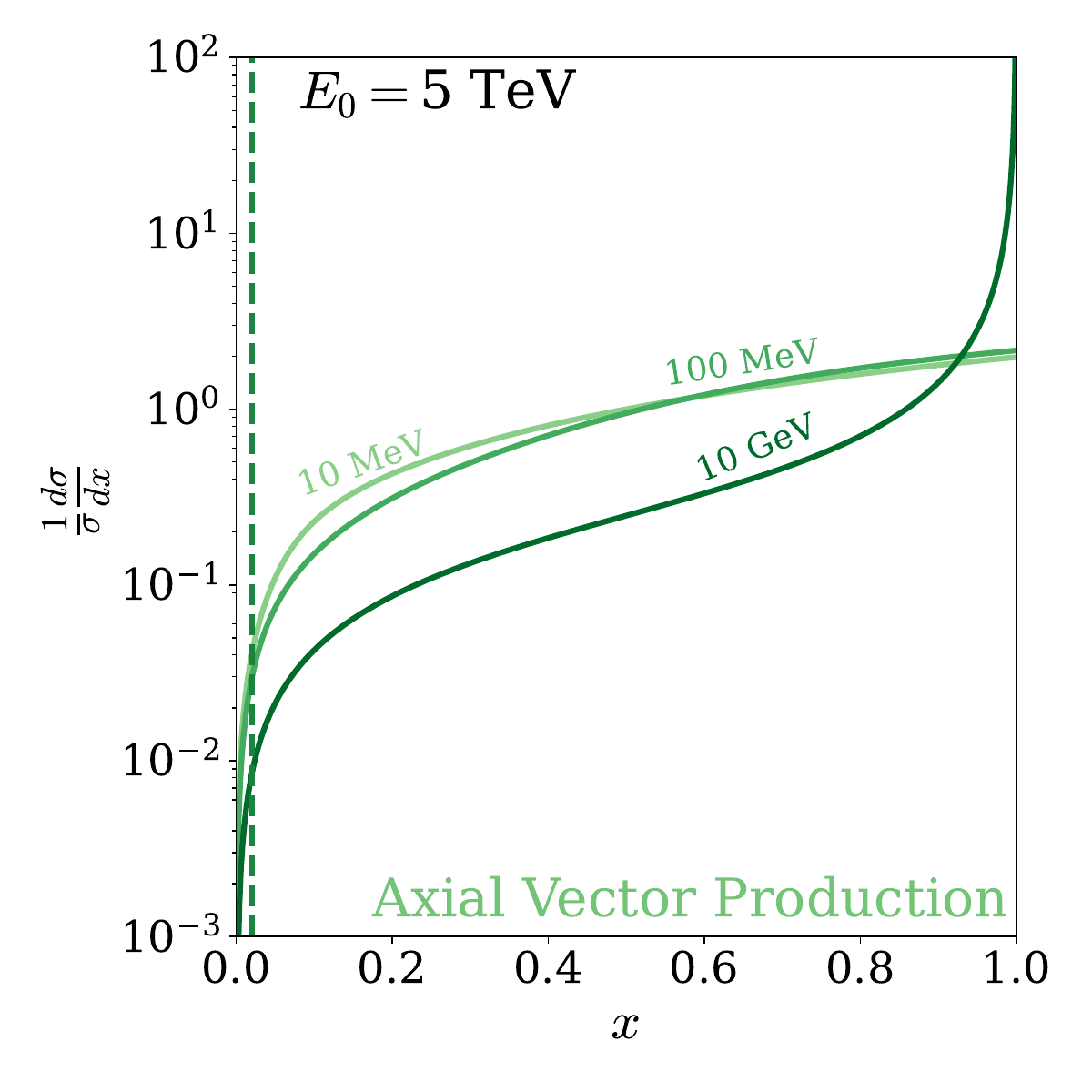}
         \caption{Axial Vector Production}
         \label{fig:xSecAV}
     \end{subfigure}
        \caption{The normalized differential cross sections in fractional energy carried off by the new particle $x$ for (a) scalar, (b) pseudoscalar, (c) vector, and (d) axial vector new physics scenario. We show the distributions for a variety of new particle masses ranging from 10 MeV to 10 GeV. The dashed line is the threshold at which the 10 GeV is produced with a boost of $\gamma = 100$. From this indicator, it is evident all particles are being produced highly relativistically and the WW approximation is valid.}
        \label{fig:xsec}
\end{figure}

We then numerically compute the differential cross section for the various new particle masses and beam configurations. 
We show some of these distributions in \Fig{fig:xsec}. 
For concreteness, we show the distributions with a muon beam energy of $E_0= 5$ TeV corresponding to the full 10 TeV MuC, but the curves are qualitatively similar for all energies considered over smaller mass windows.\footnote{More plots at different values of $E_0$ can be found in Appendix~\ref{app:plots}.}

Finally, we calculate the decay widths for the various species into two-fermion final states. 
The widths, computed again using the interaction terms given in Eq.~\ref{eq:lagrangians}, are
%
 %
\begin{equation}
\begin{aligned}
\Gamma_{\phi \rightarrow l^+l^-} & = g_S^2 \frac{m_\phi}{8\pi} \left(1-\frac{4m_l^2}{m_\phi^2}\right)^{3/2} \\
\Gamma_{a  \rightarrow l^+l^- } & = g_P^2 \frac{m_\phi}{8\pi} \left( 1 - \frac{4m_l^2}{m_\phi^2}\right)^{1/2}\\
\Gamma_{V  \rightarrow l^+l^- } & = g_V^2 \frac{m_\phi}{12 \pi} \left( 1 + \frac{2 m_l^2}{m_\phi^2} \right) \left( 1 - \frac{4 m_l^2 }{m_\phi^2}\right)^{1/2}\\
\Gamma_{A  \rightarrow l^+l^- } & = g_A^2 \frac{m_\phi}{12 \pi} \left(1-\frac{4m_l^2}{m_\phi^2}\right)^{3/2}. 
\end{aligned}
\end{equation}
For some of the models we consider, hadronic decays are also possible, and we discuss their affects on the branching ratio further in Sec.~\ref{sec:results}. 
Note that the spin-0 (spin-1) particles could also decay to di-photon (tri-photon \cite{Landau:1948kw,PhysRev.77.242,Zhemchugov:2014dza}) states via fermion loops. 
However, for the mass range of interest, these channels are highly suppressed and we restrict our attention to the tree-level processes. 

The final state of interest for signal is either dielectron or dimuon. 
We treat other possible final states as effectively invisible, although one could potentially improve the sensitivity of the experiment by including these final states in the analysis strategy.
Thus the final piece of Eq.~\ref{eq:FoM} is 
\begin{equation}
\mathcal{BR}  = \frac{\Gamma_{X\rightarrow e^+e^-} + \Gamma_{X\rightarrow \mu^+\mu^-} }{\Gamma_\text{tot}}.
\end{equation}
We do not consider missing energy or hadronic signatures for experimental ease, although including these channels could potentially improve the reach. 
We expect that signal detection could be accomplished with minimal instrumentation, such as a calorimeter and potentially some tracking or timing information for further background mitigation. 
%

\section{Results}
\label{sec:results}

In this section we summarize the sensitivity to the various models with the exploratory experimental parameters summarized in Table~\ref{tab:Summ}. 
We organize the results by model, where each model may have several relevant particle species. 
The contours are drawn for the detection of 5 signal events. 
We choose this number to account for the possibility of some background events that would require most advanced simulations to fully quantify. 
In order to spare the reader of plot fatigue, we present only a subset of the possible plots for the variety of configurations, and include additional plots in Appendix~\ref{app:plots}.

For many of the models, there are existing constraints from a variety of experiments. 
Many of these constraints come from previous proton or electron beam-dump and fixed-target experiments, such as E774 at Fermilab \cite{Bross:1989mp},  E141 \cite{Riordan:1987aw} and E137 at SLAC \cite{Bjorken:1988a, Batell:2014mga, Marsicano:2018krp}, and CHARM \cite{CHARM:1985nku, Gninenko:2012eq}, COMPASS \cite{COMPASS:2007rjf} and NuCal \cite{Blumlein:1990ay, Blumlein:2011mv, Blumlein:2013cua} at CERN. 
Strong constraints in this region are also set by previous dilepton resonant searches (into either $e^+e^-$ or $\mu^+\mu^-$) from NA48 \cite{NA482:2015wmo}, LHCb \cite{LHCb:2019vmc}, and FASER \cite{FASER:2023tle} at CERN, KLOE \cite{KLOE-2:2011hhj, KLOE-2:2012lii, KLOE-2:2014qxg, KLOE-2:2016ydq}, and the A1 Experiment at the Mainz Microtron \cite{Merkel:2014avp}.
Additionally there are relevant astrophysical constraints for long-lived particles set by Supernova 1978A (SN1978A) \cite{Chang:2016ntp}.


\subsection{Muonphilic}
\begin{figure}[t!]
\centering
     \begin{subfigure}[b]{0.49\textwidth}
         \centering
         \includegraphics[width=.95\textwidth]{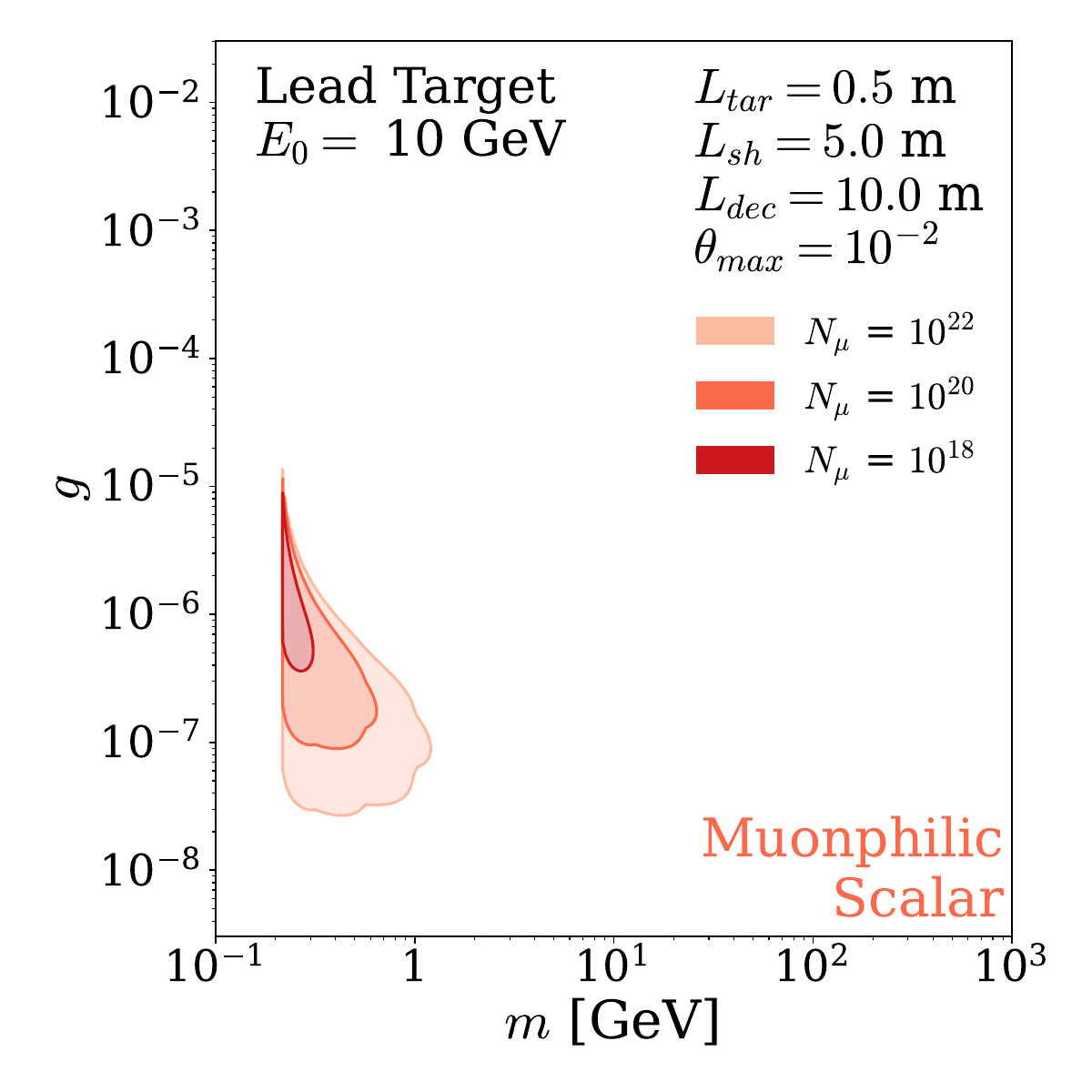}
         \caption{Scalar Production}
     \end{subfigure}
     \hfill
     \begin{subfigure}[b]{0.49\textwidth}
         \centering
         \includegraphics[width=.95\textwidth]{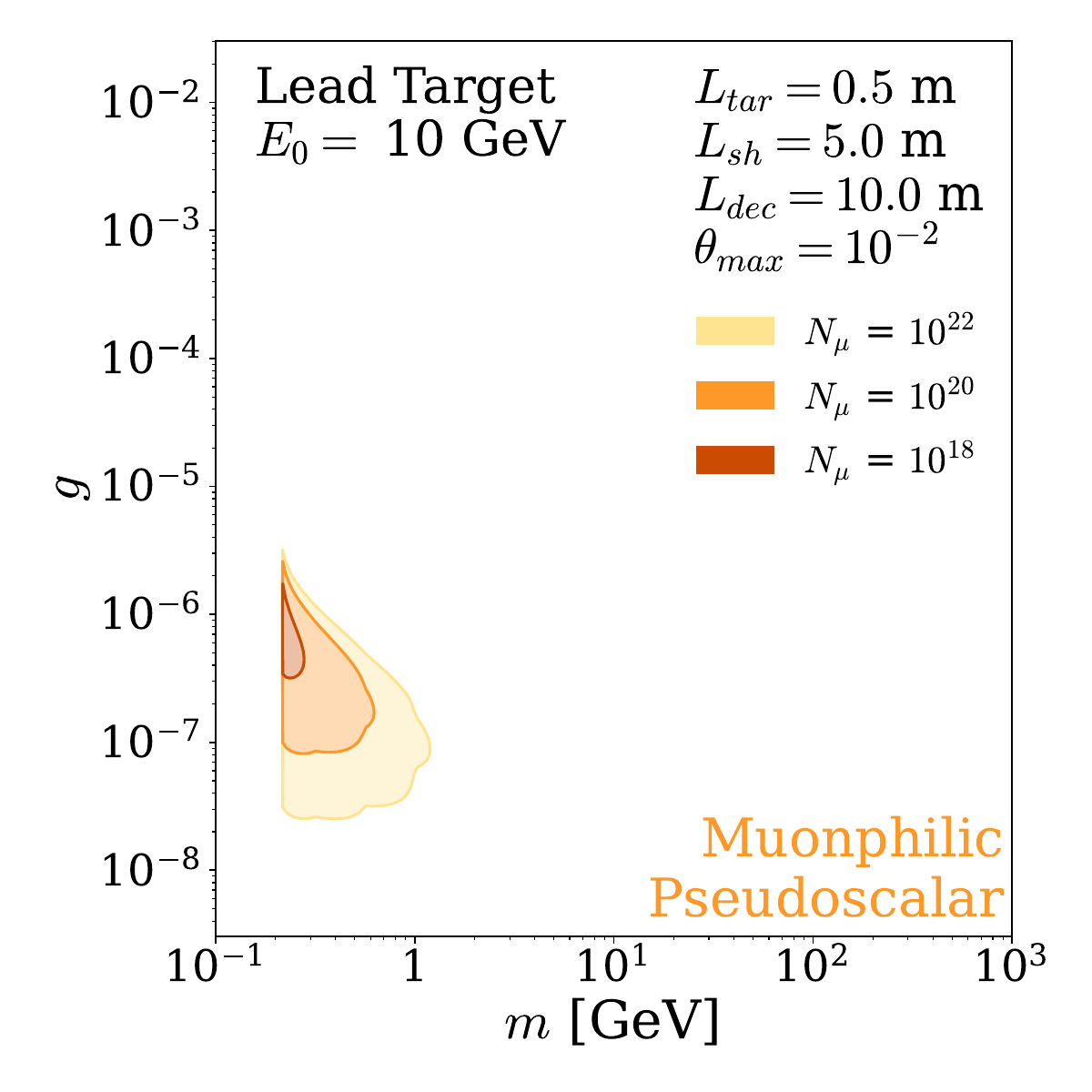}
         \caption{Pseudoscalar Production}
     \end{subfigure}
     \hfill
     \begin{subfigure}[b]{0.49\textwidth}
         \centering
         \includegraphics[width=.95\textwidth]{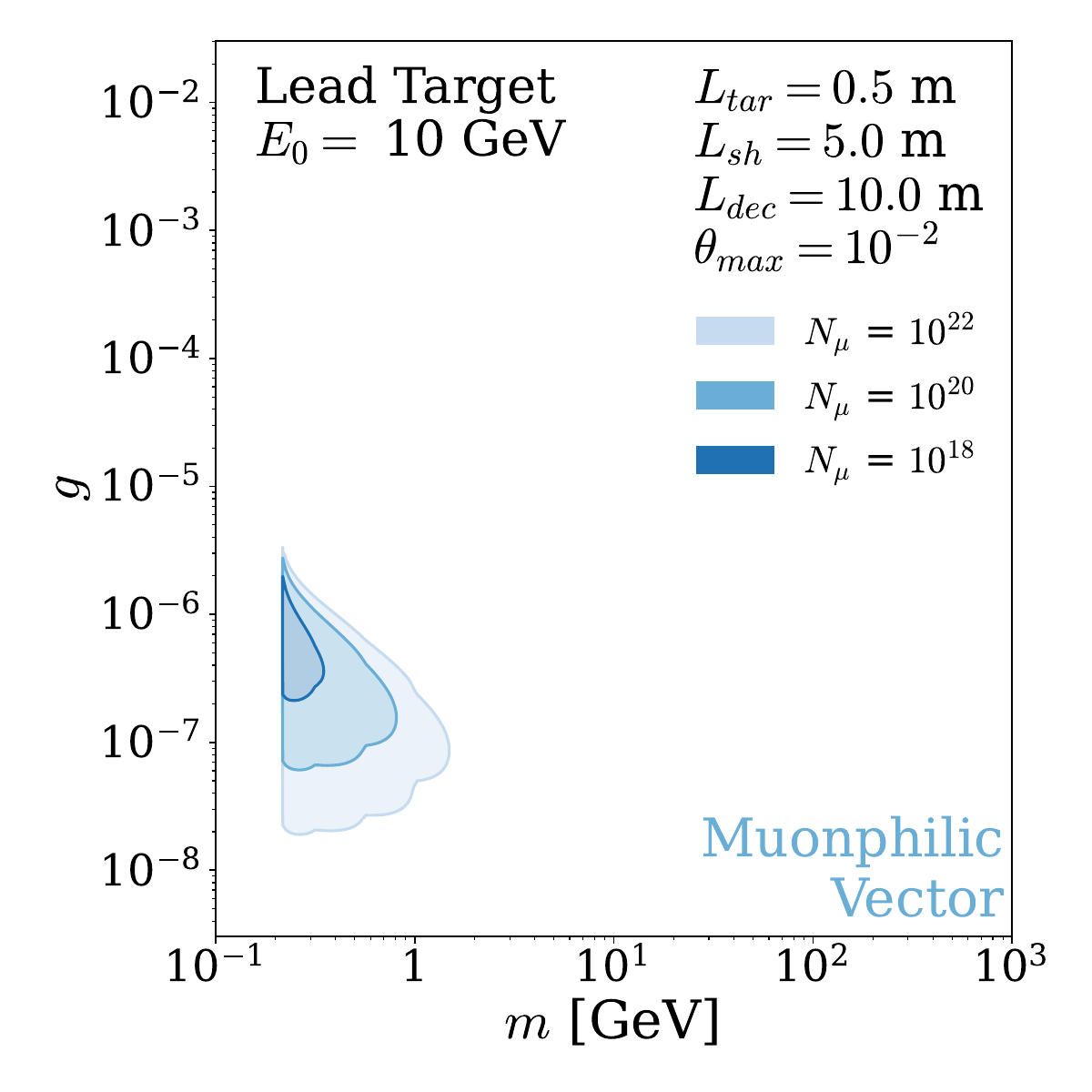}
         \caption{Vector Production}
     \end{subfigure}
     \begin{subfigure}[b]{0.49\textwidth}
         \centering
         \includegraphics[width=.95\textwidth]{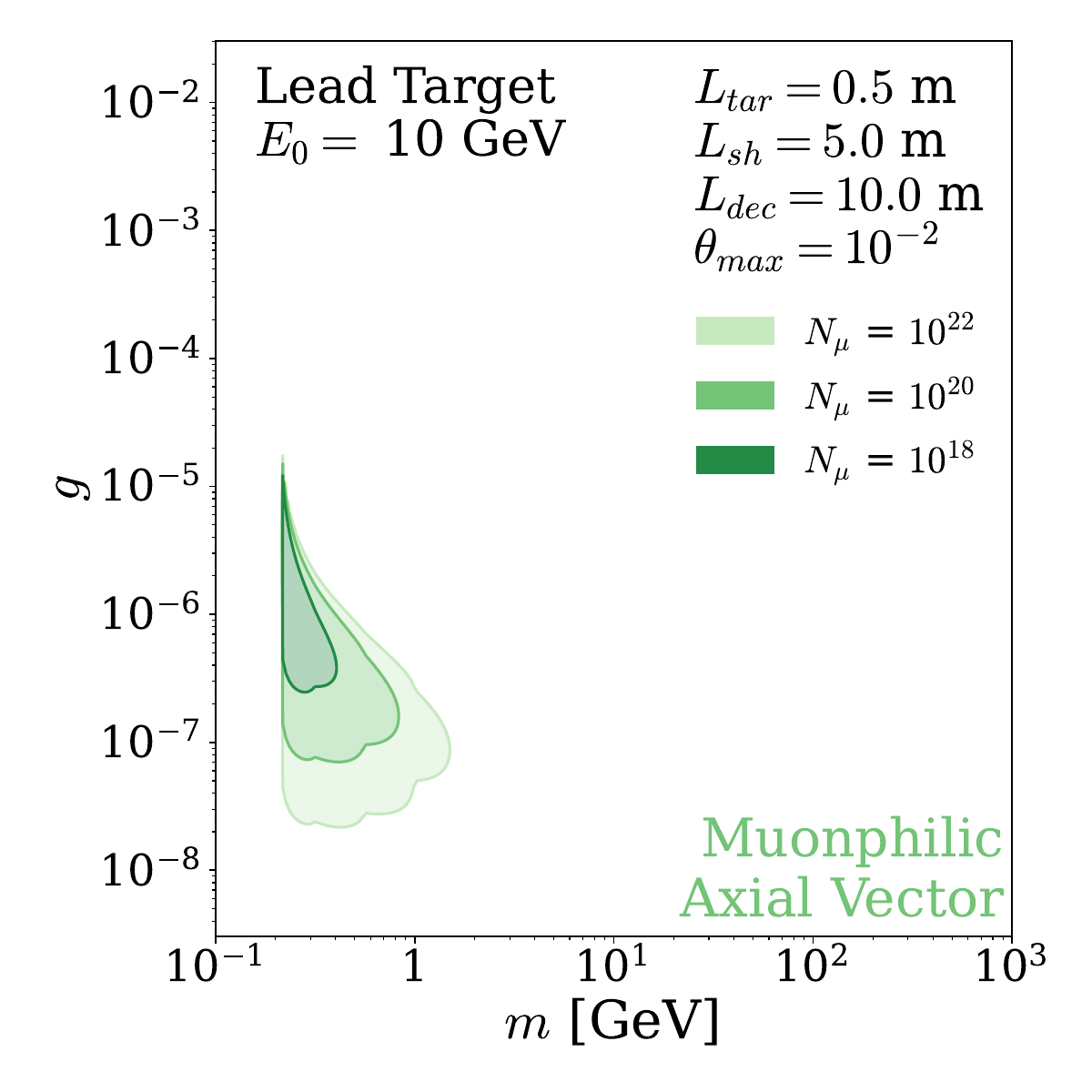}
         \caption{Axial Vector Production}
     \end{subfigure}
        \caption{The reach plots for a 10 GeV muon beam on a lead target for the various particles species with a muonphilic coupling as defined in \Eq{eq:muPhil}. The contours are $N_\mu$ = $10^{18}$, $10^{20}$, and $10^{22}$.  }
        \label{fig:muphil10GeV}
\end{figure}
%
%
\begin{figure}[h!]
\centering
     \begin{subfigure}[b]{0.49\textwidth}
         \centering
         \includegraphics[width=.95\textwidth]{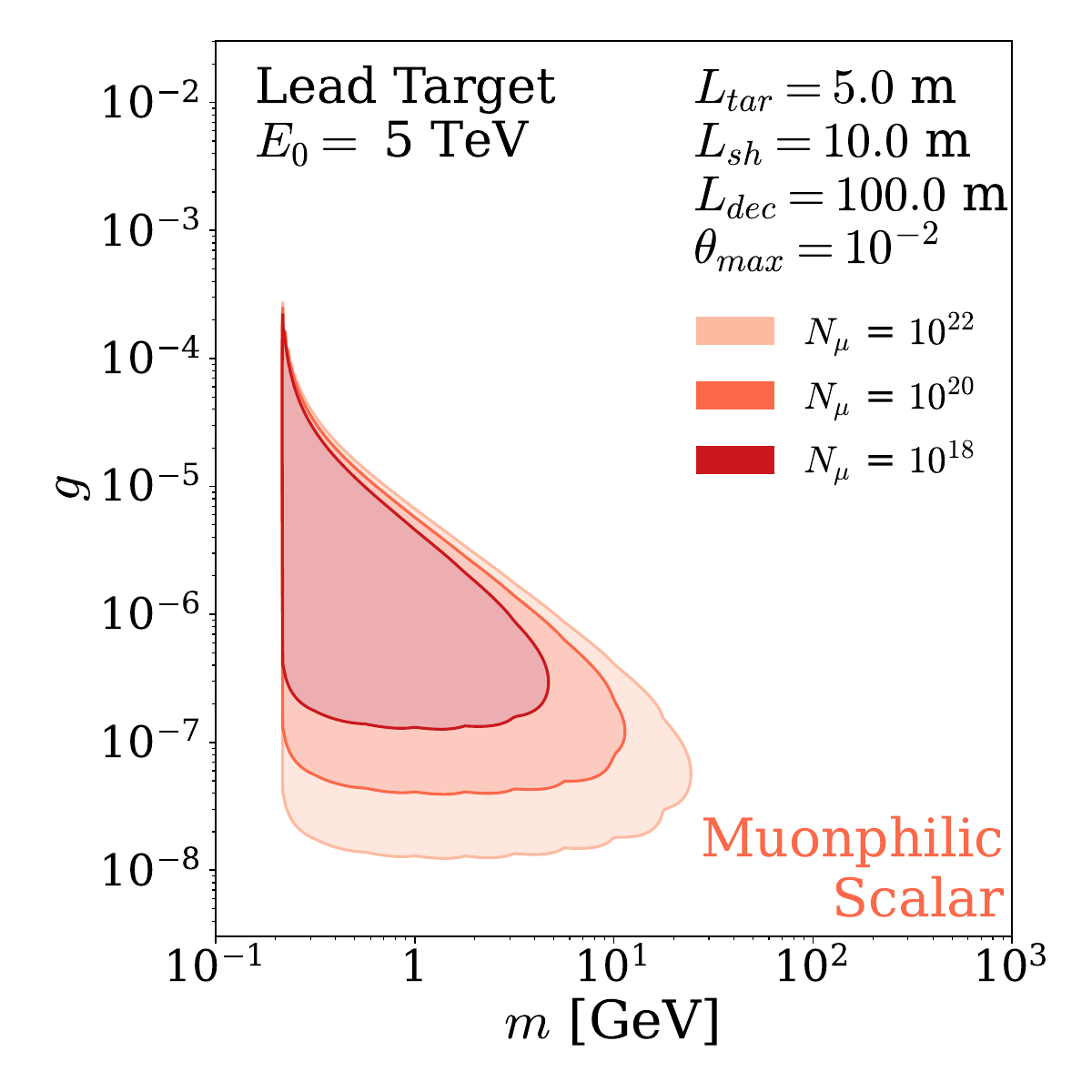}
         \caption{Scalar Production}
     \end{subfigure}
     \hfill
     \begin{subfigure}[b]{0.49\textwidth}
         \centering
         \includegraphics[width=.95\textwidth]{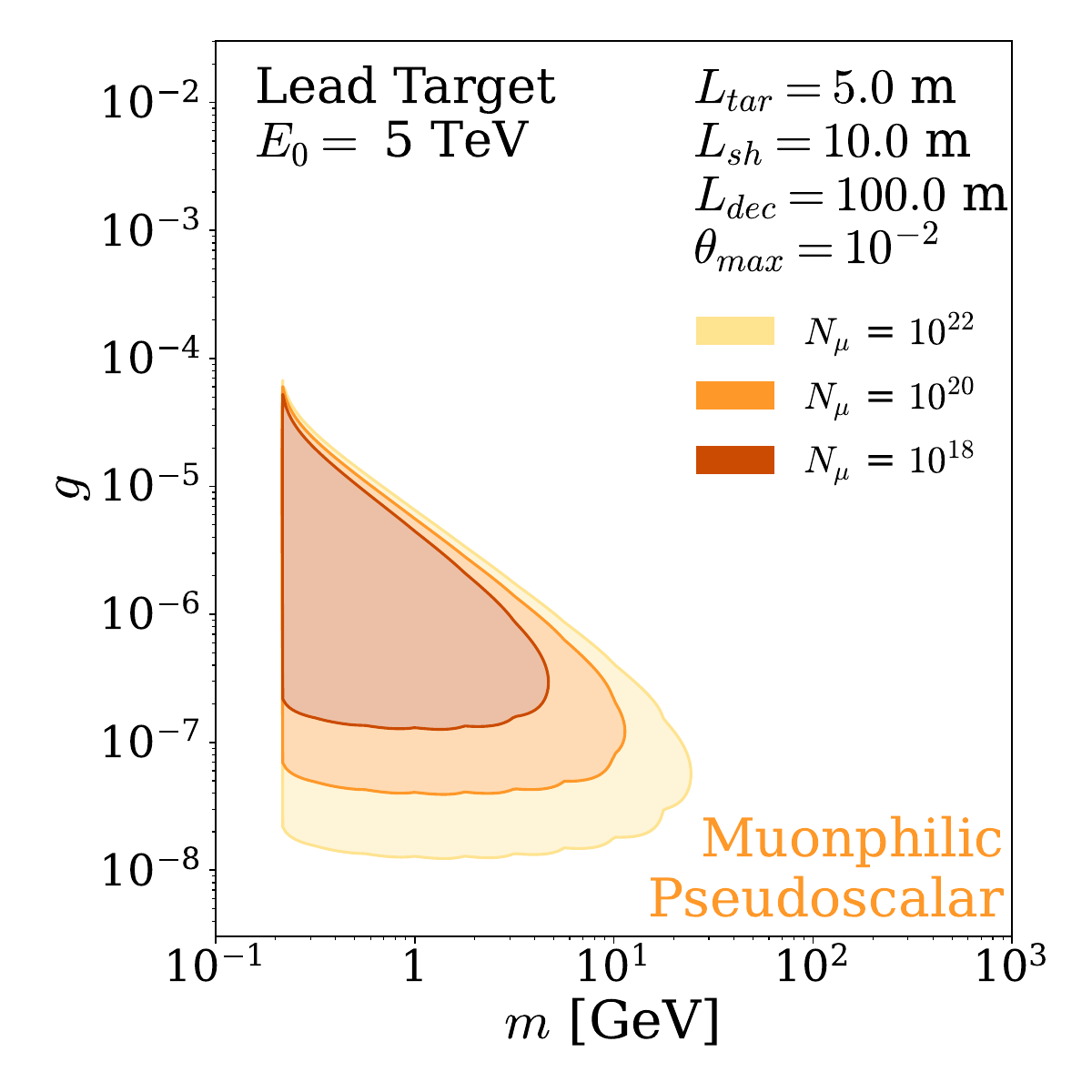}
         \caption{Pseudoscalar Production}
     \end{subfigure}
     \hfill
     \begin{subfigure}[b]{0.49\textwidth}
         \centering
         \includegraphics[width=.95\textwidth]{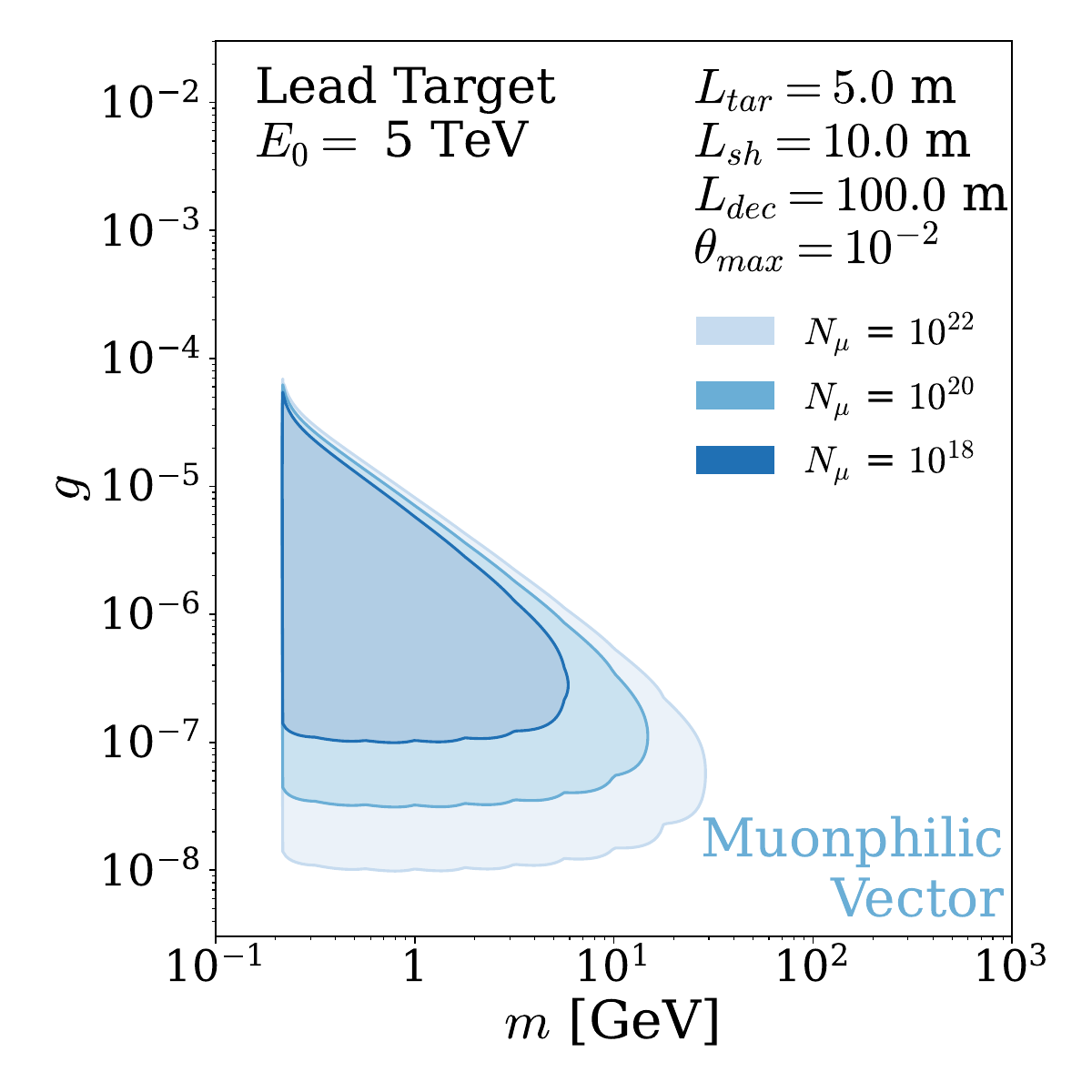}
         \caption{Vector Production}
     \end{subfigure}
     \begin{subfigure}[b]{0.49\textwidth}
         \centering
         \includegraphics[width=.95\textwidth]{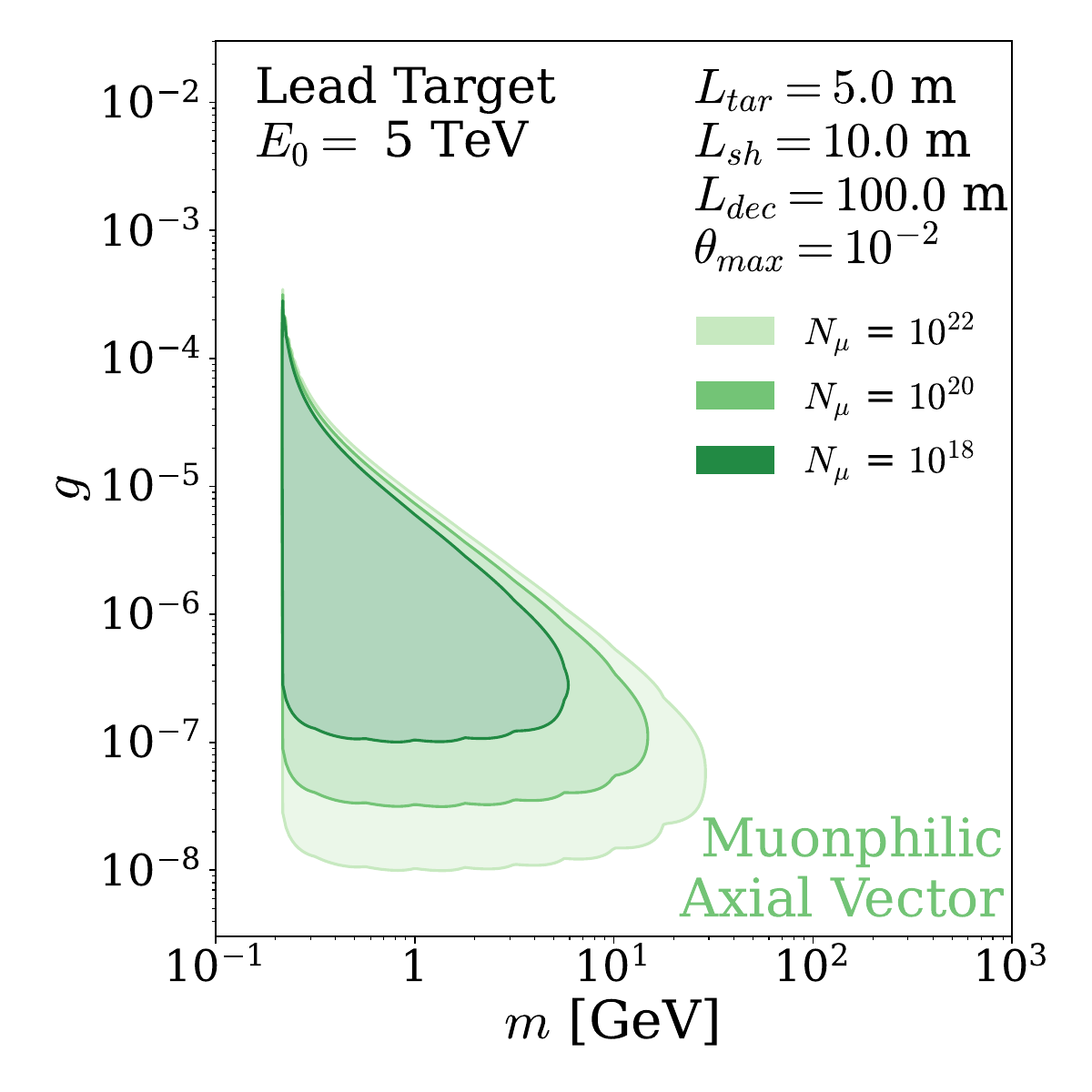}
         \caption{Axial Vector Production}
     \end{subfigure}
        \caption{The same as Fig.~\ref{fig:muphil10GeV} but with a beam energy of 5 TeV. }
        \label{fig:muphil5TeV}
\end{figure}
The first model of interest that we consider is one with muonphilic couplings.
In this phenomenologically minimal model, the only nonzero coupling of the new particle is to muons. 
This model has been previously explored for different experimental set-ups, for example Refs.~\cite{Chen:2017awl, Forbes:2022bvo, Krnjaic:2019rsv, Rella:2022len, Sieber:2023nkq}. 
While this model is a bit contrived, one can conceive of UV dynamics that result in such couplings---for example, integrating out heavy new fermions above the electroweak scale as shown in Refs.~\cite{Chen:2015vqy, Krnjaic:2019rsv}.
This model is also a useful benchmark as it provides the most optimistic scenario for producing signal events for this experimental configuration. 

For all particle species we consider this extremal model where 
\begin{equation}
g_\mu \neq 0, \text{ else } g = 0.
\label{eq:muPhil}
\end{equation}
The parameter space for these models is wide open due to the lack of dedicated muon sources (see Fig.~\ref{fig:muphil10GeV} and Fig.~\ref{fig:muphil5TeV}). 
We find that at both high and low energies there is nontrivial new coverage of parameter space to be made with the various configurations of muons on target.

\subsection{Leptophilic}
Another minimal model is a leptophilic coupling.
 In this model, we consider only couplings to the leptons, but with each coupling proportional to the SM Yukawa coupling:
 \begin{equation}
 g_{l} = g \frac{m_l}{v}
 \label{eq:leptoCoupling}
 \end{equation}
 with the operators given in \Eq{eq:lagrangians}. 
This coupling structure is consistent with the Minimal Flavour Violation (MFV) hypothesis, which assumes that flavor symmetry violation from new physics has the same structure as that from the SM---the Yukawa couplings \cite{Chivukula:1987py, Hall:1990ac, Buras:2000dm, DAmbrosio:2002vsn, Cirigliano:2005ck}. 
In this model, there is no coupling to quarks and therefore no tree-level hadronic final states. 
\begin{figure}
\centering
     \begin{subfigure}[b]{0.49\textwidth}
         \centering
         \includegraphics[width=.95\textwidth]{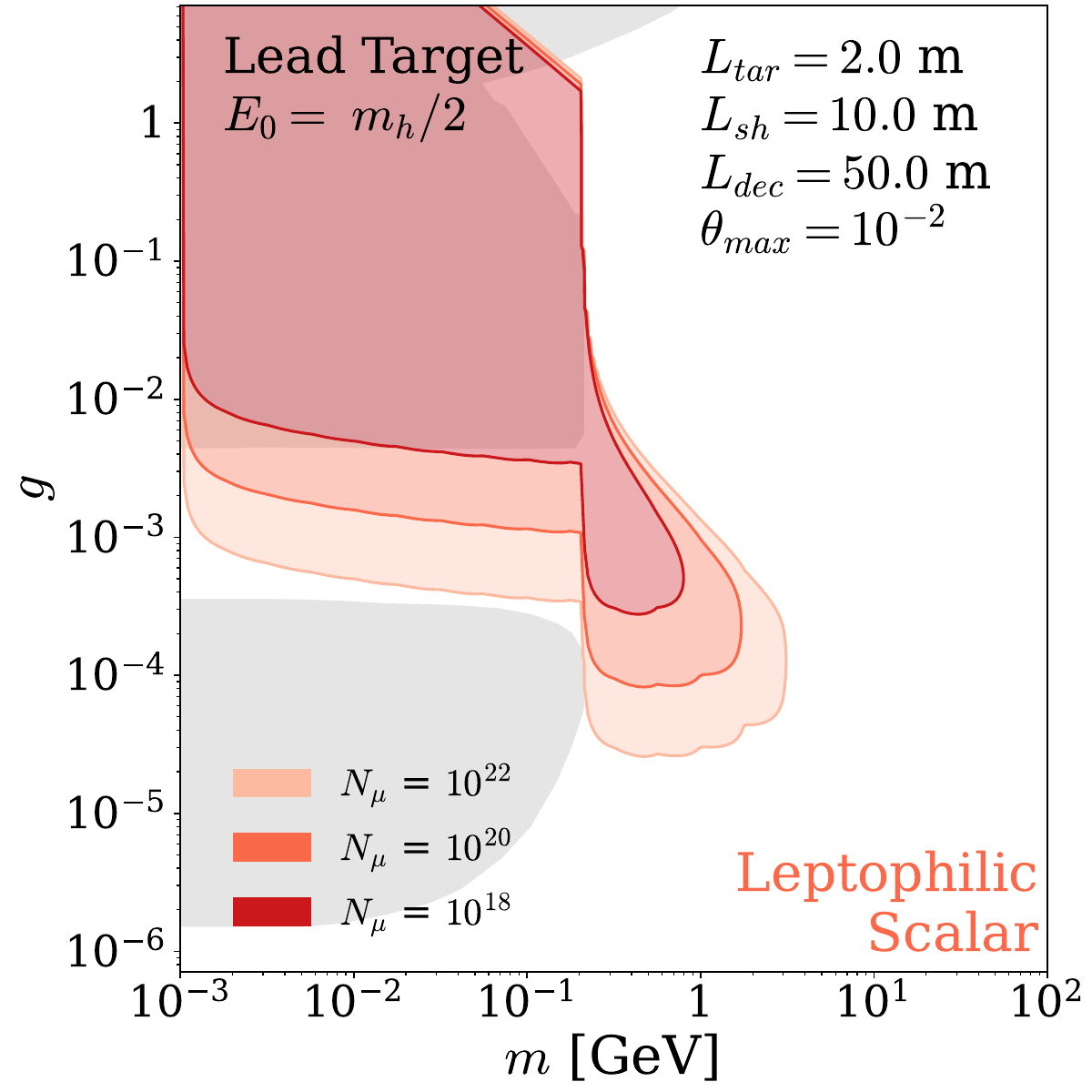}
         \caption{Scalar Production}
     \end{subfigure}
     \hfill
     \begin{subfigure}[b]{0.49\textwidth}
         \centering
         \includegraphics[width=.95\textwidth]{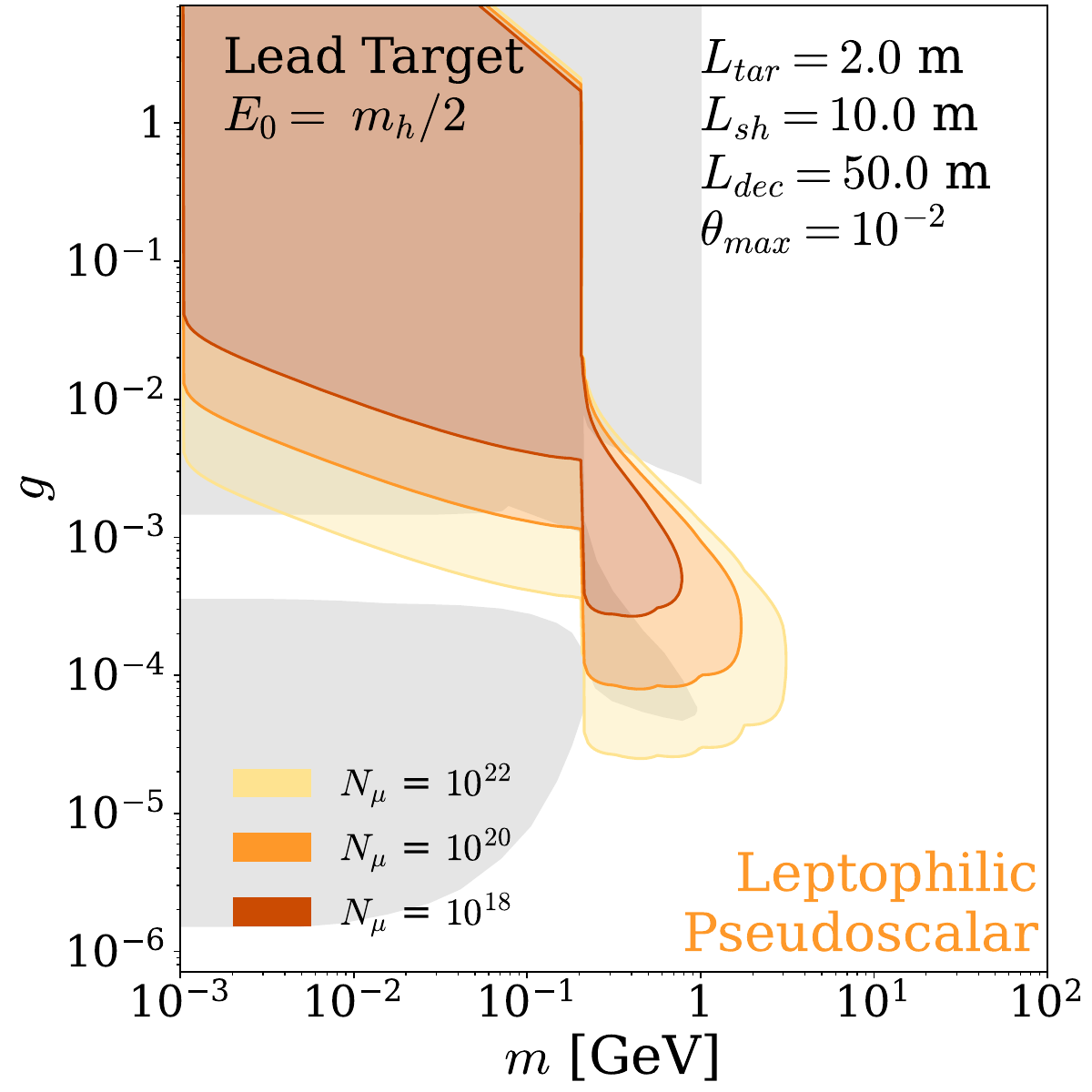}
         \caption{Pseudoscalar Production}
     \end{subfigure}
        \caption{The reach plots for a 63 GeV muon beam ($m_h$/2) on a lead target for the various particles species with a leptophilic coupling as defined in \Eq{eq:leptoCoupling}. We consider only the scalar and pseudoscalars out of theoretical motivation. The 10 GeV scenario is not shown as it does not cover new parameter space.}
        \label{fig:leptoHiggs}
\end{figure}
\begin{figure}
\centering
     \begin{subfigure}[b]{0.49\textwidth}
         \centering
         \includegraphics[width=.95\textwidth]{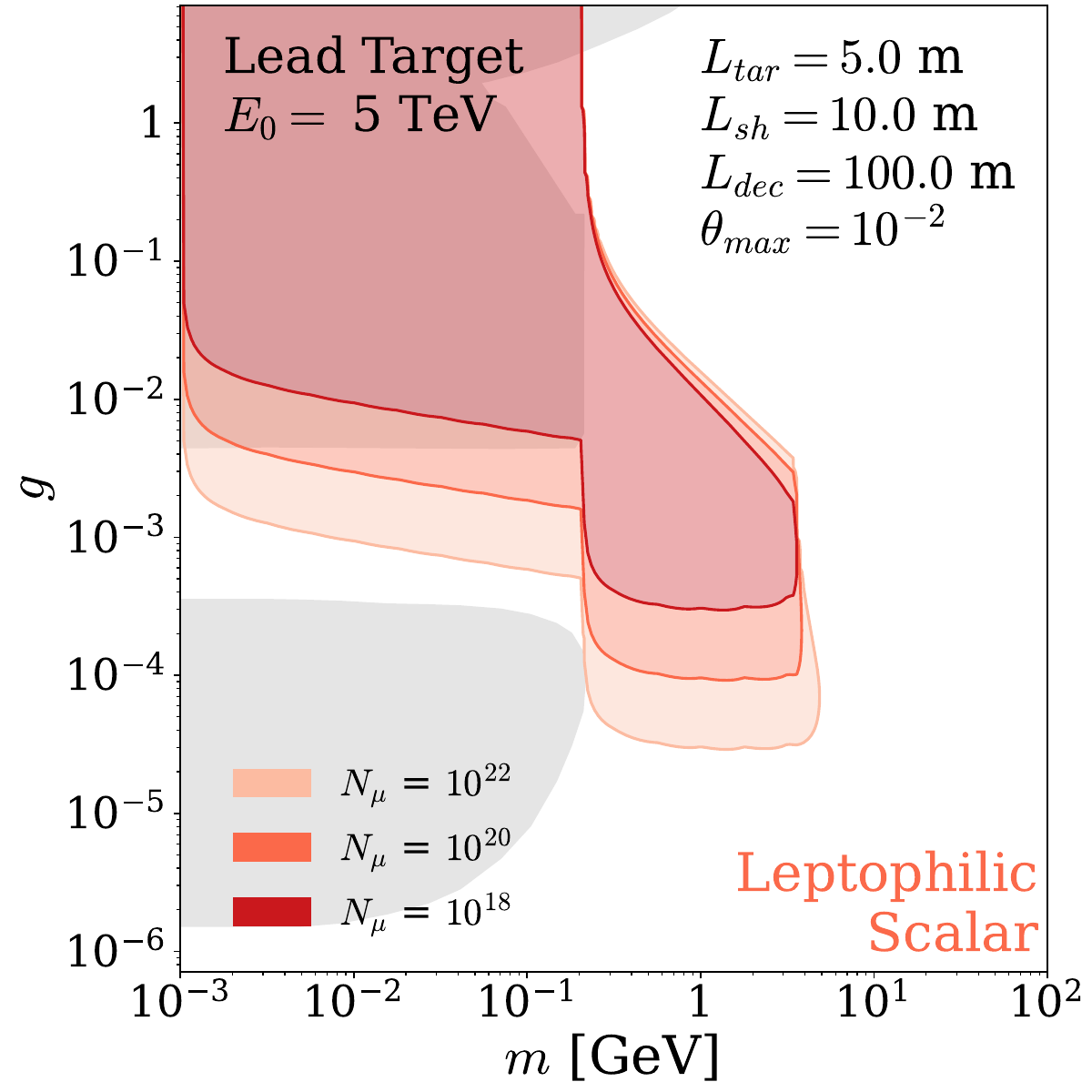}
         \caption{Scalar Production}
     \end{subfigure}
     \hfill
     \begin{subfigure}[b]{0.49\textwidth}s
         \centering
         \includegraphics[width=.95\textwidth]{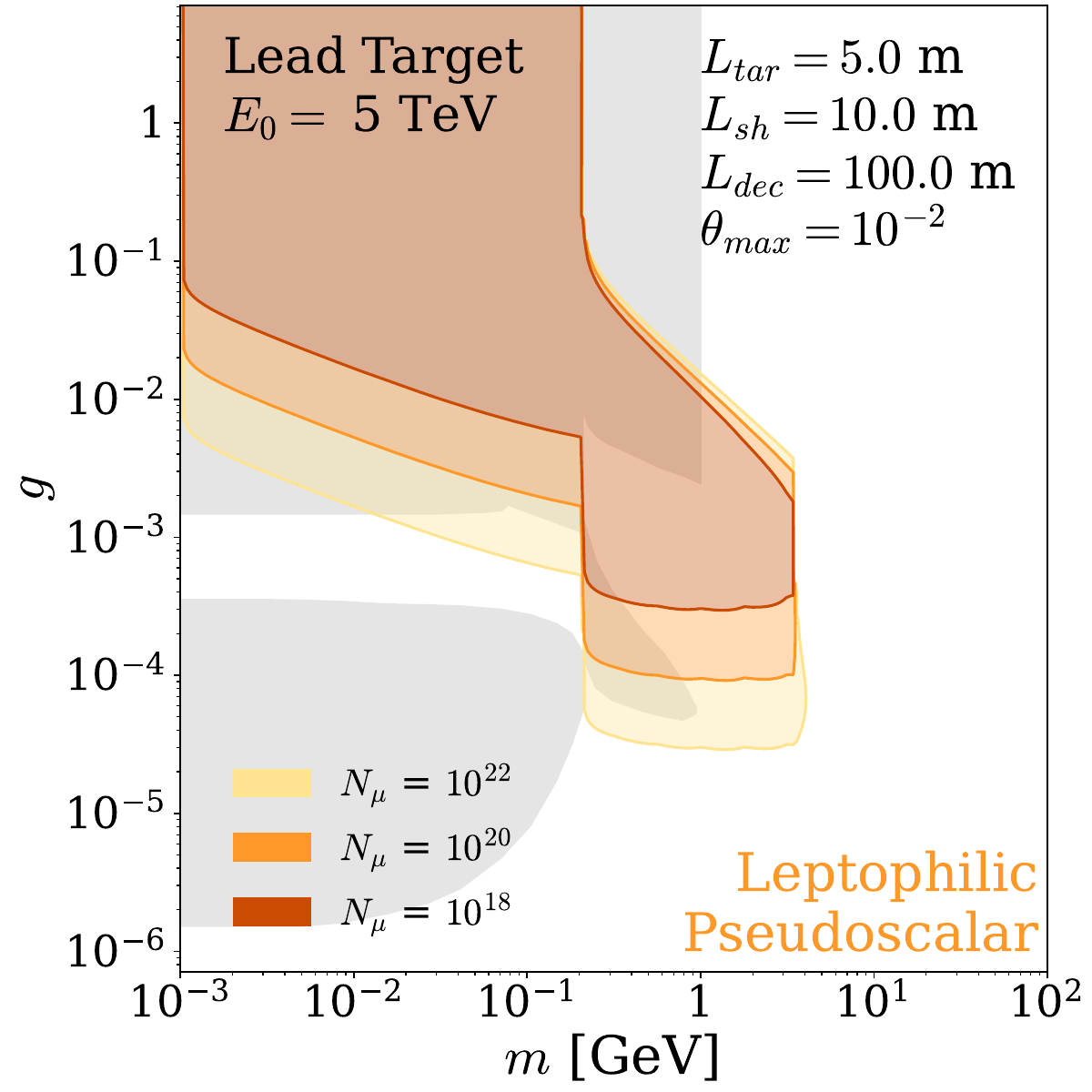}
         \caption{Pseudoscalar Production}
     \end{subfigure}
        \caption{The same as Fig.~\ref{fig:leptoHiggs} but with a beam energy of 5 TeV.}
        \label{fig:lepto5TeV}
\end{figure}

We only consider new physics particles that are either scalars or pseudoscalars. 
A scalar portal with mass-like coupling structure can arise straightforwardly from mixing with the Higgs. 
To accomplish strictly leptophilic couplings, additional complexity must be added, such as appending the SM with `lepton-specific' two Higgs Doublet Models, as detailed in Refs.~\cite{Batell:2016ove, Su:2009fz, Marshall:2010qi, Branco:2011iw, Chun:2015hsa, Chun:2016hzs}. 
Pseudoscalars of this variety can arise from an approximate symmetry breaking at a high scale $\Lambda$ such that the coupling to SM particles will go as $g \sim m_l / \Lambda$ \cite{Essig:2010gu}.
This new physics scenario appears in many phenomenological models, such as extensions of the seesaw mechanism \cite{Alonso:2011jd}, as possible dark matter portals \cite{Kopp:2009et}, and to address the ongoing $g-2$ anomaly (for example, Refs.~\cite{ PhysRevD.93.035006, Marsicano:2018vin}).

We present the results for this model at beam energies of $E_0 = m_h/2$, $5$ TeV as lower energies are already excluded by previous experiments\footnote{Note that for the leptophilic pseudoscalar the existing constraints from meson decays discussed in Ref.~\cite{Essig:2010gu} is the bound with the sharp cut-off at 1 GeV. As the authors refer to themselves as non-experts reinterpreting this constraint, we do not attempt to demonstrate our further removal from expert status by extending the curve, although it likely cuts into the sensitivity of the higher-energy beam further than what is plotted.} in Fig.~\ref{fig:leptoHiggs} and Fig.~\ref{fig:lepto5TeV}. 
However, there is still non-trivial coverage by a low energy muon beam as sensitivity to this model is parametrically increased by the utilization of a muon beam rather than a proton or electron beam.

\subsection{$L_\mu - L_\tau$}
\begin{figure}
\centering
     \begin{subfigure}[b]{0.49\textwidth}
         \centering
         \includegraphics[width=.95\textwidth]{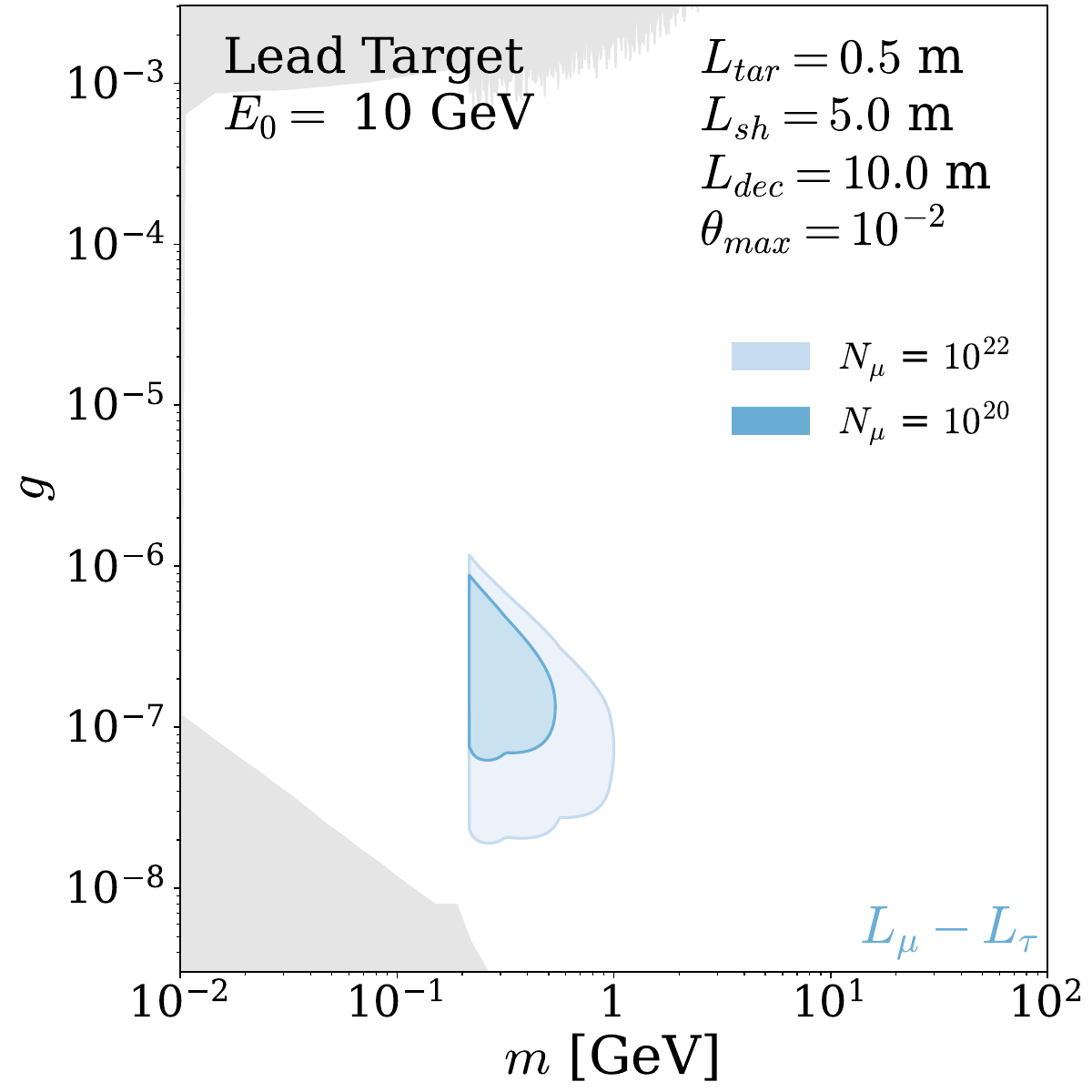}
         \caption{10 GeV Beam}
     \end{subfigure}
     \hfill
     \begin{subfigure}[b]{0.49\textwidth}
         \centering
         \includegraphics[width=.95\textwidth]{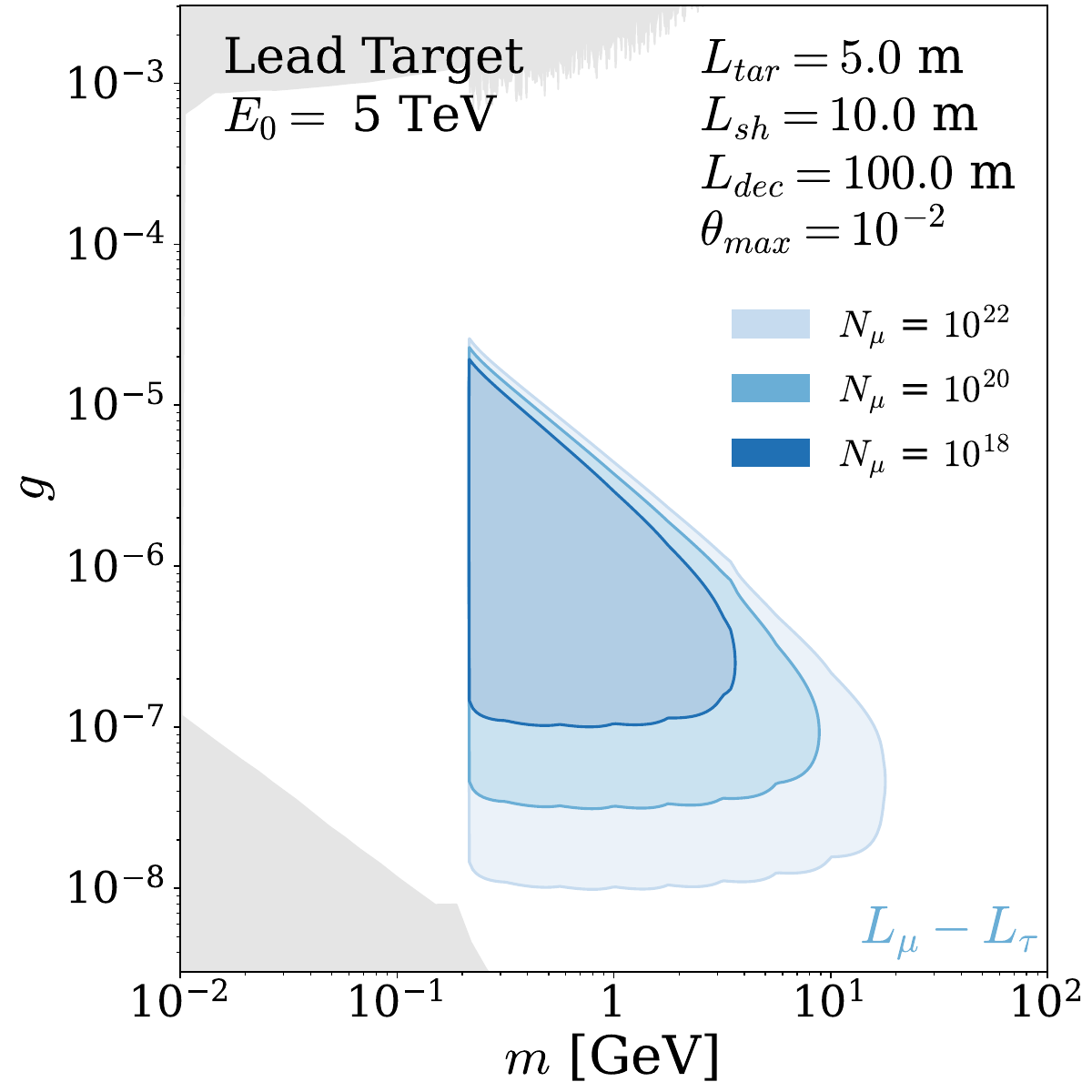}
         \caption{5 TeV Beam}
     \end{subfigure}
        \caption{Sensitivity to the vector boson of the gauged $L_\mu-L_\tau$ symmetry for a low-energy beam (a) and high-energy beam (b). While the low energy requires high luminosity to achieve sensitivity, all of the possible reach is unconstrained for both energy considerations.}
        \label{fig:gaugeLmuLtau}
\end{figure}

Another new physics model of interest is a gauged $L_\mu - L_\tau$ symmetry. 
In this model, we gauge the anomaly-free lepton number symmetry \cite{PhysRevD.43.R22, Foot:1990mn, PhysRevD.44.2118, Ma:2001md, Huang:2021nkl, Dasgupta:2023zrh, Bauer:2018onh}.
The new gauge boson, which we will call $Z'$, only couples to muons and muon neutrinos with charge $+1$, and taus and tau neutrinos with charge $-1$: 
\begin{equation}
\mathcal{L} \supset \frac{1}{2} m_{Z'}^2 Z'^\mu Z'_\mu  - i g \bigg[ \left( \bar{\mu} \gamma^\rho \mu  + \nu_\mu^\dagger \bar{\sigma}^\rho \nu_\mu \right) - \left( \bar{\tau} \gamma^\rho \tau  + \nu_\tau^\dagger \bar{\sigma}^\rho \nu_\tau \right) \bigg] Z'_\rho.
\label{eq:Llmultau}
\end{equation}
This model is interesting not only because it is largely unconstrained, but also because it could explain ongoing mysteries of the SM, such as the near maximal mixing between muon and tau neutrinos \cite{Ma:2001md}. 
The sensitivity to a vector boson of the gauged $L_\mu - L_\tau$ is shown in Fig.~\ref{fig:gaugeLmuLtau}.
The reach is significantly improved by a high-energy muon beam, as the strongest previous constraints came from couplings to neutrinos.  
%
\subsection{Dark Photon}
Finally, we consider a dark photon model. 
The generic dark photon (also denoted here as $Z'$) model is a common target for beam-dump experiments. 
In this model, we assume that the SM is extended by a $U(1)'$ gauge symmetry with a gauge boson $Z'$.
While SM particles are not explicitly charged under this new symmetry, the $Z'$ can mix with the photon and couple to the SM charged current. 
Without loss of generality, we can write the Lagrangian in the following basis: 
\begin{equation}
\mathcal{L} \supset \frac{1}{2} m_{Z'}^2 Z'^\mu Z'_\mu  - \sum_{l \in e, \mu, \tau} i \epsilon e \left(   \bar{l} \gamma^\mu l \right) Z'_\mu
\end{equation}
where $e$ is the SM electromagnetic charge. 

Note that since the $Z'$ couples to all charged particles, it will also interact with the quarks and decay into hadronic final states. 
We compute the width into hadrons using the parameter $ R \equiv \sigma (e^+ e^- \rightarrow \text{hadrons}) / \sigma (e^+e^- \rightarrow \mu^+\mu^-)$ \cite{Workman:2022ynf}. 
While we could construct a dark photon model that only couples to the charged leptons, for example see Ref.~\cite{Fox:2008kb}, we consider the more generic case as is often searched for at beam dumps. 
However, the bounds that could be set on this more complex theory would be stronger than the generic dark photon. 

Because the $Z'$ couples to all charged particles, many previous experimental set-ups have been able to probe the parameter space. 
The advantage of using a muon collider or muon collider staging would not be due to muons specifically, but rather the high energy and high intensity of the beam. 

\begin{figure}
\centering
     \begin{subfigure}[b]{0.49\textwidth}
         \centering
         \includegraphics[width=.95\textwidth]{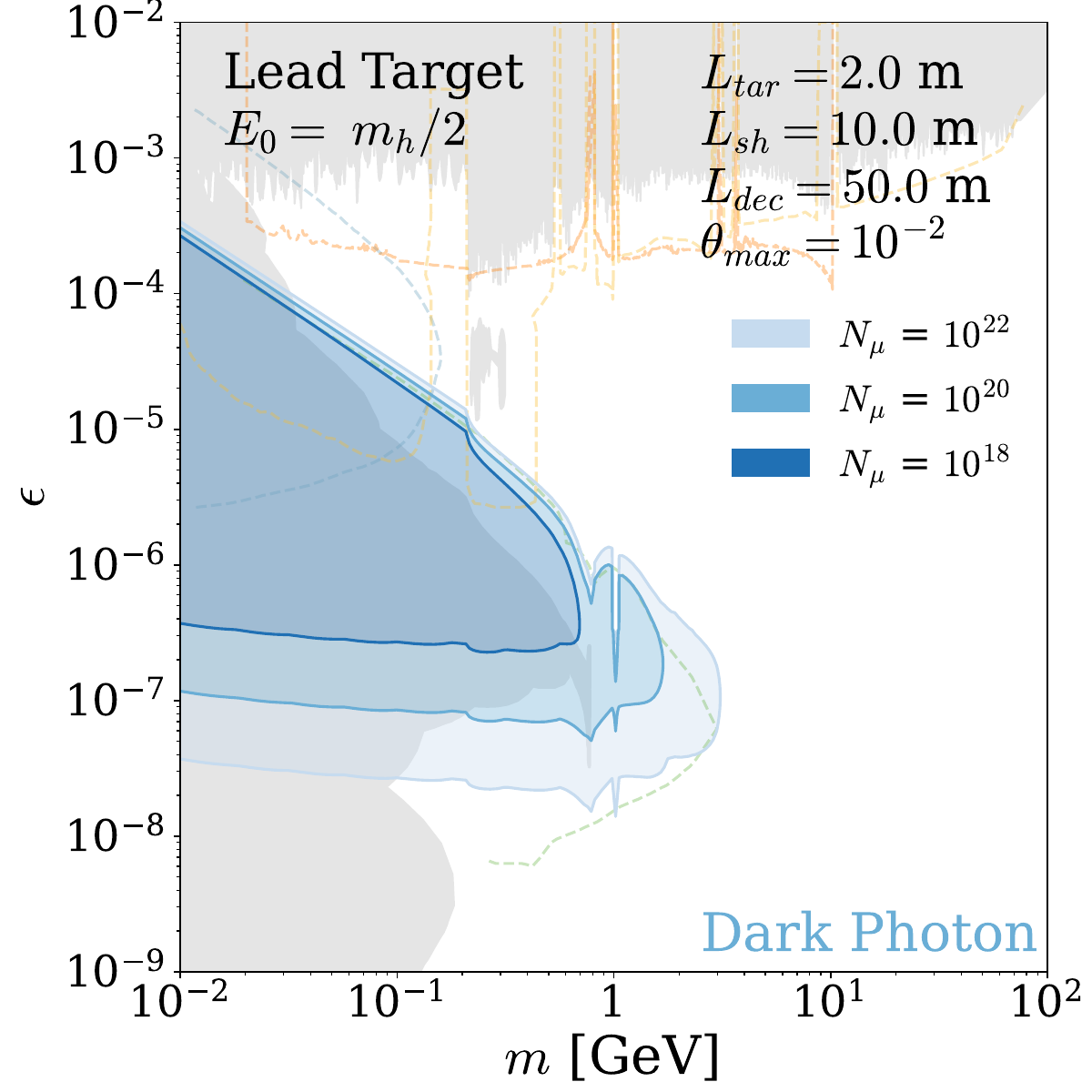}
         \caption{Dark photon limits at $E_0 = m_h/2$.}
         \label{fig:DPlow}
     \end{subfigure}
     \hfill
     \begin{subfigure}[b]{0.49\textwidth}
         \centering
         \includegraphics[width=.95\textwidth]{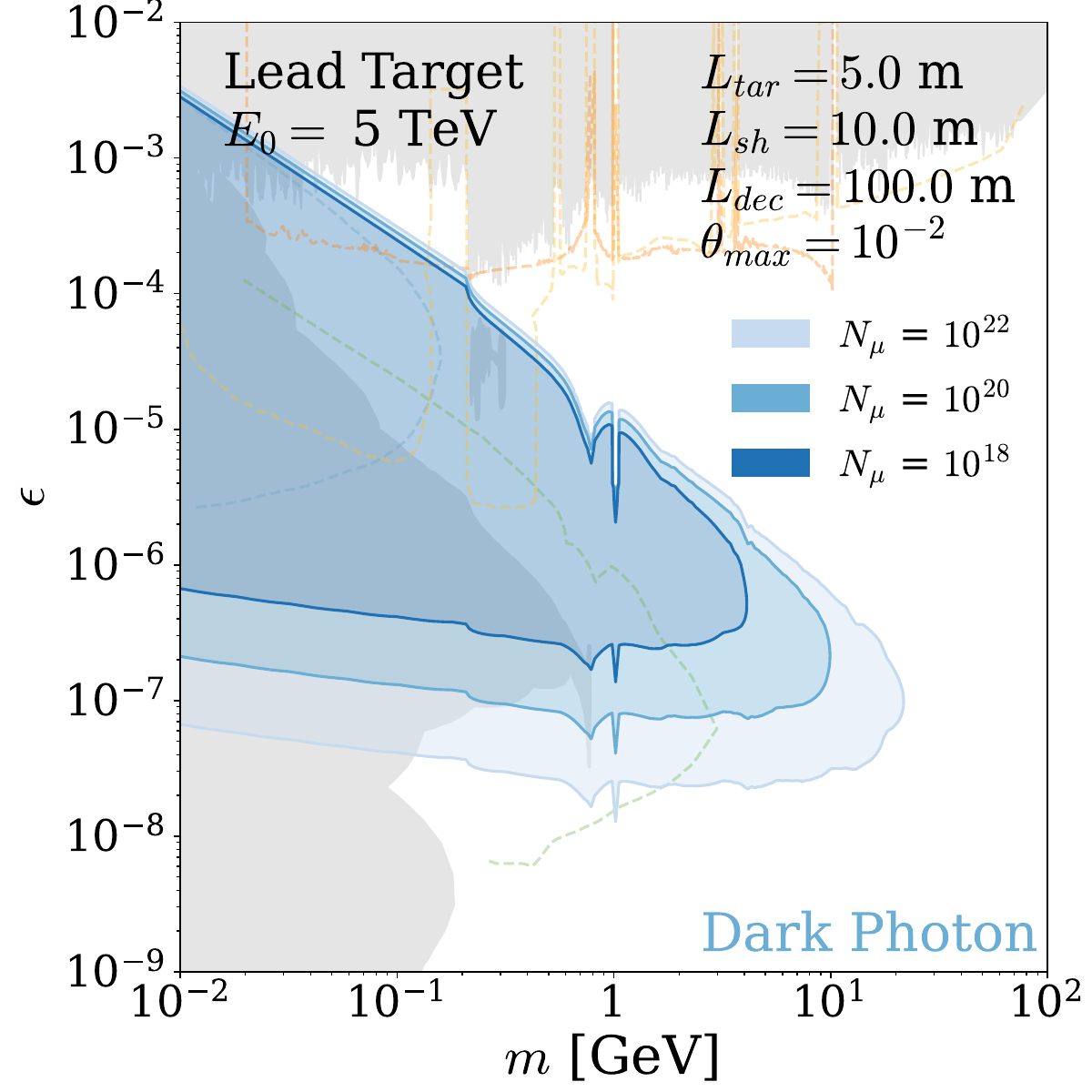}
         \caption{Dark photon limits at $E_0 = 5$ TeV.}
         \label{fig:DPhigh}
     \end{subfigure}
        \caption{Limits on a dark photon model for low (a) and high (b) energy beams. We do not present the 10 GeV scenario as no new parameter space is covered. An in-depth description of the various existing constraints can be found in Ref.~\cite{Batell:2022dpx}. }
        \label{fig:darkPhoton}
\end{figure}

We present the dark photon limits for the full 10 TeV collider as well as for a benchmark study at $E_0 = m_h/2$ in Fig.~\ref{fig:darkPhoton}. 
While the improvement for the high energy beam is clearly much more dramatic, one should note that even intense charged-particle beams with energy near 100 GeV could improve the reach. 
Such a beam configuration, which would provide too low of luminosity to be an extended collider run, is still a conceivable staging option given the required R\&D studies to be conducted. 
%

\section{Conclusions}
\label{sec:conc}
As we consider the future of high-energy particle physics experiments, it is imperative that we consider how to optimize the physics output and discovery potential of any new collider. 
A muon collider offers several unique advantages to probing unexplored regions of parameter space due to the unprecedented energy and second-generation particle beam. 
Utilizing the beam for not only particle-antiparticle collisions but also for beam-dump experiments is an economical and efficient way to expand the new physics reach of a muon collider in complementary regimes.
Moreover, since many R\&D studies on the muon beam must occur in the decades leading up to the full 10 TeV collider, there will be other, lower energy muon beams that can be used for beam dumps on much shorter time scales. 

We show that even with moderately low-energy beams ($\gtrsim \mathcal{O}(10)$ GeV), we can improve the sensitivity to many robust new physics models. 
The muon beam enables us to be sensitive to either specifically muonphilic models or models with couplings that are stronger for the higher generation particles, such as the gauged $L_\mu - L_\tau$ symmetry or a new scalar with Yukawa-like couplings.
As discussed, these models are of interest as they could explain open questions in the SM, such as the persistent $g-2$ anomaly \cite{Muong-2:2023cdq}. 
For the high-energy scenario, we can extend the reach on even well-searched new physics scenarios such as a dark photon: the large $\gamma$ extends the lifetime of the particles and we can thus measure signals of short-lived particles, and the increased center-of-mass energy of the scattering process allows the on-shell creation of heavier particles. 

In this work we focus on new physics scenarios where a particle is radiated off the muons, but these are not the only interesting particles produced at beam dumps.
We mentioned previously that with a $\sim 10$ GeV muon beam, one could construct a muon storage ring for short-baseline neutrino oscillation experiments \cite{nuSTORM:2022div}.
Additionally, beam dumps and fixed-target experiments are often used or proposed to measure precision observables of the SM, such as rare meson decays (for example, \cite{NA62:2017rwk, Bernhard:2019jqz, Dery:2021mct}).
The production of muons demands an intense proton beam on target, which will also generate copious amounts of kaons. 
In the spirit of maximizing returns on the way to a future collider, we encourage exploration towards probing rare SM processes with the same experimental infrastructure to expand the full physics program. 

We conclude by reiterating the results of this work are intended to be taken as a proof-of-concept rather than an experimental prescription. 
In order to fully account for backgrounds and optimize the geometry of the setup, $\textsc{GEANT}$ simulations should be run for a more sophisticated understanding of the signal and background. 
Many elements of the experiment are presented in a simplified manner here, such as the target, which will likely need dedicated targetry studies to ensure its endurance in the beam, as well as the shielding region to reject SM background from the target, bent beam, and dumped beam after being swept out of acceptance. 
We also take a naive efficiency of 100\% which we attempt to compensate for by drawing the detection contours at 5 events. 
Certainly this number is optimistic and the results should only be taken as order-of-magnitude projections. 
Note that we also report signal for only the $e^+e^-$ and $\mu^+\mu^-$ final states. 
To identify the boosted charged leptons and reconstruct their individual energies to perform a missing-mass analysis, likely a magnetic field or tracker would be necessary in addition to a calorimeter. 
With more instrumentation, it may be possible to consider other, more complicated final states such as $\tau^+ \tau^- \rightarrow \mu^+ \mu^- + \slashed{E}$ or hadrons. 
This could also improve the sensitivity of the experiments considerably and is worth investigating in future studies.

In the current era of particle physics, it is unclear where the next breakthrough will occur. 
It is therefore of the utmost importance to the community to search broadly and robustly for signs of new physics to better our chances of discovery. 
The discovery of a new extremely weakly-coupled particle at a beam-dump experiment could have huge implications for physics in the UV \cite{Arkani-Hamed:2006emk} as well as provide novel insight to outstanding problems such as dark matter or the hierarchy problem that have persisted for decades. 
Thus a beam-dump auxiliary program is not only a responsible application of the full-scale collider and staging facilities, but also a powerful probe towards uncovering completely new phenomena in the natural world.

\section*{Code}

The code to perform all numeric calculations presented in this paper is available publicly at \url{https://github.com/rikab/MuonBeamDump}. 
In particular, the results and plots of Section~\ref{sec:csc} can be reproduced with \url{https://github.com/rikab/MuonBeamDump/blob/main/cross_section_plots.ipynb}, and the results and plots of Section~\ref{sec:results} can be reproduced with \url{https://github.com/rikab/MuonBeamDump/blob/main/models.ipynb}.

\section*{Acknowledgements}
We would like to thank Samuel Alipour-Fard for early discussions and contributions to this work.
Additionally we thank Nima Arkani-Hamed, Dario Buttazzo, Nathaniel Craig, Karri DiPetrillo, Samuel Homiller, Sergo Jindariani, Yoni Kahn, Gordan Krnjaic, Patrick Meade, Federico Meloni, Rashmish Mishra, Juli\'an Mu\~noz, Matthew Reece, John Stout, Jesse Thaler, Nhan Tran, and Andrea Wulzer for useful feedback and discussions.
C.C. and R.G. are supported by the U.S. Department of Energy (DOE) Office of High Energy Physics under Grant Contract No. DE-SC0012567.
C.C. would like to thank KITP, IAS, and Cornell University for their hospitality during the completion of this work.

\appendix
\section{Radiative Losses and Material Effects}
\label{app:materials}

The average rate of energy loss per unit length as a particle traverses a distance $x$ in a material is given by is \emph{mass stopping power}, $\rho\expval{-\frac{dE}{dx}}$, where $\rho$ is the material's mass density.
For muons at relativistic energies ($E \gtrsim 1$ GeV), the mass stopping power can be written in the following form~\cite{Workman:2022ynf, RevModPhys.24.133}:
\begin{align}
    \frac{dE}{dx} &= a(E) + b(E)E \label{eq:stopping_power},
\end{align}
where $a(E)$ is the energy loss due to ionization, and $b(E)$ is the energy loss due to bremsstrahlung, $e^+e^-$ pair production, and photonuclear interactions.
These functions are slowly-varying functions of $E$ and can be approximated as constant across a small energy range.
Experimental values of $a(E)$ and $b(E)$ are well-tabulated for a variety of materials across many different energies~ \cite{AtomicNuclearProperties}.

\begin{figure}[ht!]
    \subfloat[]{
        \includegraphics[width=0.45\textwidth]{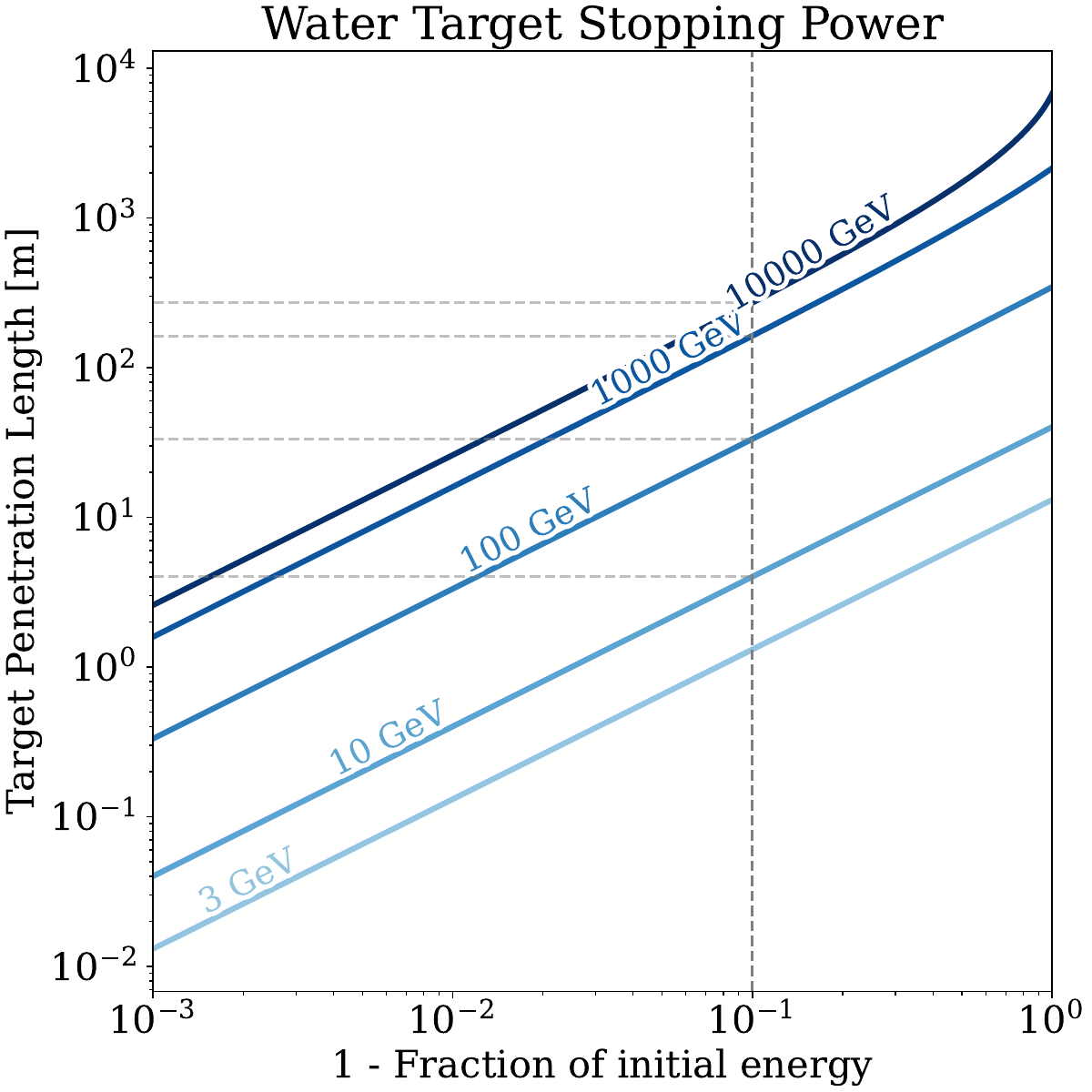}
        \label{fig:water_lengths}
    }
    \hfill
    \subfloat[]{
        \includegraphics[width=0.45\textwidth]{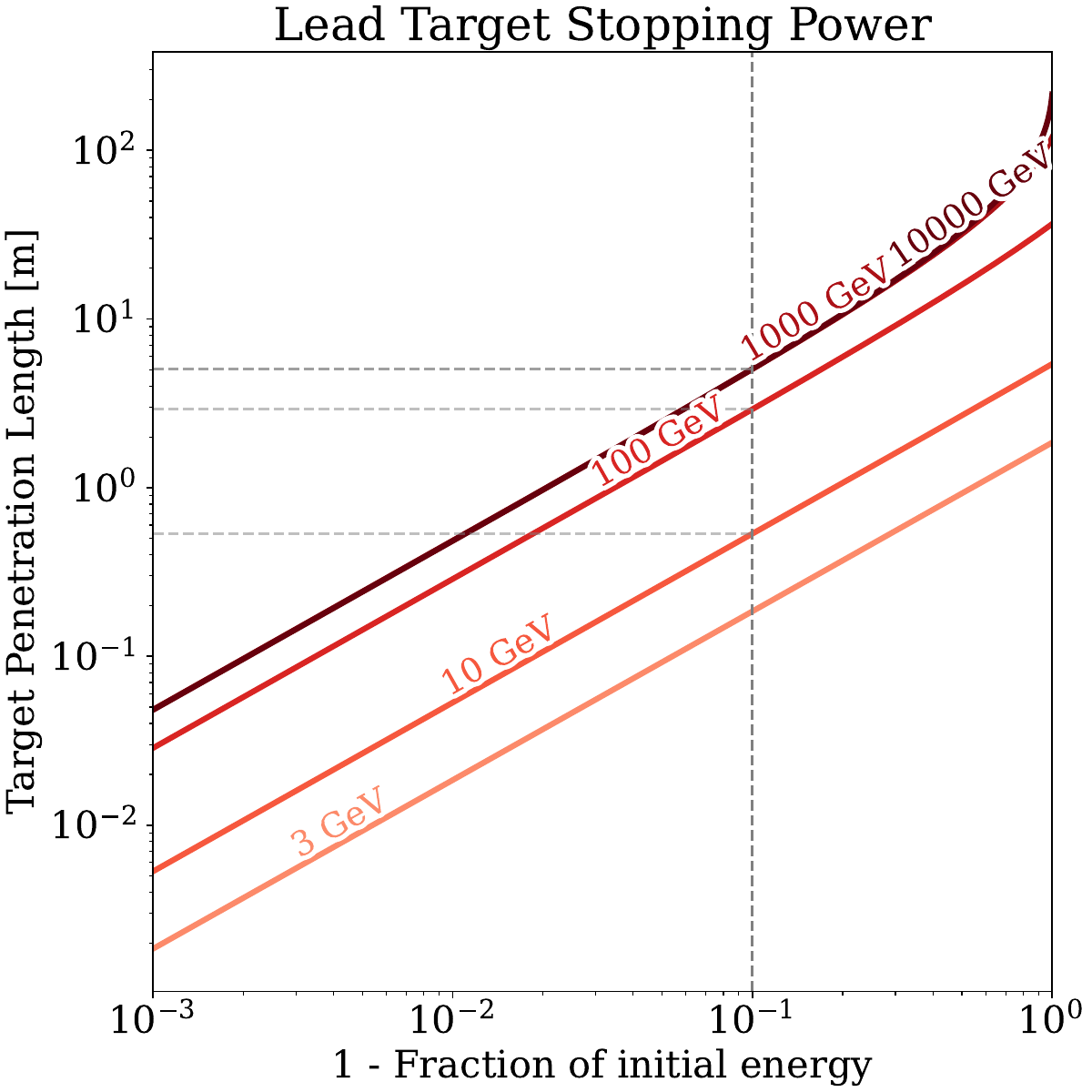}
        \label{fig:lead_lengths}
    }
    \caption{ The penetration depth $R(\lambda)$ as a function of energy fraction $\lambda$ of muons in (a) water and (b) lead, for a variety of initial beam energies.
        The vertical line at $1 - \lambda = 0.1$ corresponds to how far a muon can travel before losing $10\%$ of its energy, as tabulated in Tables \ref{tab:water_length} and \ref{tab:lead_length}.
    }
    \label{fig:radiation_lengths}
\end{figure}

The penetration length $R(\lambda)$ that a muon with initial energy $E_0$ can travel while maintaining a fraction $\lambda$ of its initial energy can be calculated by integrating Eq.~\eqref{eq:stopping_power}:
\begin{align}
    R(\lambda) & = \frac{1}{\rho b(E_0)} \log \left(\frac{a(E_0) + b(E_0) E_0}{a(E_0) + \lambda b(E_0) E_0}\right). \label{eq:radiation_length}
\end{align}
This is most reliable when $\lambda \sim 1$ -- that is, when the final energy $E_f = \lambda E_0$ is close to the initial energy $E_0$ so that $a(E)$ and $b(E)$ do not change significantly.\footnote{The quantity $R(0)$ is commonly called the \emph{Continuous Slowing-Down Approximation} (CSDA) range in the literature, which is the average distance a muon with energy $E_0$ travels \cite{Workman:2022ynf}. }
In tables \ref{tab:water_length} and \ref{tab:lead_length}, we tabulate the values of $a(E)$ and $b(E)$ for a variety of beam energies in liquid water and lead (the two target materials considered in this paper)~\cite{AtomicNuclearPropertiesWater,AtomicNuclearPropertiesLead}, as well as the 90\% penetration depth $R(0.9)$ for each beam energy.
We also show the penetration length $R(\lambda)$ as a function of $1-\lambda$ for a variety of beam energies in Fig.~\ref{fig:radiation_lengths} for both liquid water and lead.

\begin{table}[h]
    \centering
    \begin{tabular}{|c|c c|c|}
        \hline\hline
        Energy [GeV] & $a(E)$ [MeV g/cm$^2$] & $b(E)$ [10$^6$ g/cm$^2$] & 90\% Depth [m] \\
        \hline\hline
        3            & 2.287                 & $\sim1.000$              & 1.31           \\
        10           & 2.482                 & 1.4380                   & 4.01           \\
        100          & 2.781                 & 2.2780                   & 33.36          \\
        1000         & 3.325                 & 2.9575                   & 163.04         \\
        10000        & 3.634                 & 3.4961                   & 271.60         \\
        \hline\hline
    \end{tabular}
    \caption{\label{tab:water_length} PDG values for $a(E)$, $b(E)$~\cite{AtomicNuclearPropertiesWater}, and the 90\% penetration depth for muons (as calculated using Eq.~\eqref{eq:radiation_length}) in liquid water for a variety of initial beam energies.
    }
\end{table}

\begin{table}[h]
    \centering
    \begin{tabular}{|c|c c|c|}
        \hline\hline
        Energy [GeV] & $a(E)$ [MeV g/cm$^2$] & $b(E)$ [10$^6$ g/cm$^2$] & 90\% Depth [m] \\
        \hline\hline
        3            & 1.442                 & $\sim3.000$              & 0.18           \\
        10           & 1.615                 & 6.7899                   & 0.53           \\
        100          & 1.860                 & 12.6448                  & 2.92           \\
        1000         & 2.051                 & 16.4724                  & 5.05           \\
        10000        & 2.251                 & 18.3613                  & 5.06           \\
        \hline\hline
    \end{tabular}
    \caption{\label{tab:lead_length} PDG values for $a(E)$, $b(E)$~\cite{AtomicNuclearPropertiesLead}, and the 90\% penetration depth for muons (as calculated using Eq.~\eqref{eq:radiation_length}) in lead for a variety of initial beam energies.
    }
\end{table}

Using this, we may select suitable target lengths, such that we may use the thin target approximation and reasonably assume $E' = E_0$, as discussed in Sec.~\ref{sec:csc}.
We choose target lengths such that at (on average) muons will lose no more than $10\%$ of their initial energy.
These target lengths are tabulated in Table~\ref{tab:target_lengths} for a variety of beam energies.
We note that the target lengths for lead are shorter than those for water, due to the significantly higher density of lead.
We consider only energies above 10 GeV, as the mass stopping power has significantly different behavoir at low energies, leading to even higher energy loss (this is in addition to the WW approximation being more accurate at higher energies).

\begin{table}[h]
    \centering
    \begin{tabular}{|c|c c|}
        \hline\hline
        Energy [GeV] & Water Target Length [m] & Lead Target Length [m] \\
        \hline\hline
        10           & 4.0                 & 0.5                              \\
        63          & 10.0                 & 2.0                             \\
        1500         & 10.0                 & 5.0                             \\
        5000        & 10.0                 & 5.0                             \\
        \hline\hline
    \end{tabular}
    \caption{\label{tab:target_lengths} Selected target lengths for a variety of beam energies, for both water and lead targets.
    }
\end{table}

Given these target lengths, we can also calculate the total energy deposited into the target per year using $L_{tar} = R(\lambda)$.
This is shown in Table~\ref{tab:target_energy_consumption} for a variety of beam energies for both water and lead targets, assuming $10^{20}$ muons on target per year.
We also calculate the average power deposited into the target, most of which goes into heating the target.
This heat must be dissapated in order to prevent target failure, which should be taken into consideration when designing the target. 

\begin{table}[h]
    \centering
    \begin{tabular}{|c|c c|c c|}
        \hline\hline
         & Water &  & Lead &  \\
        Energy [GeV] & Energy Deposit [$10^9 \frac{\text{J}}{\text{Year}}$] & Power [kW] & Energy Deposit [$10^9 \frac{\text{J}}{\text{Year}}$] & Power [kW] \\
        \hline\hline
        10 & 15.99 & 0.51 & 1.35 & 0.04 \\
        63 & 44.49 & 1.42 & 7.70 & 0.25 \\
        1500 & 125.14 & 3.99 & 214.80 & 6.84 \\
        5000 & 311.04 & 9.91 & 707.42 & 22.53 \\
        \hline\hline
    \end{tabular}
    \caption{\label{tab:target_energy_consumption} Selected target lengths for a variety of beam energies, for both water and lead targets.
    }
\end{table}

\section{Additional Plots}
\label{app:plots}
In this appendix we put a variety of plots supplemental to the body of the text. 
We do not endeavor to describe the plots much as they are simply variations on plots already discussed. 
This section is meant to be used to consider the intermediate energies as well as alternative target materials. 
The behavior is qualitatively similar to the plots shown in the body of the text. 

In Appendix \ref{app:cross_sections}, we show differential cross sections for a variety of beam energies below the 5 TeV shown in Fig.~\ref{fig:xsec}.
In Appendices \ref{app:muonphilic}-\ref{app:darkphoton}, we show model sensitivity plots for lead target beam-dump experiments for more energies beyond those shown in Section \ref{sec:results}.
For a water target, all of these results are very similar.

\subsection{Cross Sections}\label{app:cross_sections}

\begin{figure}[t!]
     \centering
     \begin{subfigure}[b]{0.49\textwidth}
         \centering
         \includegraphics[width=.95\textwidth]{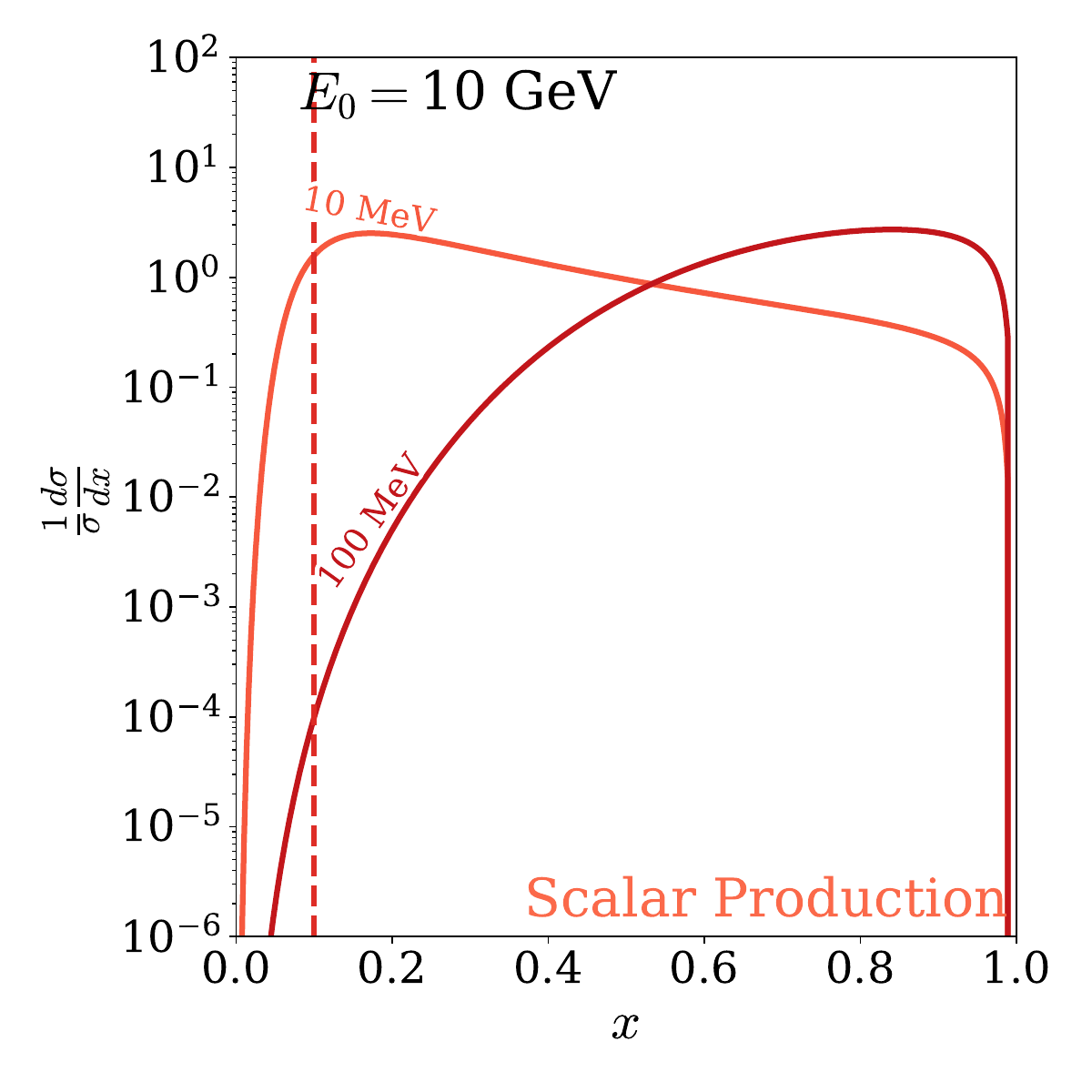}
         \caption{Scalar Production}
         \label{fig:xsecScalar10}
     \end{subfigure}
     \hfill
     \begin{subfigure}[b]{0.49\textwidth}
         \centering
         \includegraphics[width=.95\textwidth]{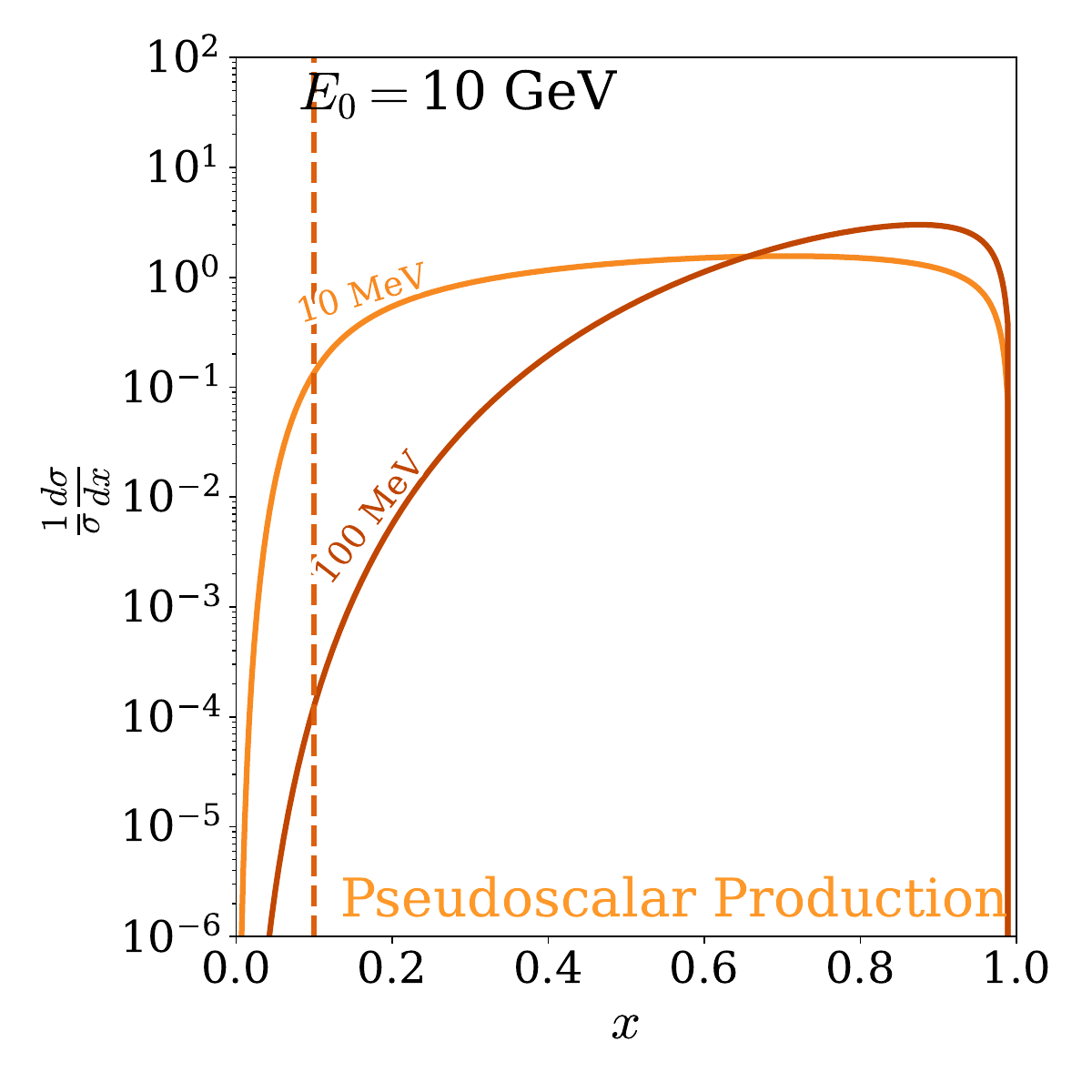}
         \caption{Pseudoscalar Production}
         \label{fig:xSecPS10}
     \end{subfigure}
     \hfill
     \begin{subfigure}[b]{0.49\textwidth}
         \centering
         \includegraphics[width=.95\textwidth]{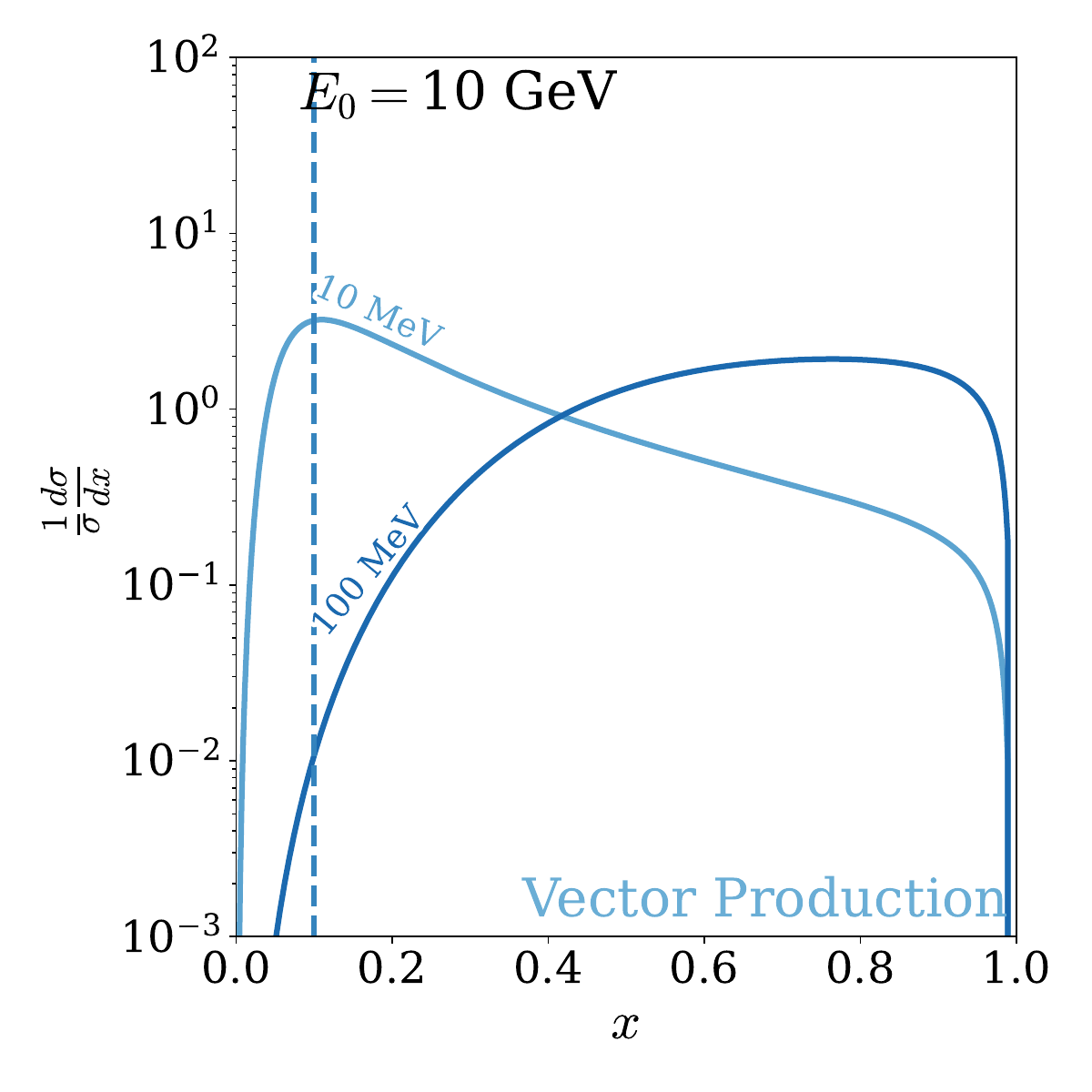}
         \caption{Vector Production}
         \label{fig:xSecV10}
     \end{subfigure}
     \begin{subfigure}[b]{0.49\textwidth}
         \centering
         \includegraphics[width=.95\textwidth]{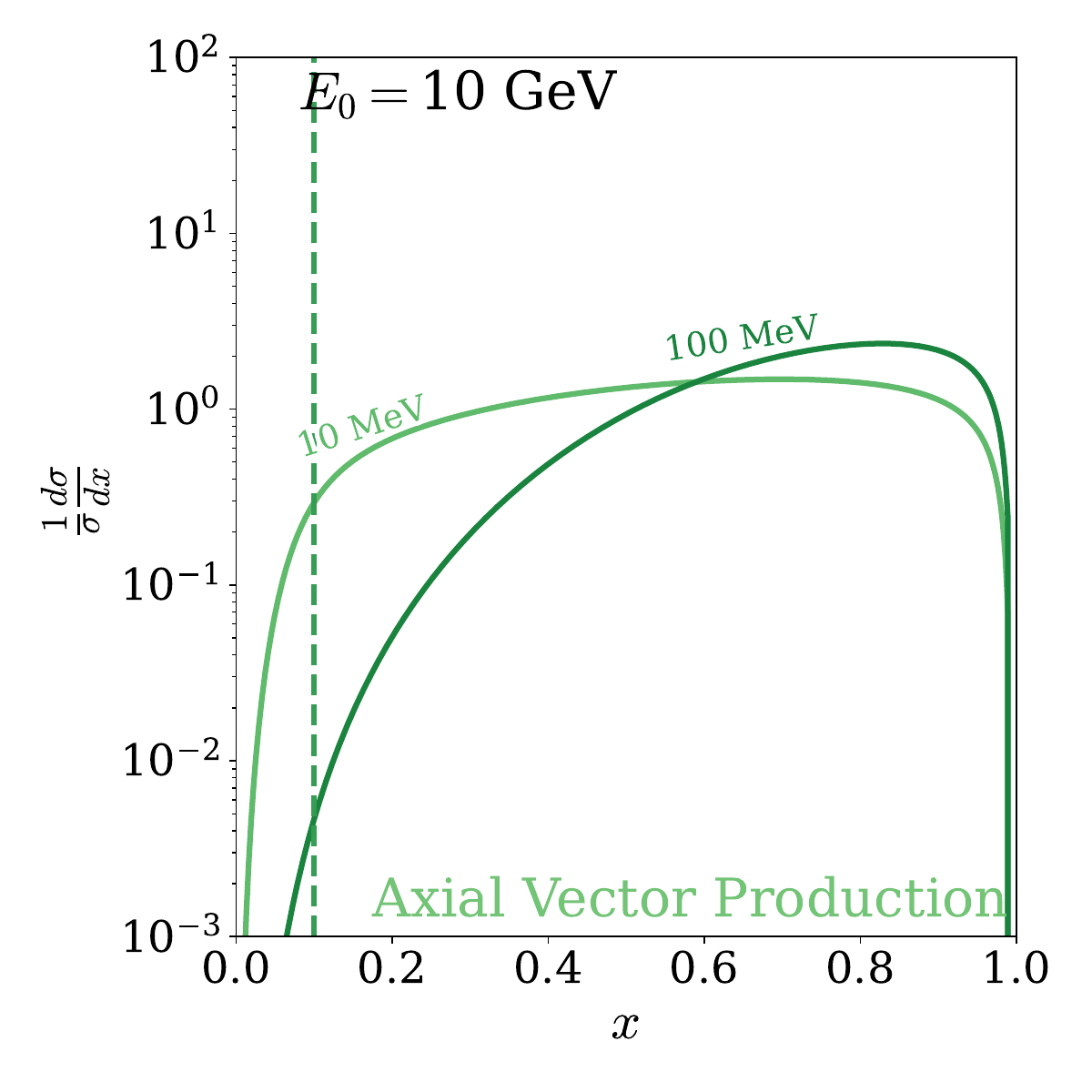}
         \caption{Axial Vector Production}
         \label{fig:xSecAV10}
     \end{subfigure}
        \caption{The same as Fig.~\ref{fig:xsec}, but for a beam energy of 10 GeV. Here, the dashed line corresponds to the threshold at which a 100 MeV particle is produced with $\gamma = 100$.}
        \label{fig:xsec10}
\end{figure}

\begin{figure}[t!]
     \centering
     \begin{subfigure}[b]{0.49\textwidth}
         \centering
         \includegraphics[width=.95\textwidth]{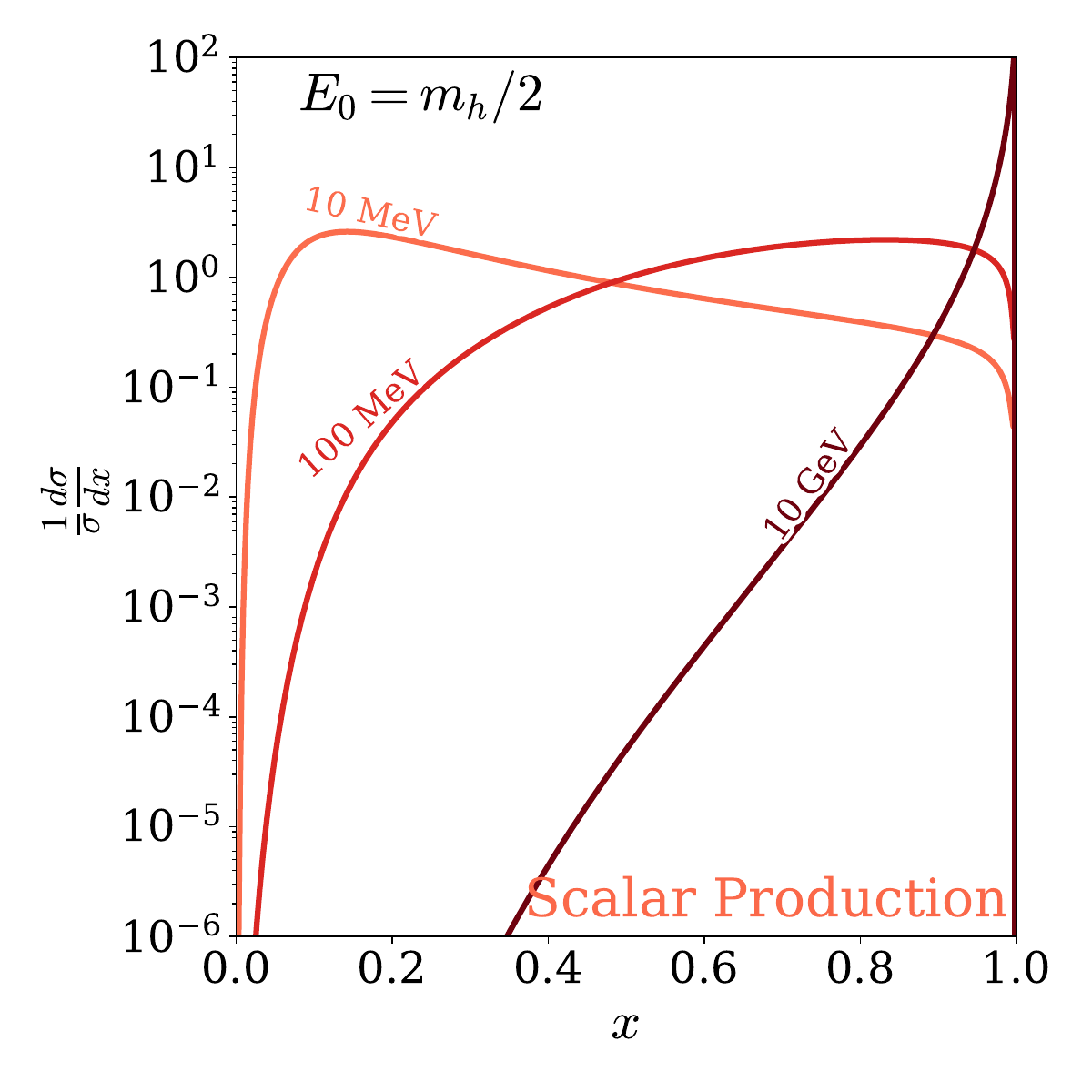}
         \caption{Scalar Production}
         \label{fig:xsecScalar63}
     \end{subfigure}
     \hfill
     \begin{subfigure}[b]{0.49\textwidth}
         \centering
         \includegraphics[width=.95\textwidth]{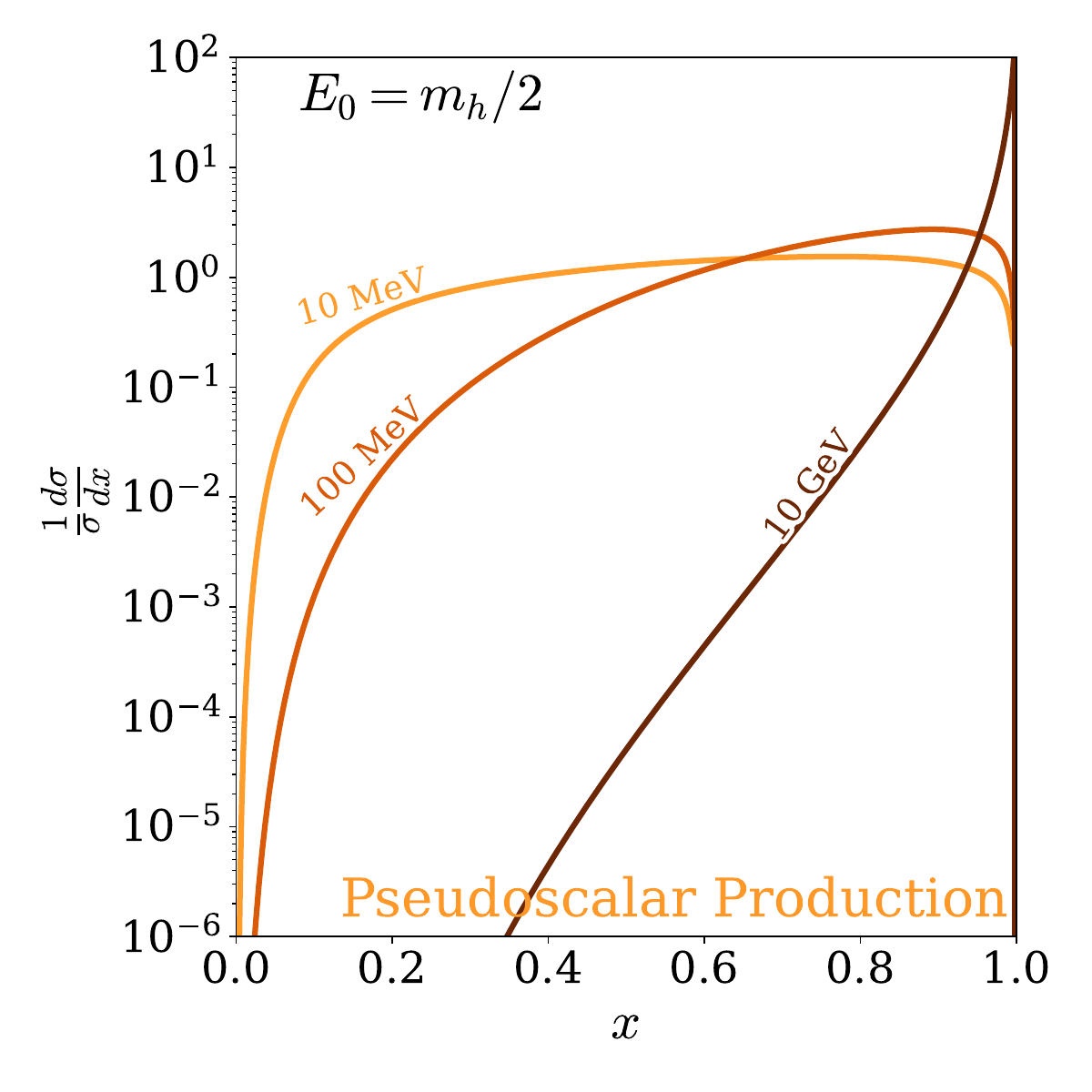}
         \caption{Pseudoscalar Production}
         \label{fig:xSecPS63}
     \end{subfigure}
     \hfill
     \begin{subfigure}[b]{0.49\textwidth}
         \centering
         \includegraphics[width=.95\textwidth]{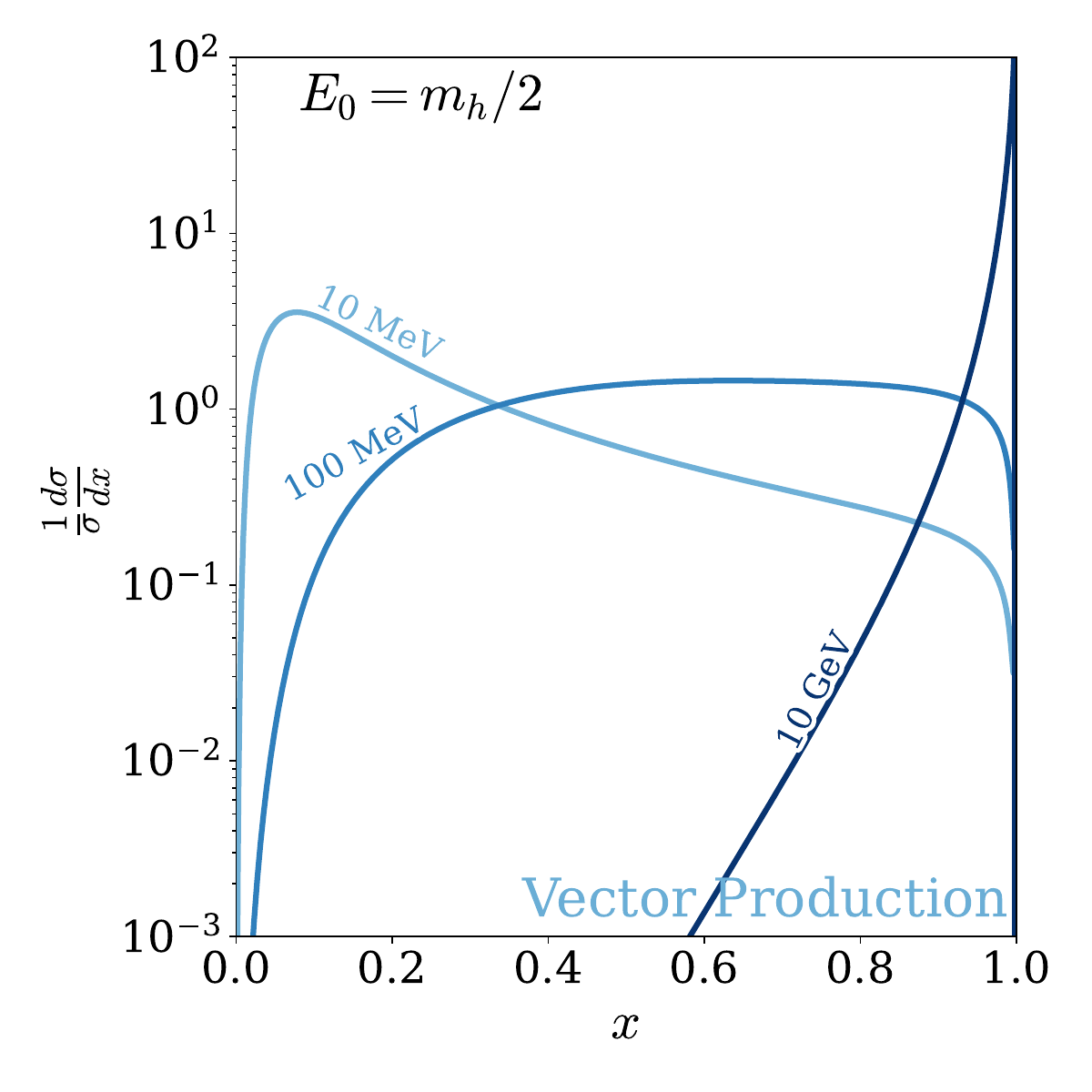}
         \caption{Vector Production}
         \label{fig:xSecV63}
     \end{subfigure}
     \begin{subfigure}[b]{0.49\textwidth}
         \centering
         \includegraphics[width=.95\textwidth]{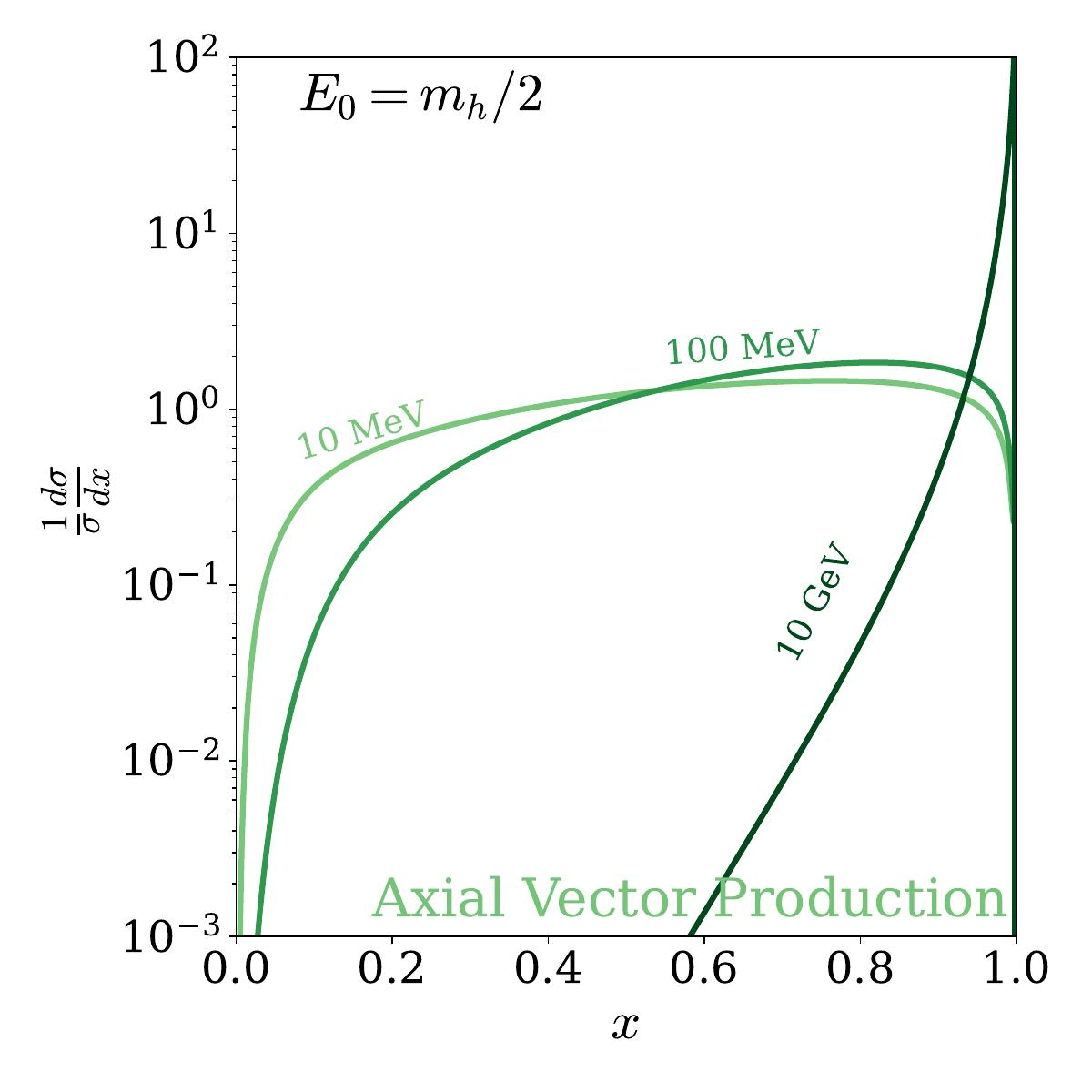}
         \caption{Axial Vector Production}
         \label{fig:xSecAV63}
     \end{subfigure}
        \caption{The same as Fig.~\ref{fig:xsec}, but for a beam energy of $m_h /2$. }
        \label{fig:xsec63}
\end{figure}

\begin{figure}[t!]
     \centering
     \begin{subfigure}[b]{0.49\textwidth}
         \centering
         \includegraphics[width=.95\textwidth]{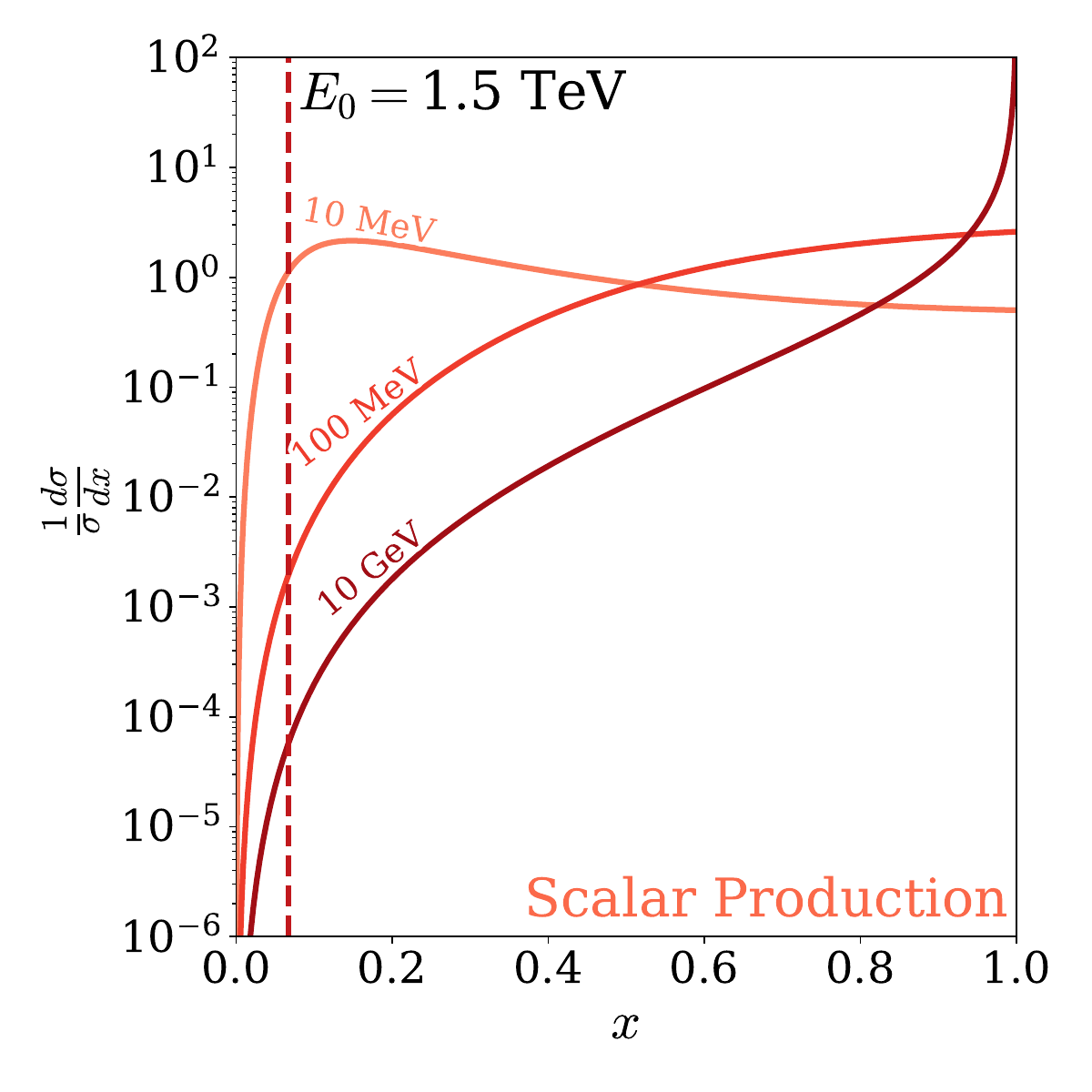}
         \caption{Scalar Production}
         \label{fig:xsecScalar1500}
     \end{subfigure}
     \hfill
     \begin{subfigure}[b]{0.49\textwidth}
         \centering
         \includegraphics[width=.95\textwidth]{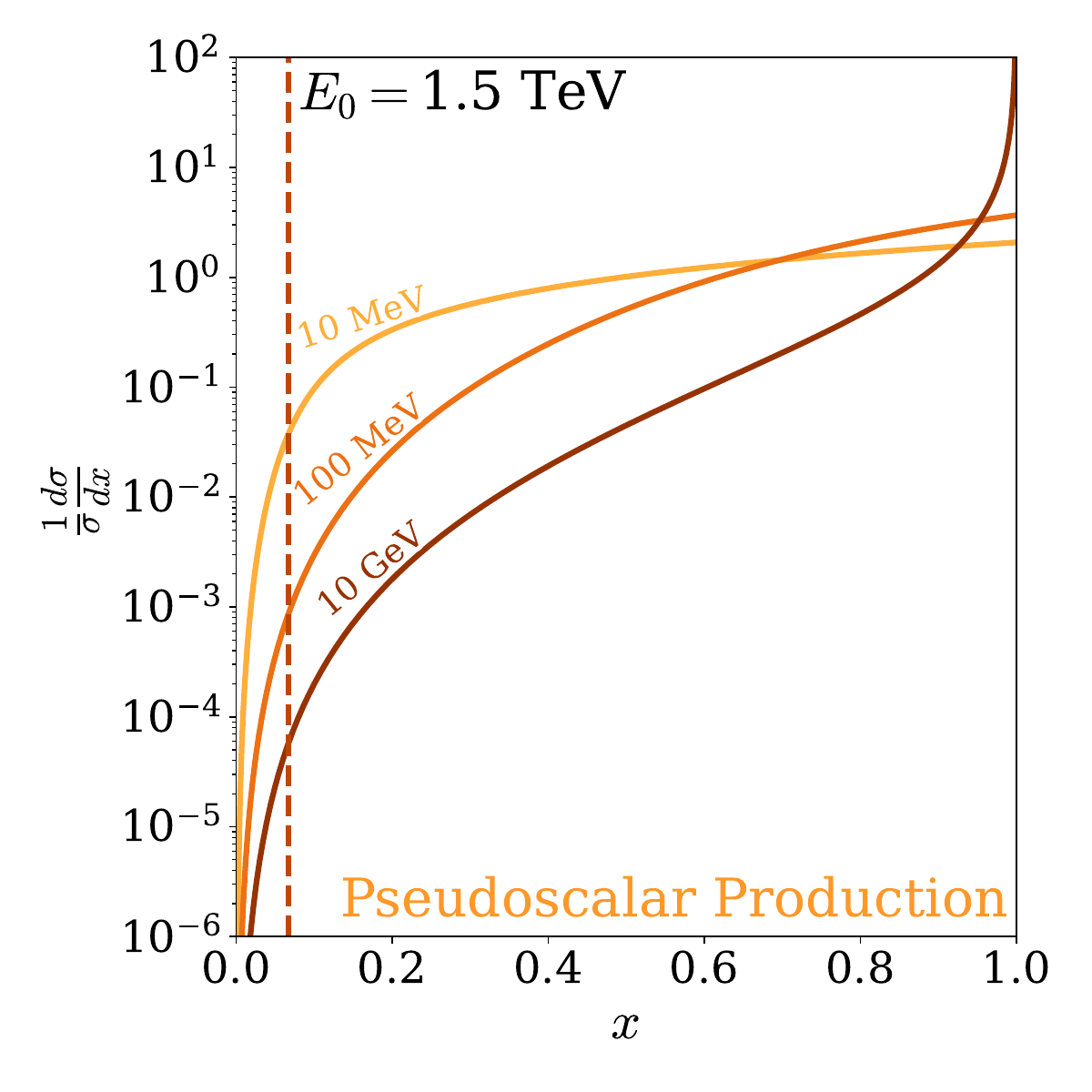}
         \caption{Pseudoscalar Production}
         \label{fig:xSecPS1500}
     \end{subfigure}
     \hfill
     \begin{subfigure}[b]{0.49\textwidth}
         \centering
         \includegraphics[width=.95\textwidth]{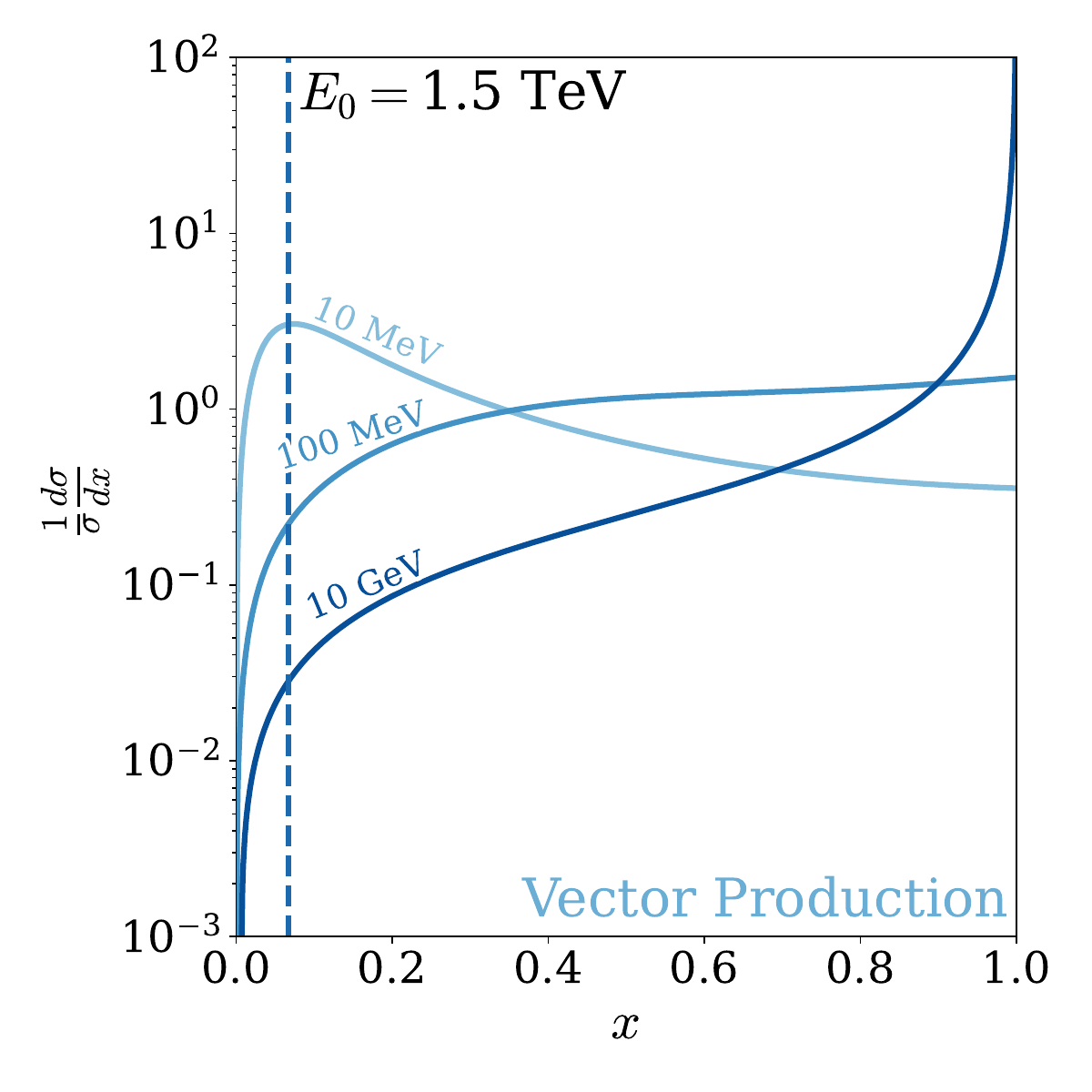}
         \caption{Vector Production}
         \label{fig:xSecV1500}
     \end{subfigure}
     \begin{subfigure}[b]{0.49\textwidth}
         \centering
         \includegraphics[width=.95\textwidth]{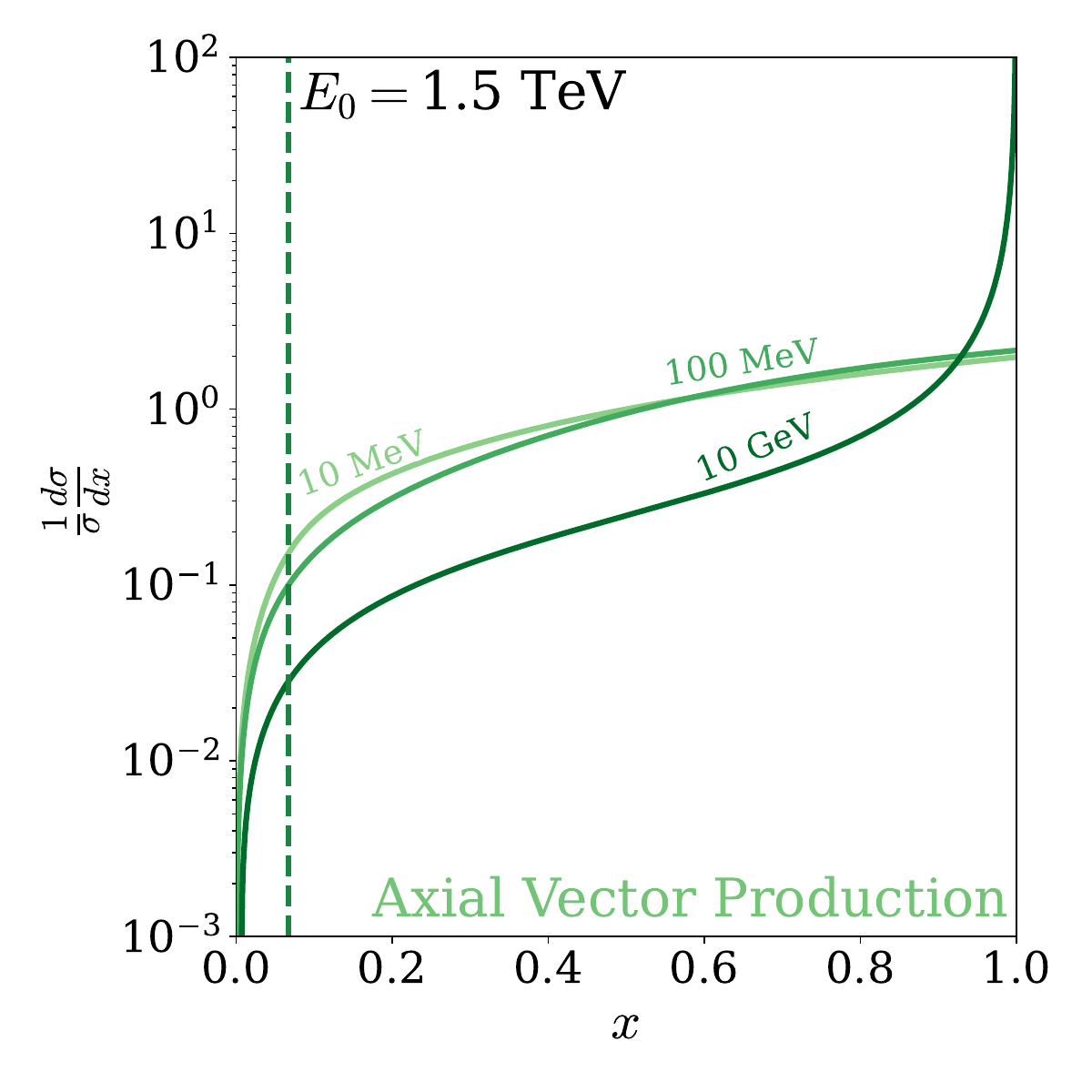}
         \caption{Axial Vector Production}
         \label{fig:xSecAV1500}
     \end{subfigure}
        \caption{The same as Fig.~\ref{fig:xsec}, but for a beam energy of 1.5 TeV. }
        \label{fig:xsec1500}
\end{figure}

Here, we show the normalized differential cross sections for each of the four production modes at beam energies of $E_0 = $ 10 GeV (Fig.~\ref{fig:xsec10}), $m_h/2$ (Fig.~\ref{fig:xsec63}), and 1.5 TeV (Fig.~\ref{fig:xsec1500}).
Note that for the first two cases, we use the full WW approximation rather than the IWW approximation to perform these calculations. 

\clearpage

\clearpage

\clearpage

\subsection{Muonphilic}\label{app:muonphilic}

\begin{figure}[h!]
\centering
     \begin{subfigure}[b]{0.49\textwidth}
         \centering
         \includegraphics[width=.95\textwidth]{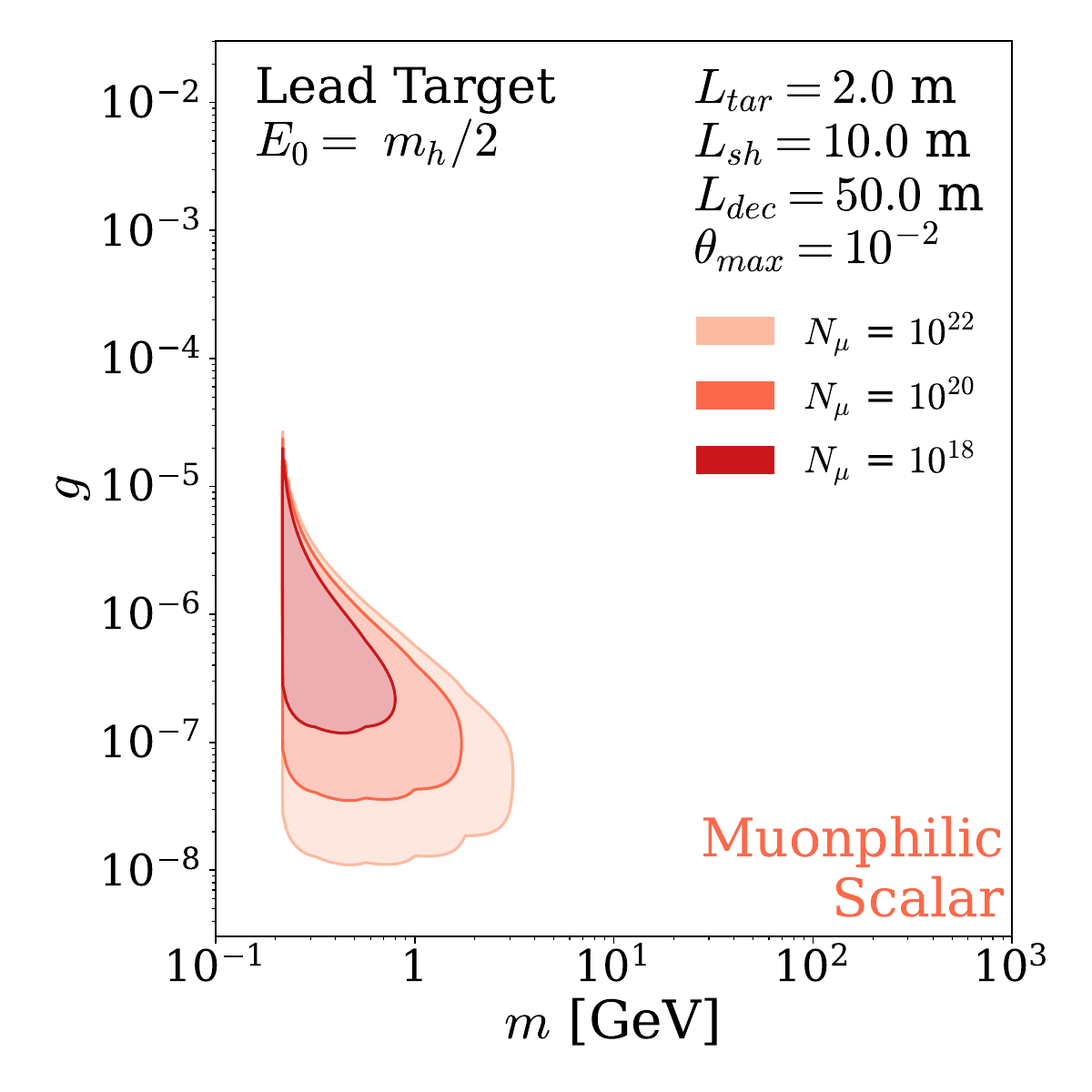}
         \caption{Scalar Production}
     \end{subfigure}
     \hfill
     \begin{subfigure}[b]{0.49\textwidth}
         \centering
         \includegraphics[width=.95\textwidth]{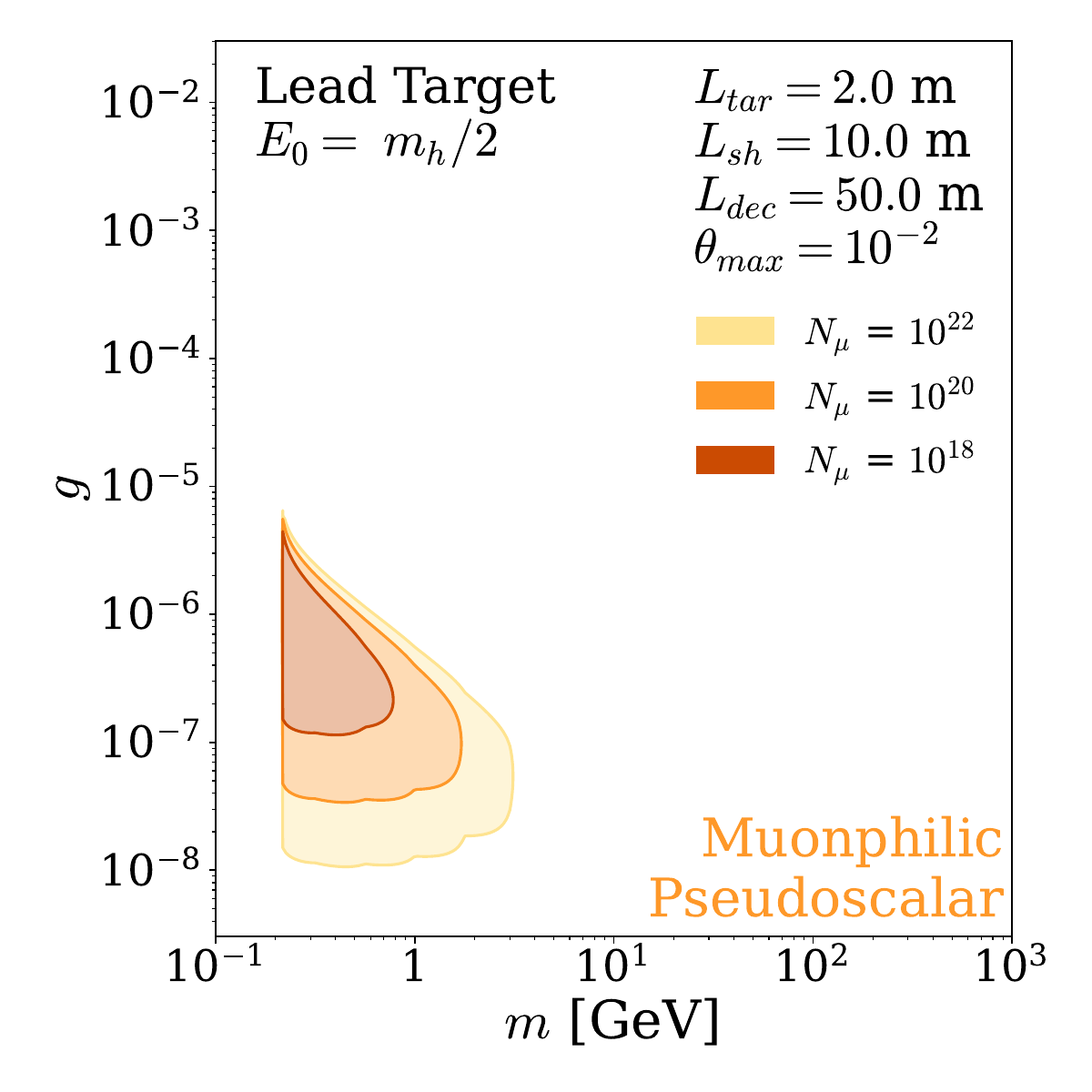}
         \caption{Pseudoscalar Production}
     \end{subfigure}
     \hfill
     \begin{subfigure}[b]{0.49\textwidth}
         \centering
         \includegraphics[width=.95\textwidth]{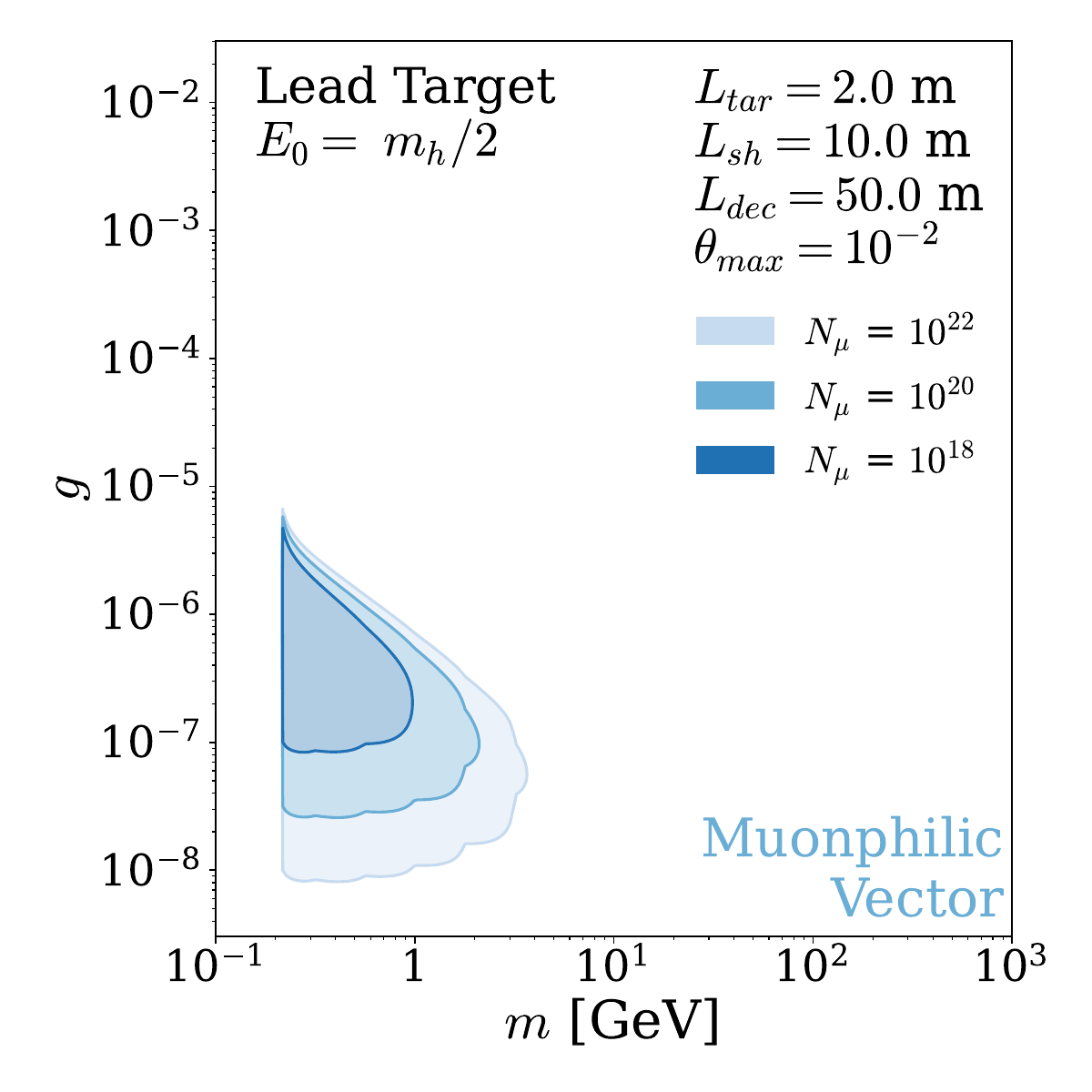}
         \caption{Vector Production}
     \end{subfigure}
     \begin{subfigure}[b]{0.49\textwidth}
         \centering
         \includegraphics[width=.95\textwidth]{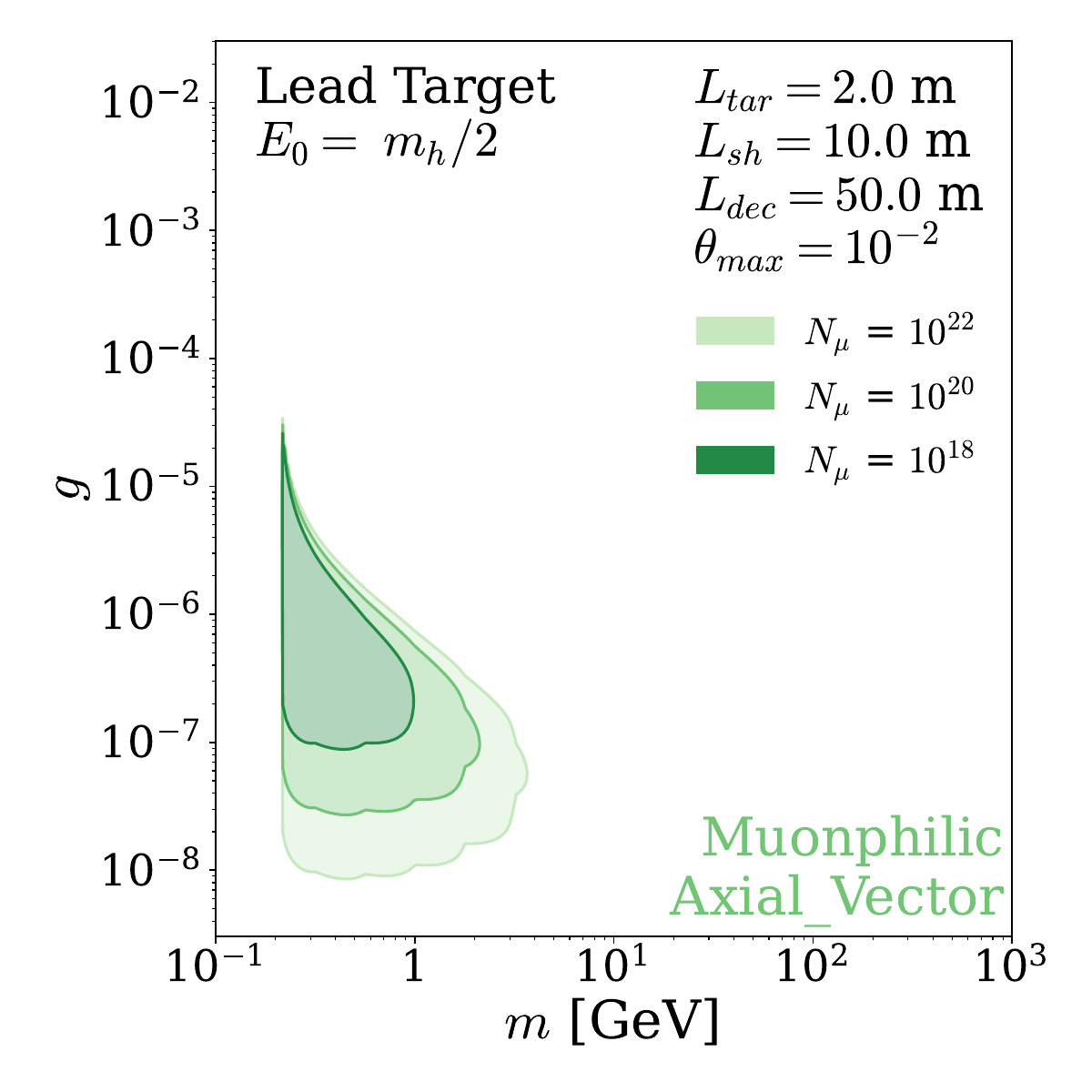}
         \caption{Axial Vector Production}
     \end{subfigure}
        \caption{The same as Fig.~\ref{fig:muphil10GeV} but with a beam energy of $m_h/2$. }
        \label{fig:muphil63}
\end{figure}

\begin{figure}[h!]
\centering
     \begin{subfigure}[b]{0.49\textwidth}
         \centering
         \includegraphics[width=.95\textwidth]{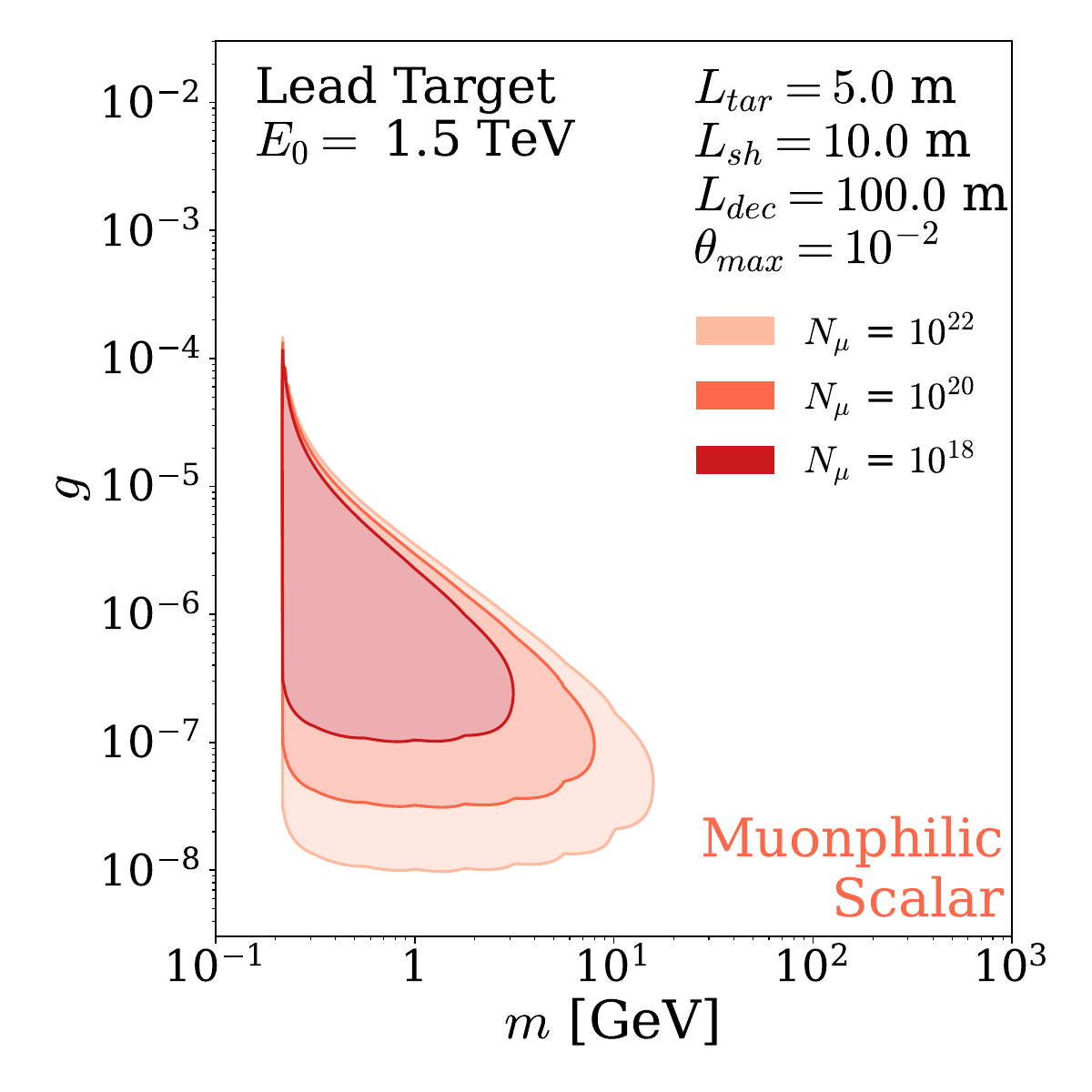}
         \caption{Scalar Production}
     \end{subfigure}
     \hfill
     \begin{subfigure}[b]{0.49\textwidth}
         \centering
         \includegraphics[width=.95\textwidth]{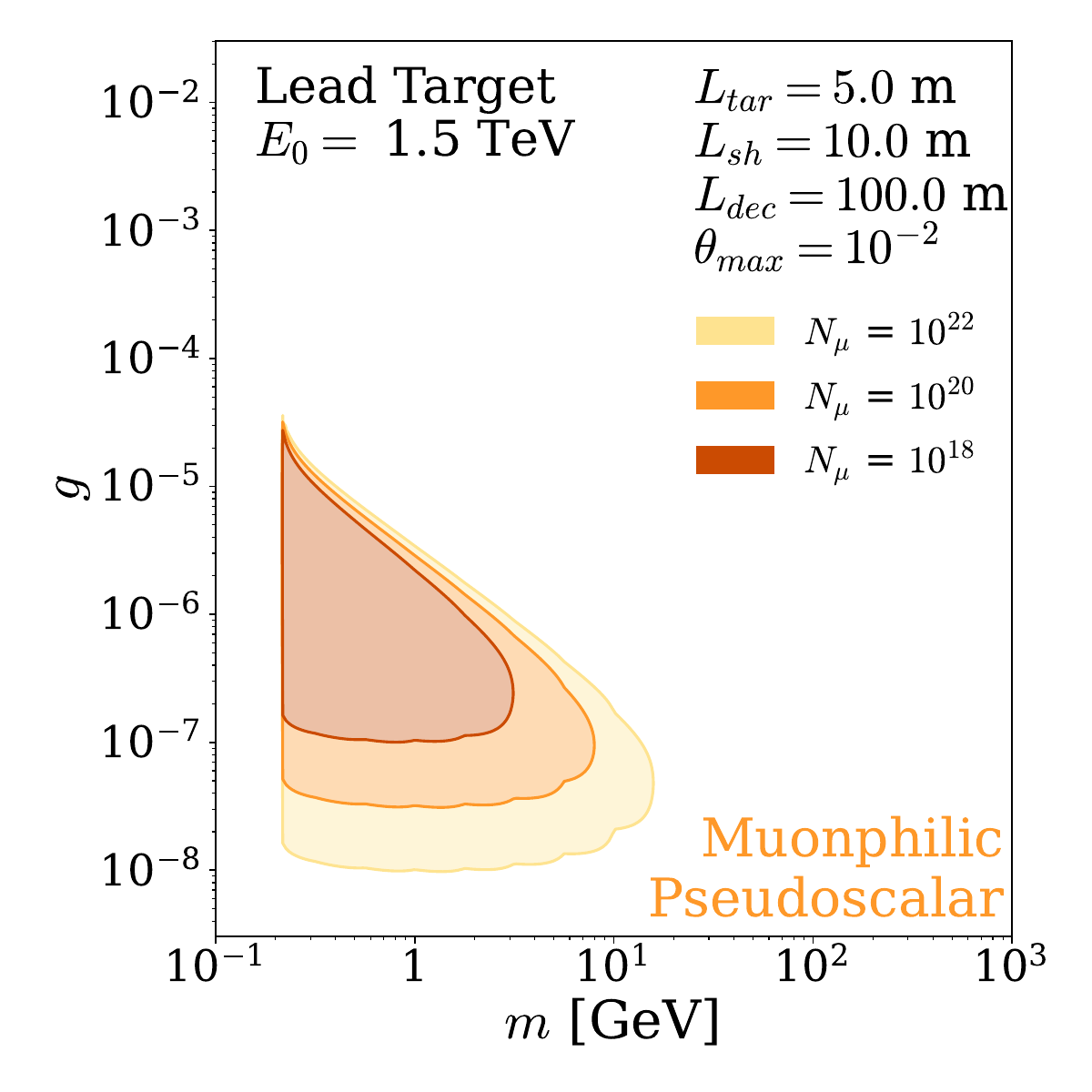}
         \caption{Pseudoscalar Production}
     \end{subfigure}
     \hfill
     \begin{subfigure}[b]{0.49\textwidth}
         \centering
         \includegraphics[width=.95\textwidth]{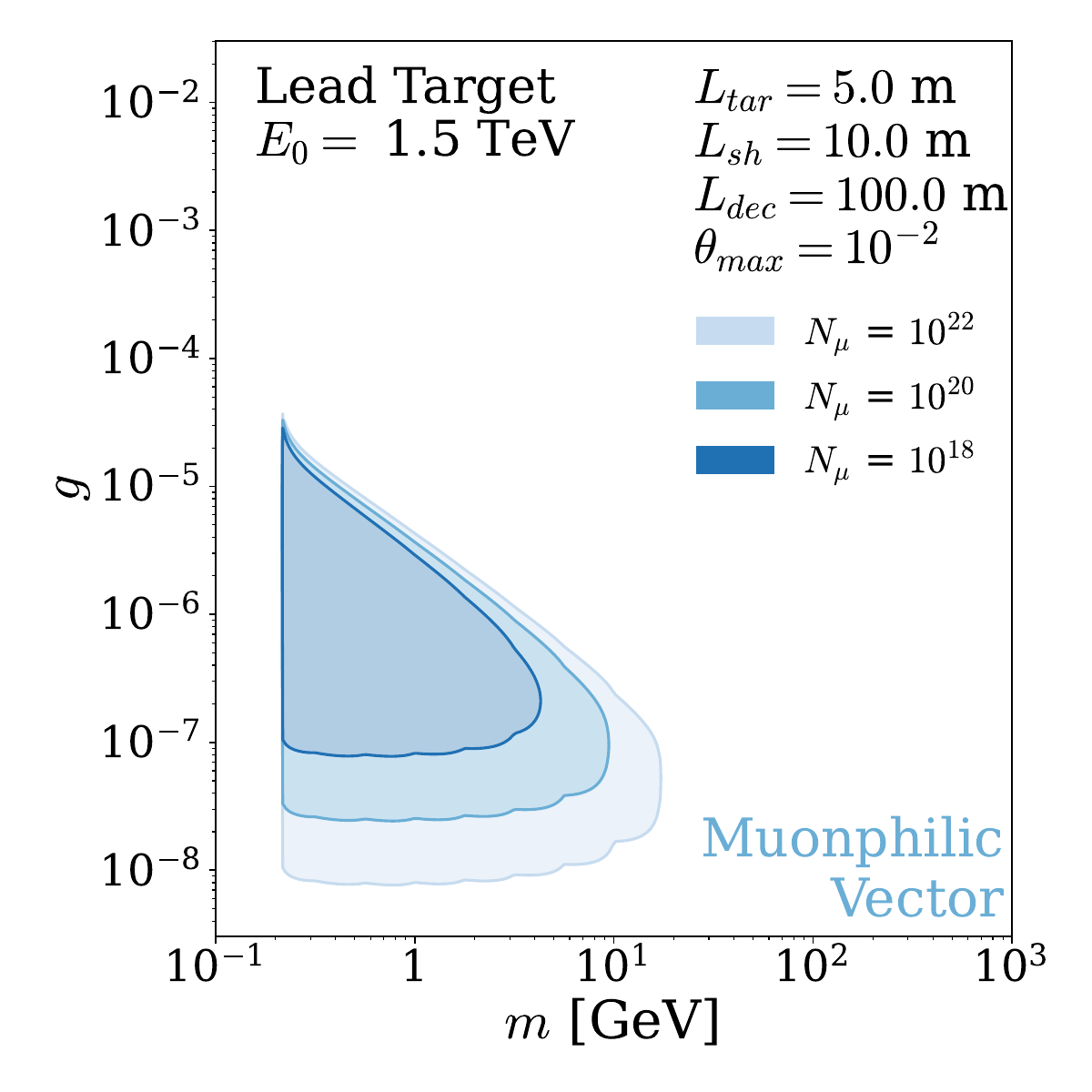}
         \caption{Vector Production}
     \end{subfigure}
     \begin{subfigure}[b]{0.49\textwidth}
         \centering
         \includegraphics[width=.95\textwidth]{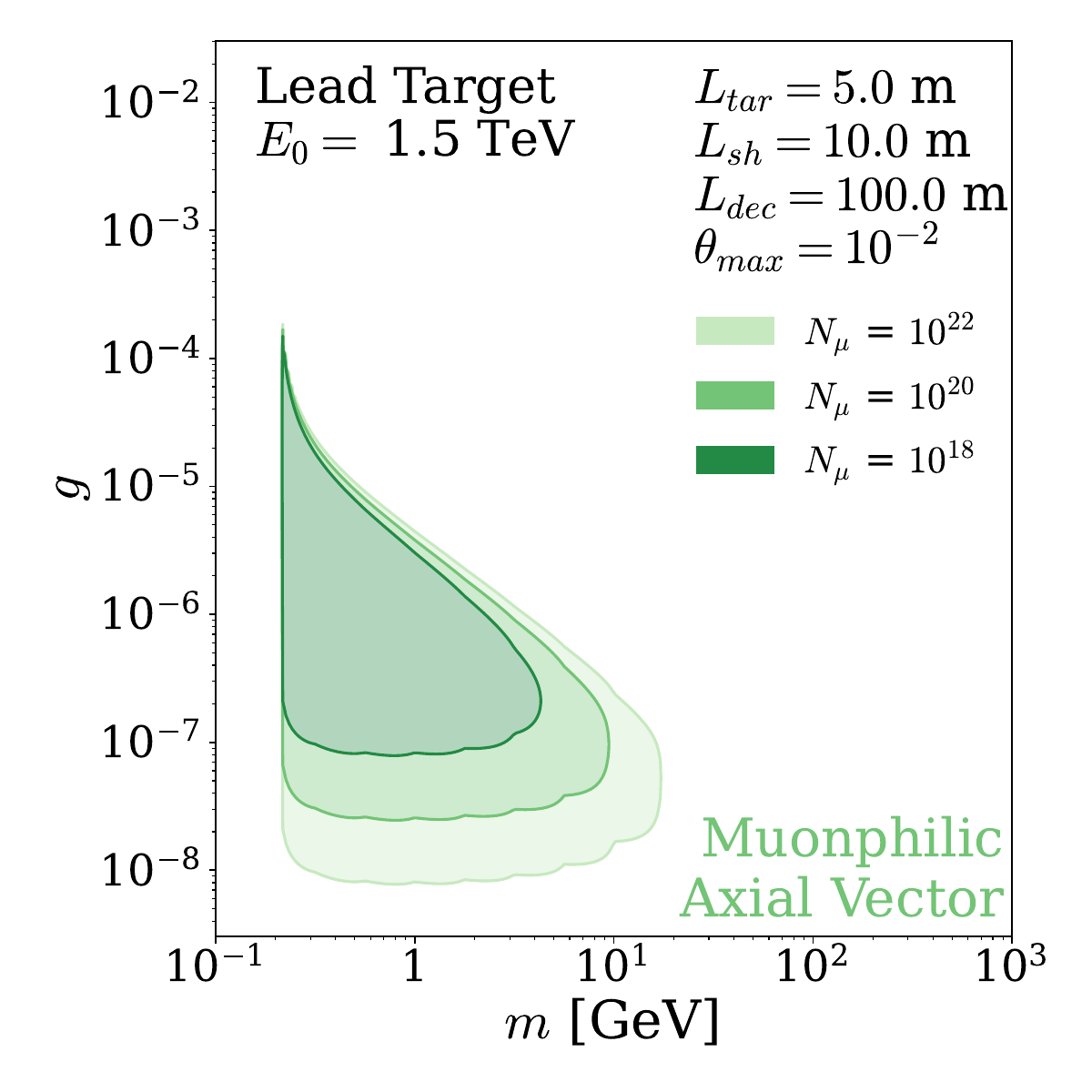}
         \caption{Axial Vector Production}
     \end{subfigure}
        \caption{The same as Fig.~\ref{fig:muphil10GeV} but with a beam energy of 1.5 TeV. }
        \label{fig:muphil1500}
\end{figure}

Here, we show the expected sensitivity to a muonphilic model for beam energies of $m_h/2$ (Fig.~\ref{fig:muphil63}) and 1.5 TeV (Fig.~\ref{fig:muphil1500}).

\clearpage

\clearpage

\subsection{Leptophilic}\label{app:leptophilic}

\begin{figure}
\centering
     \begin{subfigure}[b]{0.49\textwidth}
         \centering
         \includegraphics[width=.95\textwidth]{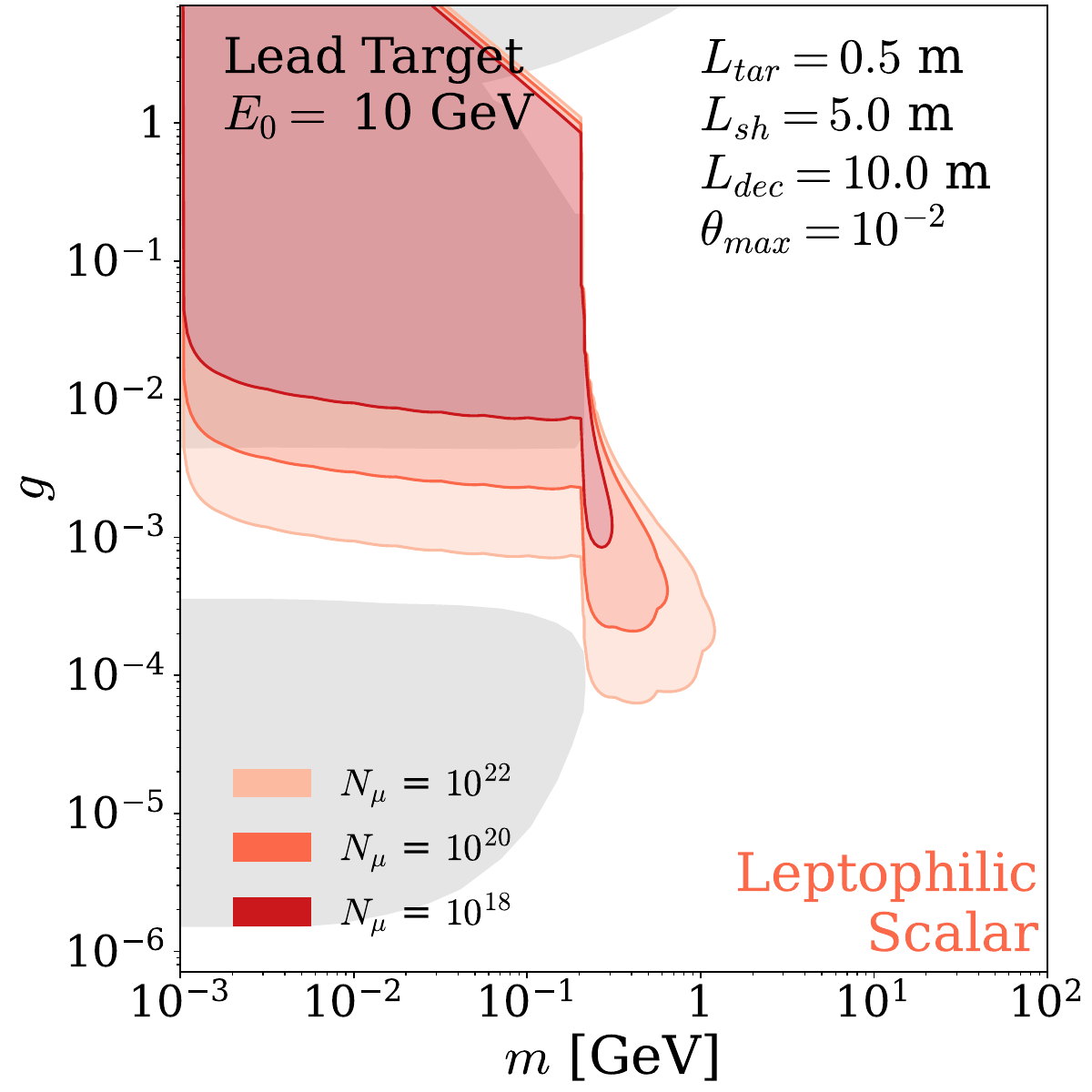}
         \caption{Scalar Production}
     \end{subfigure}
     \hfill
     \begin{subfigure}[b]{0.49\textwidth}
         \centering
         \includegraphics[width=.95\textwidth]{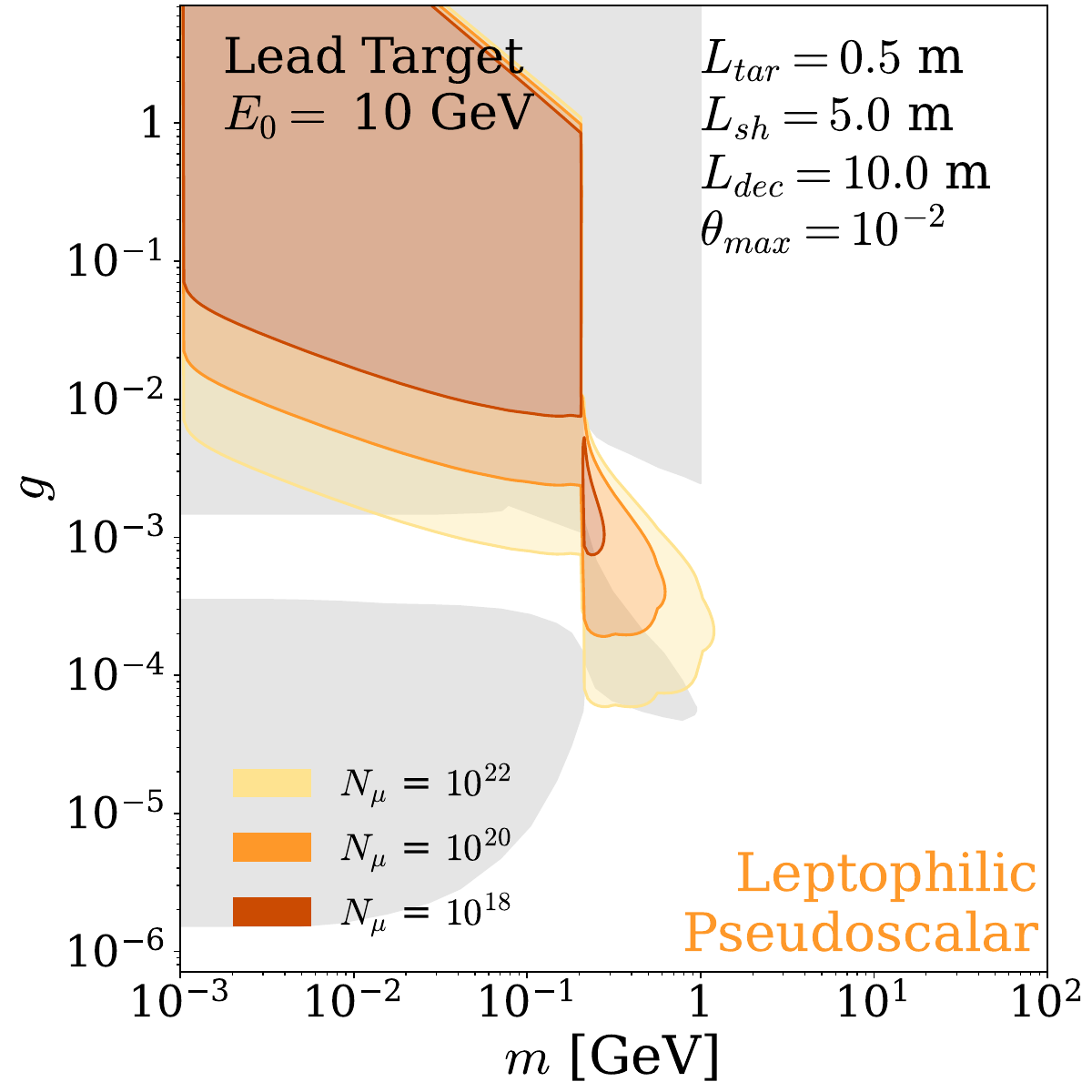}
         \caption{Pseudoscalar Production}
     \end{subfigure}
        \caption{The same as Fig.~\ref{fig:leptoHiggs} but with a beam energy of 10 GeV.}
        \label{fig:lepto10}
\end{figure}

\begin{figure}
\centering
     \begin{subfigure}[b]{0.49\textwidth}
         \centering
         \includegraphics[width=.95\textwidth]{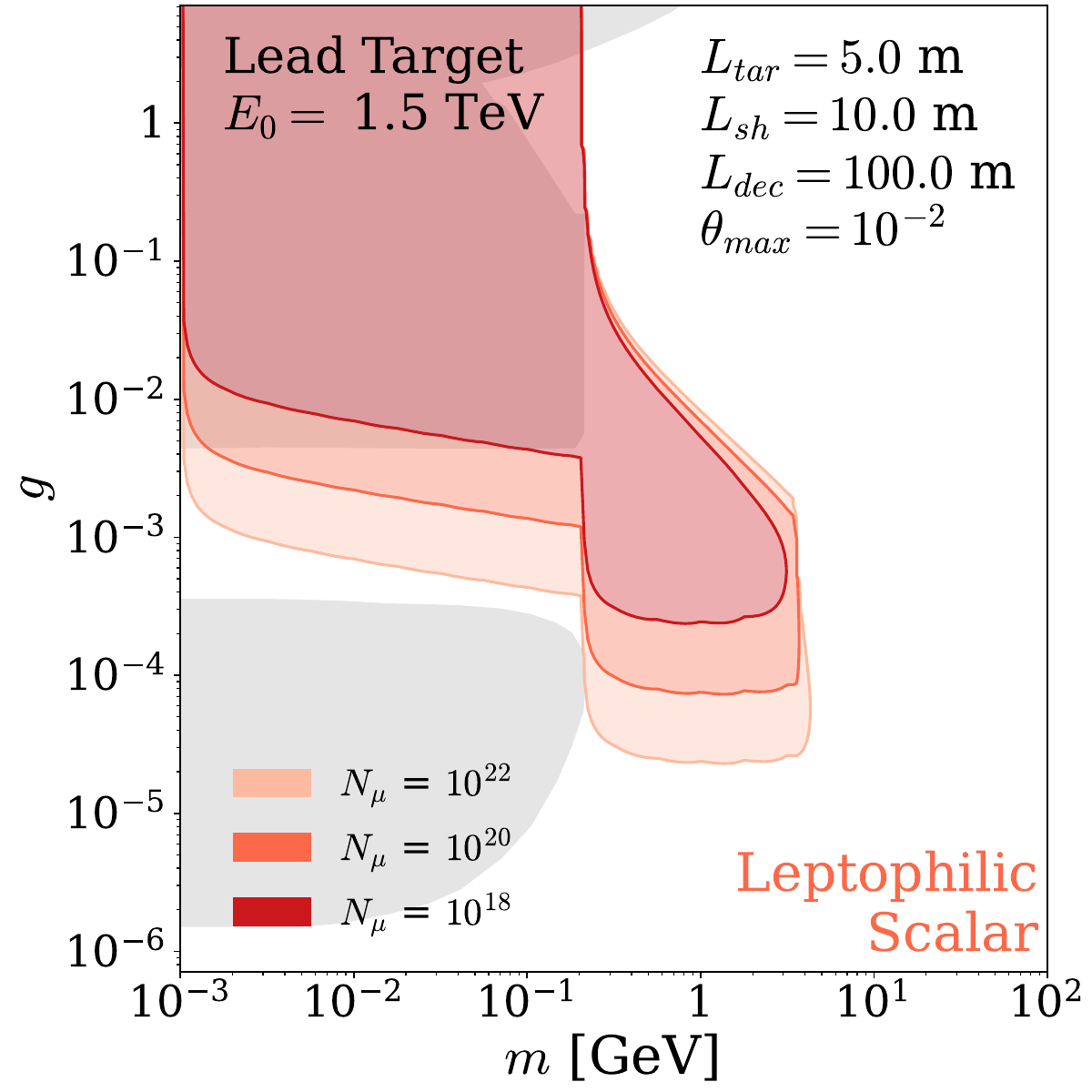}
         \caption{Scalar Production}
     \end{subfigure}
     \hfill
     \begin{subfigure}[b]{0.49\textwidth}
         \centering
         \includegraphics[width=.95\textwidth]{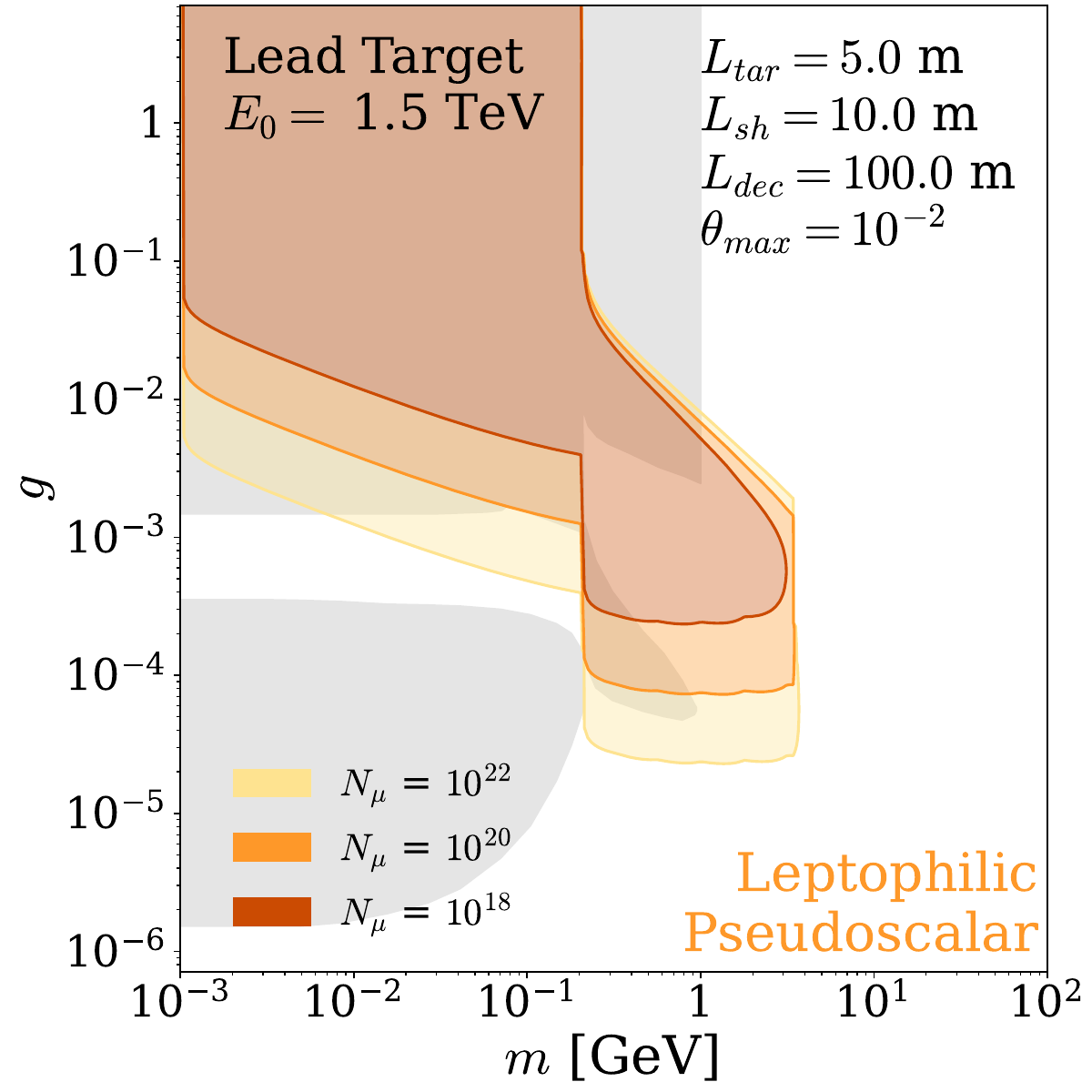}
         \caption{Pseudoscalar Production}
     \end{subfigure}
        \caption{The same as Fig.~\ref{fig:leptoHiggs} but with a beam energy of 1.5 TeV.}
        \label{fig:lepto1500}
\end{figure}

Here, we show the expected sensitivity to a leptophilic scalar or pseudoscalr for beam energies of 10 GeV (Fig.~\ref{fig:lepto10}) and 1.5 TeV (Fig.~\ref{fig:lepto1500}).

\clearpage

\subsection{$L_\mu - L_\tau$}\label{app:lmu}

\begin{figure}
\centering
     \begin{subfigure}[b]{0.49\textwidth}
         \centering
         \includegraphics[width=.95\textwidth]{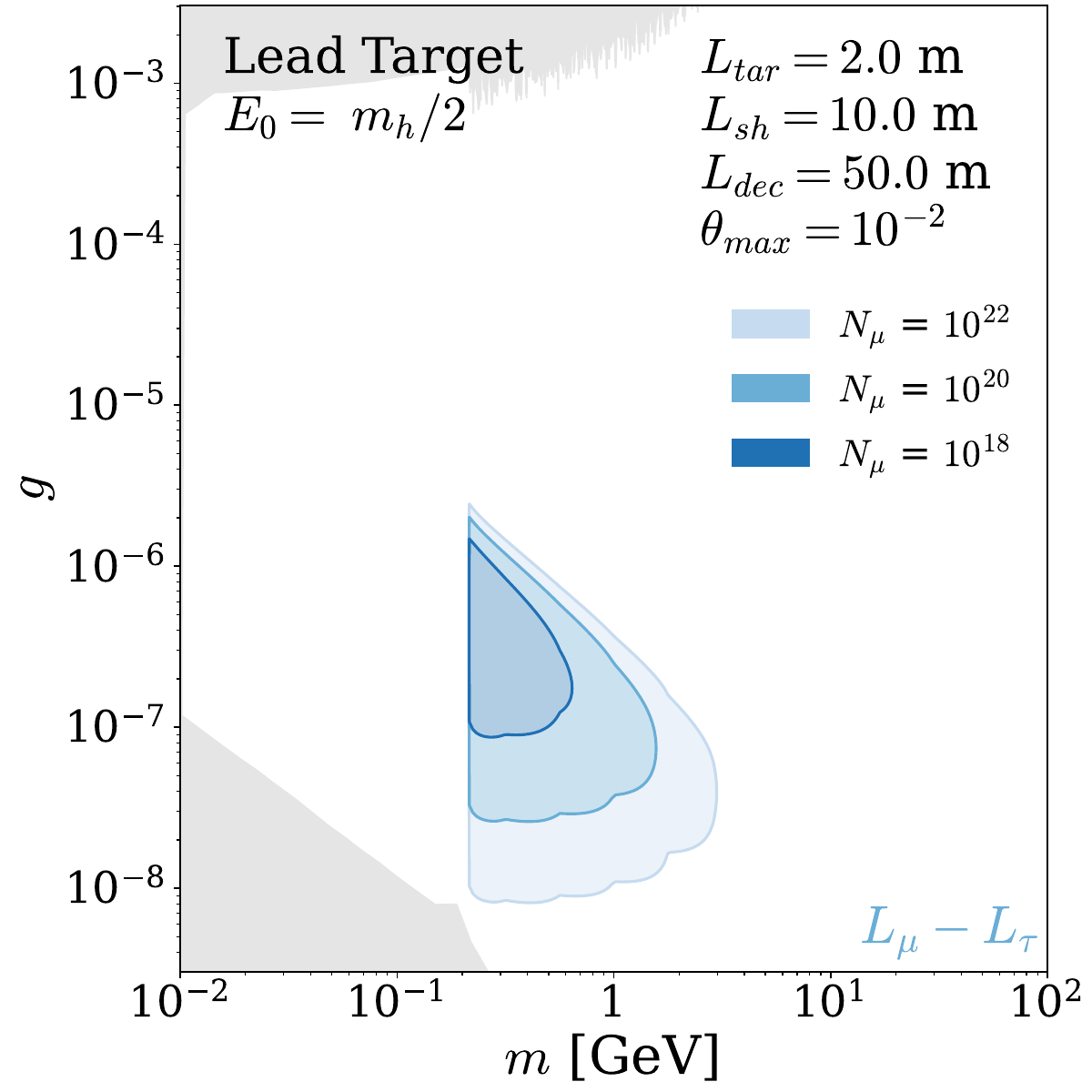}
         \caption{$m_h/2$ Beam}
         \label{fig:LMLT63}
     \end{subfigure}
     \hfill
     \begin{subfigure}[b]{0.49\textwidth}
         \centering
         \includegraphics[width=.95\textwidth]{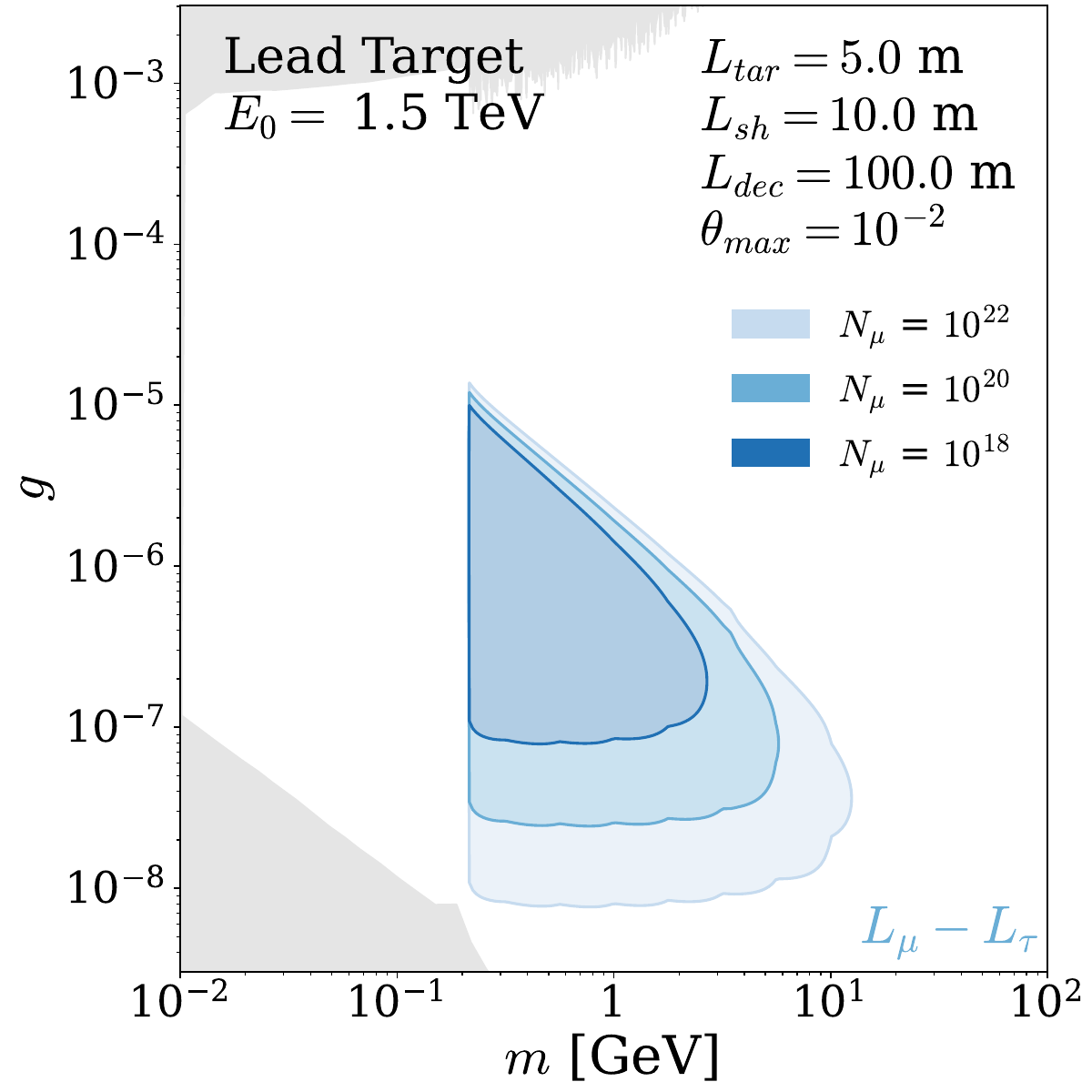}
         \caption{1.5 TeV Beam}
          \label{fig:LMLT1500}
     \end{subfigure}
        \caption{The same as Fig.~\ref{fig:gaugeLmuLtau} but with a beam energy of (a) $m_h/2$ and (b) 1500 GeV.}
        \label{fig:lmu_ltau1500}
\end{figure}

Here, we show the expected sensitivity to an $L_\mu - L_\tau$ vector boson for beam energies of $m_h/2$ (Fig.~\ref{fig:LMLT63}) and 1.5 TeV (Fig.~\ref{fig:LMLT1500}).

\clearpage
\subsection{Dark Photon}\label{app:darkphoton}

\begin{figure}
\centering
     \begin{subfigure}[b]{0.49\textwidth}
         \centering
         \includegraphics[width=.95\textwidth]{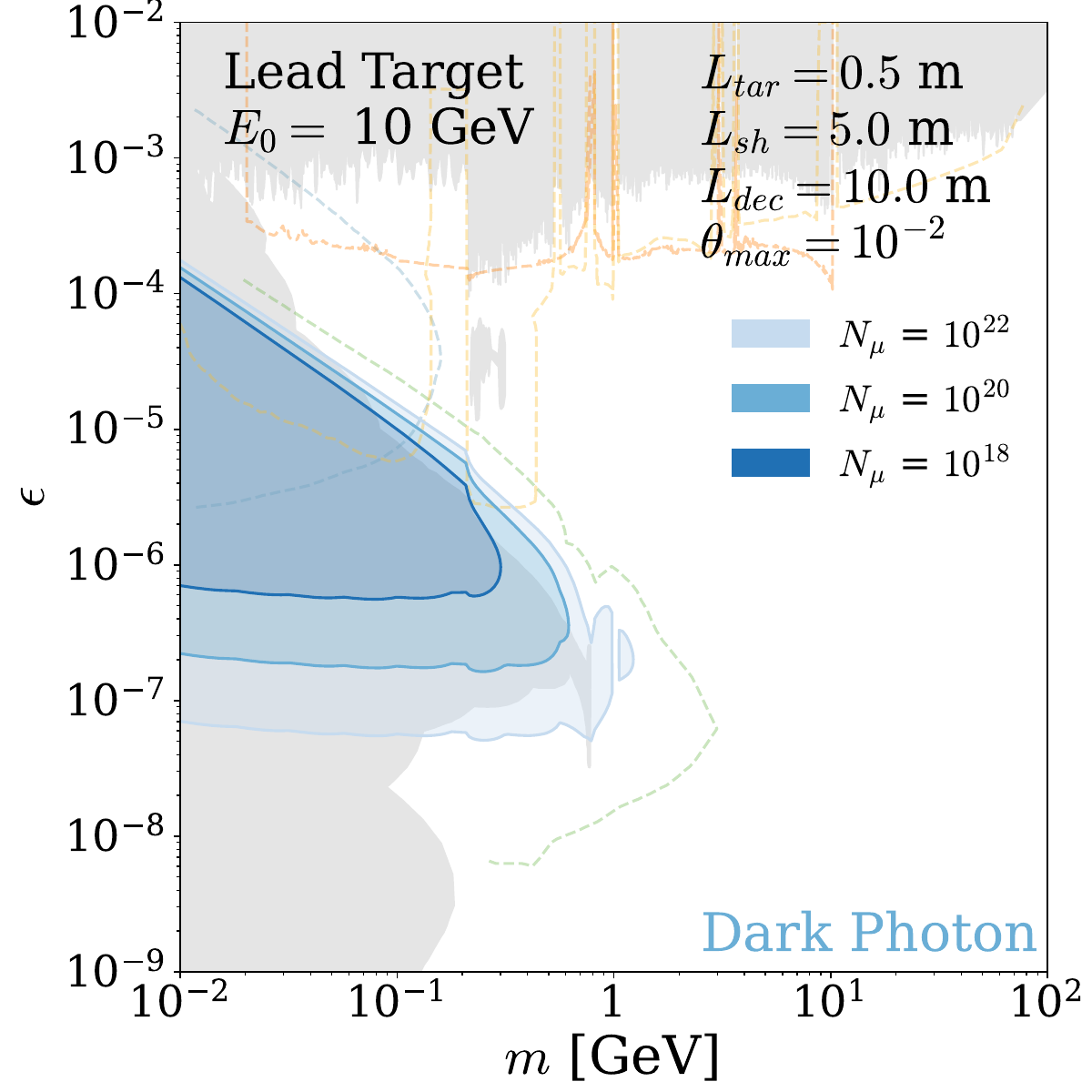}
         \caption{Dark photon limits at $E_0 = 10$ GeV.}
         \label{fig:DP10}
     \end{subfigure}
     \hfill
     \begin{subfigure}[b]{0.49\textwidth}
         \centering
         \includegraphics[width=.95\textwidth]{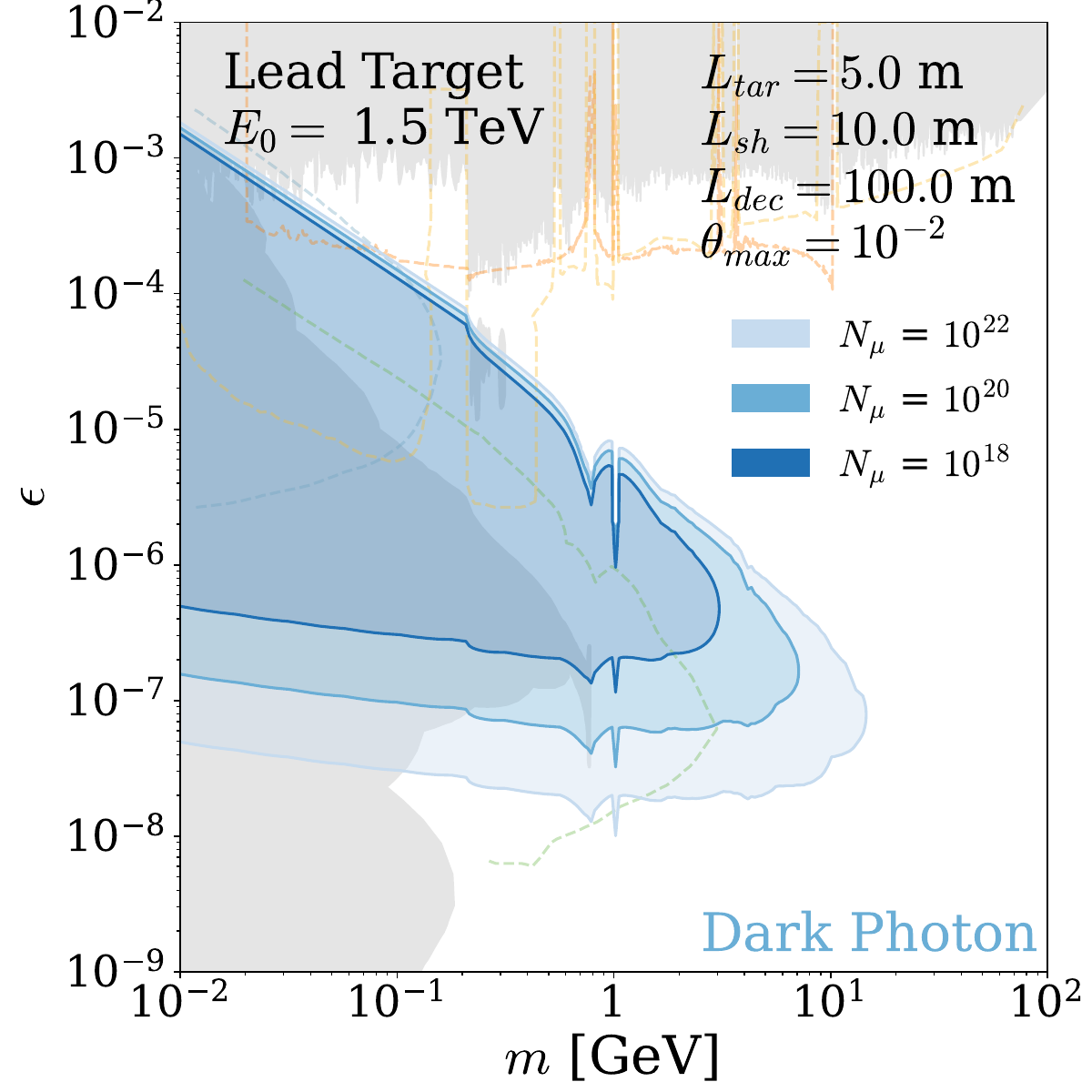}
         \caption{Dark photon limits at $E_0 = 1.5$ TeV.}
         \label{fig:DP1500}
     \end{subfigure}
        \caption{The same as Fig.~\ref{fig:darkPhoton} but with a beam energy of (a) 10 GeV and (b) 1500 GeV.}
        \label{fig:darkPhoton1500}
\end{figure}

Here, we show the expected sensitivity to a dark photon for beam energies of 10 GeV (Fig.~\ref{fig:DP10}) and 1.5 TeV (Fig.~\ref{fig:DP1500}).

\clearpage
\bibliographystyle{utphys}
\bibliography{beamdump}

\end{document}